\documentclass[zpreprint,zbstepj]{zeus_paper}

\usepackage[english]{babel}

\chardef\usc=95
\chardef\til=126
\catcode`\@=11 
\DeclareRobustCommand\xdotspace{\futurelet\@let@token\@xdotspace}
\def\@xdotspace{%
  \ifx\@let@token.\else
  \ifx\@let@token\bgroup.\else
  \ifx\@let@token\egroup.\else
  \ifx\@let@token\/.\else
  \ifx\@let@token\ .\else
  \ifx\@let@token~.\else
  \ifx\@let@token!.\else
  \ifx\@let@token,.\else
  \ifx\@let@token:.\else
  \ifx\@let@token;.\else
  \ifx\@let@token?.\else
  \ifx\@let@token/.\else
  \ifx\@let@token'.\else
  \ifx\@let@token).\else
  \ifx\@let@token-.\else
  \ifx\@let@token\@xobeysp.\else
  \ifx\@let@token\space.\else
  \ifx\@let@token\@sptoken.\else
   .\space
   \fi\fi\fi\fi\fi\fi\fi\fi\fi\fi\fi\fi\fi\fi\fi\fi\fi\fi}
\catcode`\@=12 

\newcommand{\stru}[2]{%
   \relax\ifmmode\hbox{\vrule height#1 depth#2 width0pt}%
   \else\vrule height#1 depth#2 width0pt\fi}

\newcommand{\Ronum}[1]{\uppercase\expandafter{\romannumeral#1}}
\newcommand{\ronum}[1]{\expandafter{\romannumeral#1}}
\DeclareRobustCommand{\LaTeXZ}{%
  \LaTeX\kern-.05em4\kern-.1em
  {\raisebox{-0.2ex}{$\scriptstyle\text{ZEUS}$}}\xspace}

\DeclareMathAlphabet{\mathbf}{OT1}{cmr}{bx}{sl}
\newcommand{\eVdist}{\kern-0.06667em}

\newcommand{\Gev}{{\text{Ge}\eVdist\text{V\/}}}

\newcommand{\pb}{\,\text{pb}}
\newcommand{\fb}{\,\text{fb}}

\newcommand{\Tesla}{\,\text{T}}

\newcommand{\slashfrac}[2]{%
  \raisebox{0.5ex}{\ensuremath #1}\kern-0.12em/\kern-0.08em
  \raisebox{-.8ex}{\ensuremath #2}}

\newcommand{\sqr}[3]{%
    {\vcenter{\hrule height.#3ex\hbox{\vrule width.#2ex height#1ex
     \kern#1ex\vrule width.#3ex}\hrule height.#2ex}}}

\catcode`\@=11 
\newcommand{\parenbar}{\mathpalette\p@renb@r}
\def\p@renb@r#1#2{\vbox{%
  \ifx#1\scriptscriptstyle \dimen@.7em\dimen@ii.2em\else
  \ifx#1\scriptstyle \dimen@.8em\dimen@ii.25em\else
  \dimen@1em\dimen@ii.4em\fi\fi \offinterlineskip
  \ialign{\hfill##\hfill\cr
    \vbox{\hrule width\dimen@ii}\cr
    \noalign{\vskip-.3ex}%
    \hbox to\dimen@{$\mathchar300\hfil\mathchar301$}\cr
    \noalign{\vskip-.3ex}%
    $#1#2$\cr}}}
\catcode`\@=12 

\newcommand{\IP}{{\rm I$\kern-0.01667em$P}\xspace}

\newcommand{\F}{{\cal F}}

\mathchardef\qsm=63
\mathchardef\pls=43
\mathchardef\mns=512
\mathchardef\plm=518
\mathchardef\eql=61
\mathchardef\smallleft=300
\mathchardef\smallright=301
\mathchardef\les=316
\mathchardef\gre=318
\mathchardef\leq=532
\mathchardef\grq=533

\catcode`\@=11 
\newcounter{pict@width}
\newcounter{pict@height}
\newlength{\pict@scale}
\newcommand{\psfigadd}[4]{%
\setcounter{pict@width}{1*\ratio{#2+\pict@scale/2}{\pict@scale}}
\setcounter{pict@height}{1*\ratio{#3+\pict@scale/2}{\pict@scale}}
\setlength{\unitlength}{\pict@scale}
\hbox to #2{\hspace{-\fill}\begin{picture}(\thepict@width,\thepict@height)
\put(0,0){\psfig{figure=#1,width=#2,height=#3,clip=}}
\SetScale{0.283466457}
\SetWidth{1.763889}
{#4}
\end{picture}}
}
\newcounter{pict@widthfst}
\newcounter{pict@widthscd}
\newcounter{pict@widthtot}
\newcommand{\psfigaddtwo}[7]{%
\setcounter{pict@widthfst}{1*\ratio{#2+\pict@scale/2}{\pict@scale}}
\setcounter{pict@widthscd}{1*\ratio{#2+#4+\pict@scale/2}{\pict@scale}}
\setcounter{pict@widthtot}{1*\ratio{#2+#4+#6+\pict@scale/2}{\pict@scale}}
\setcounter{pict@height}{1*\ratio{#3+\pict@scale/2}{\pict@scale}}
\setlength{\unitlength}{\pict@scale}
\hbox{\hspace{-\fill}\begin{picture}(\thepict@widthtot,\thepict@height)
\put(0,0){\psfig{figure=#1,width=#2,height=#3,clip=}}
\put(\thepict@widthscd,0){\psfig{figure=#5,width=#6,height=#3,clip=}}
\SetScale{0.283466457}
\SetWidth{1.763889}
{#7}
\end{picture}}
}
\newcommand{\psfigror}[4]{%
\setcounter{pict@width}{1*\ratio{#2+\pict@scale/2}{\pict@scale}}
\setcounter{pict@height}{1*\ratio{#3+\pict@scale/2}{\pict@scale}}
\setlength{\unitlength}{\pict@scale}
\hbox{\begin{picture}(\thepict@width,\thepict@height)
\put(0,\thepict@height){\psfig{figure=#1,width=#3,height=#2,clip=,angle=270}}
\SetScale{0.283466457}
\SetWidth{1.763889}
{#4}
\end{picture}}
}
\newcommand{\psfigrol}[4]{%
\setcounter{pict@width}{1*\ratio{#2+\pict@scale/2}{\pict@scale}}
\setcounter{pict@height}{1*\ratio{#3+\pict@scale/2}{\pict@scale}}
\setlength{\unitlength}{\pict@scale}
\hbox{\begin{picture}(\thepict@width,\thepict@height)
\put(0,0){\psfig{figure=#1,width=#3,height=#2,clip=,angle=90}}
\SetScale{0.283466457}
\SetWidth{1.763889}
{#4}
\end{picture}}
}
\catcode`\@=12 
\newlength\listtextwidth

\catcode`\@=11 
\newlength{\@tabfninsert}
\newlength{\@tabfnwidth}
\newcommand{\tabfootnote}[2]{%
  \setlength{\@tabfninsert}{0.8em}
  \setlength{\@tabfnwidth}{\textwidth}
  \addtolength{\@tabfnwidth}{-\@tabfninsert}
  \addtolength{\@tabfnwidth}{-0.4em}
  \noindent\makebox[\@tabfninsert][r]{\footnotesize$^{#1}$\hfil}\hfill%
  \parbox[t]{\@tabfnwidth}{\footnotesize #2\hfill}}
\catcode`\@=12 

\def\JHEP{JHEP}

\def\etjet{E_T^{\rm jet}}
\def\etajet{\eta^{\rm jet}}

\def\etaJ{\eta^{\rm jet1}}
\def\etajj{\eta^{\rm jet2}}
\def\etajjj{\eta^{\rm jet3}}
\def\etJ{E_T^{\rm jet1}}
\def\etjj{E_T^{\rm jet2}}
\def\etjjj{E_T^{\rm jet3}}

\def\etaphi{\eta-\phi}

\def\etar{-1<\etajet<2.5}

\def\qg2{$\q2>125$~\g2}

\def\sq2{d\sigma/d\q2}
 
\def\q2{Q^2}

\def\cgh{\cos\gamma_h}
\def\pb1{pb$^{-1}$}
\def\fb1{fb$^{-1}$}
\def\gp{\gamma p}

\def\g2{GeV$^2$}
\def\gf2{GeV$\,^2$}

\def\F2g{F_2^{\gamma}}
\def\f2gv{F_2^{\gamma^*}}

\def\rr1{R=1}
\def\r7{R=0.7}
\def\R71{R=0.7\ {\rm and}\ 1}

\def\m3j{M^{\rm 3j}}

\def\kt{k_T}

\def\lq2{\log_{10}(\q2)}

\def\etjb{E^{\rm jet}_{T,{\rm B}}}
\def\etjbj{E^{\rm jet1}_{T,{\rm B}}}

\def\etajb{\eta^{\rm jet}_{\rm B}}
\def\etajmax{\eta^{\rm jet}_{\rm max}}

\def\xo{x_{\gamma}^{\rm obs}}

\def\ele{e^+e^-}

\def\qq{q\bar q}

\def\colab#1{#1 Coll.}

\def\z0{Z^0}
\def\mz{M_Z}

\def\ee{\'e}

\def\oo{\'o}

\def\ene{\~n}

\def\as{\alpha_s}
\def\oalphas2{{\cal O}(\alpha\as^2)}

\def\oass{{\cal O}(\as^2)}
\def\oasss{{\cal O}(\as^3)}
\def\asz{\as(\mz)}

\def\p2{P^2}

\def\mr2{\mu_R^2}
\def\mf2{\mu_F^2}

\def\etal{et al.}

\def\bet0#1#2#3#4#5#6{\beta_0 = #1\pm #2\ {\rm (stat.)}\ ^{+#4}_{-#3}\ {\rm (exp.)}\ ^{+#6}_{-#5}\ {\rm (th.)}}

\def\th{\theta_H}
\def\a34{\alpha_{23}}
\def\pksw{\beta_{\rm KSW}}

\def\figdir{./}

\begin{document}
\prepnum{{DESY--08--100}}

\title{
            Angular correlations in three-jet events
                 in {\boldmath $ep$} collisions at HERA
}                                                       

\author{ZEUS Collaboration}
\date{July 2008}

\abstract{
Three-jet production in deep inelastic $ep$ scattering and
photoproduction was investigated with the ZEUS detector
at HERA using an integrated luminosity of $127$~\pb1. Measurements of
differential cross sections are presented as functions of angular
correlations between the three jets in the final state and the
proton-beam direction. These correlations provide a stringent test of
perturbative QCD and show sensitivity to the contributions from
different colour configurations. Fixed-order perturbative QCD
calculations assuming the values of the colour factors $C_F$, $C_A$
and $T_F$ as derived from a variety of gauge groups were compared to
the measurements to study the underlying gauge group symmetry. The
measured angular correlations in the deep inelastic $ep$ scattering
and photoproduction regimes are consistent with the admixture of
colour configurations as predicted by SU(3) and disfavour other
symmetry groups, such as SU($N$) in the limit of large $N$.
}

\makezeustitle

\def\3{\ss}

\pagenumbering{Roman}

\begin{center}
{                      \Large  The ZEUS Collaboration              }
\end{center}

  S.~Chekanov,
  M.~Derrick,
  S.~Magill,
  B.~Musgrave,
  D.~Nicholass$^{   1}$,
  \mbox{J.~Repond},
  R.~Yoshida\\
 {\it Argonne National Laboratory, Argonne, Illinois 60439-4815,
   USA}~$^{n}$
\par \filbreak

  M.C.K.~Mattingly \\
 {\it Andrews University, Berrien Springs, Michigan 49104-0380, USA}
\par \filbreak

  P.~Antonioli,
  G.~Bari,
  L.~Bellagamba,
  D.~Boscherini,
  A.~Bruni,
  G.~Bruni,
  F.~Cindolo,
  M.~Corradi,
\mbox{G.~Iacobucci},
  A.~Margotti,
  R.~Nania,
  A.~Polini\\
  {\it INFN Bologna, Bologna, Italy}~$^{e}$
% ...

\newpage

\enlargethispage{5cm}

$^{\    1}$ also affiliated with University College London,
United Kingdom\\
% ...
$^{\    5}$ now at University of Bonn, Germany \\
% ...

\newpage

\begin{tabular}[h]{rp{14cm}}

$^{a}$ &  supported by the Natural Sciences and Engineering Research
Council of Canada (NSERC) \\
$^{b}$ &  supported by the German Federal Ministry for Education and
Research (BMBF), under contract numbers 05 HZ6PDA, 05 HZ6GUA, 05
HZ6VFA and 05 HZ4KHA\\
$^{c}$ &  supported in part by the MINERVA Gesellschaft f\"ur
Forschung GmbH, the Israel Science Foundation (grant no. 293/02-11.2)
and the U.S.-Israel Binational Science Foundation \\
$^{d}$ &  supported by the Israel Science Foundation\\
$^{e}$ &  supported by the Italian National Institute for Nuclear
Physics (INFN) \\
$^{f}$ &  supported by the Japanese Ministry of Education, Culture,
Sports, Science and Technology (MEXT) and its grants for Scientific
Research\\
$^{g}$ &  supported by the Korean Ministry of Education and Korea
Science and Engineering Foundation\\
$^{h}$ &  supported by the Netherlands Foundation for Research on
Matter (FOM)\\
$^{i}$ &  supported by the Polish State Committee for Scientific
Research, project no. DESY/256/2006 - 154/DES/2006/03\\
$^{j}$ &  partially supported by the German Federal Ministry for
Education and Research (BMBF)\\
$^{k}$ &  supported by RF Presidential grant N 8122.2006.2 for the
leading scientific schools and by the Russian Ministry of Education
and Science through its grant for Scientific Research on High Energy
Physics\\
$^{l}$ &  supported by the Spanish Ministry of Education and Science
through funds provided by CICYT\\
$^{m}$ &  supported by the Science and Technology Facilities Council,
UK\\
$^{n}$ &  supported by the US Department of Energy\\
$^{o}$ &  supported by the US National Science Foundation. Any
opinion, findings and conclusions or recommendations expressed in this
material are those of the authors and do not necessarily reflect the
views of the National Science Foundation.\\
$^{p}$ &  supported by the Polish Ministry of Science and Higher
Education as a scientific project (2006-2008)\\
$^{q}$ &  supported by FNRS and its associated funds (IISN and FRIA)
and by an Inter-University Attraction Poles Programme subsidised by
the Belgian Federal Science Policy Office\\
$^{r}$ &  supported by the Malaysian Ministry of Science, Technology
and Innovation/Akademi Sains Malaysia grant SAGA 66-02-03-0048\\

\end{tabular}

\newpage

\pagenumbering{arabic} 
\pagestyle{plain}

\section{Introduction}

Quantum chromodynamics (QCD) is based on the non-Abelian group SU(3)
which induces the self-coupling of the gluons. Investigations of the
triple-gluon vertex (TGV) were carried out at 
LEP~\cite{pl:b284:151,*zfp:c76:1,*epj:c27:1,*zfp:c59:357,*pl:b414:401,*pl:b449:383,*pl:b248:227,*zfp:c49:49,*zfp:c65:367,*epj:c20:601,pl:b255:466}
using angular correlations in four-jet events from $\z0$ hadronic
decays. At HERA, the effects of the different colour configurations
arising from the underlying gauge structure can be studied in a clean
way in three-jet production in neutral current (NC) deep inelastic
scattering (DIS) and photoproduction ($\gp$).

Neutral current DIS at high $\q2$ ($\q2\gg\Lambda_{\rm QCD}^2$, where
$\q2$ is the virtuality of the exchanged photon) up to leading order
(LO) in the strong coupling constant, $\as$, proceeds as in the
quark-parton model ($Vq\rightarrow q$, where $V=\gamma^*$ or $\z0$) or
via the boson-gluon fusion ($Vg\rightarrow\qq$) and QCD-Compton
($Vq\rightarrow qg$) processes. Photoproduction is studied at HERA by
means of $ep$ scattering at low four-momentum transfers 
($\q2\approx 0$). In $\gp$ reactions, two types of QCD processes
contribute to jet production at 
LO~\cite{pl:b79:83,*np:b166:413,*pr:d21:54,*zfp:c6:241,proc:hera:1987:331,*prl:61:275,*prl:61:682,*pr:d39:169,*zfp:c42:657,*pr:d40:2844}:
either the photon interacts directly with a parton in the proton (the
direct process) or the photon acts as a source of partons which
scatter off those in the proton (the resolved process). 

\begin{figure}[p]
\vfill
\setlength{\unitlength}{1.0cm}
\begin{picture} (18.0,15.0)
\put (2.8,9.0){\epsfig{figure=\figdir 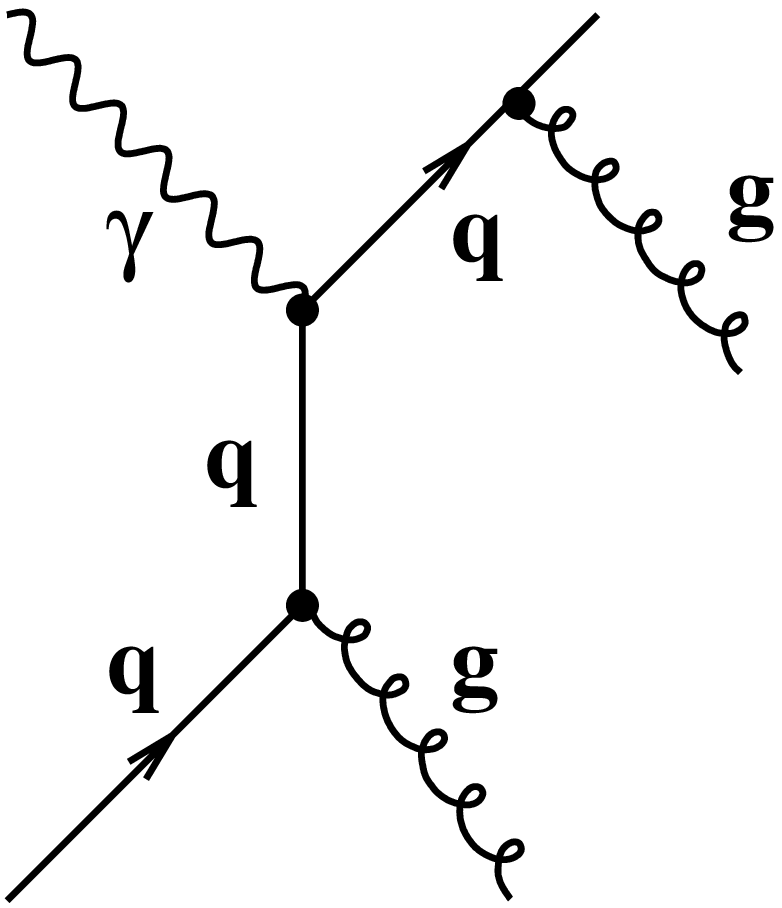,width=5cm}}
\put (10.5,8.0){\epsfig{figure=\figdir 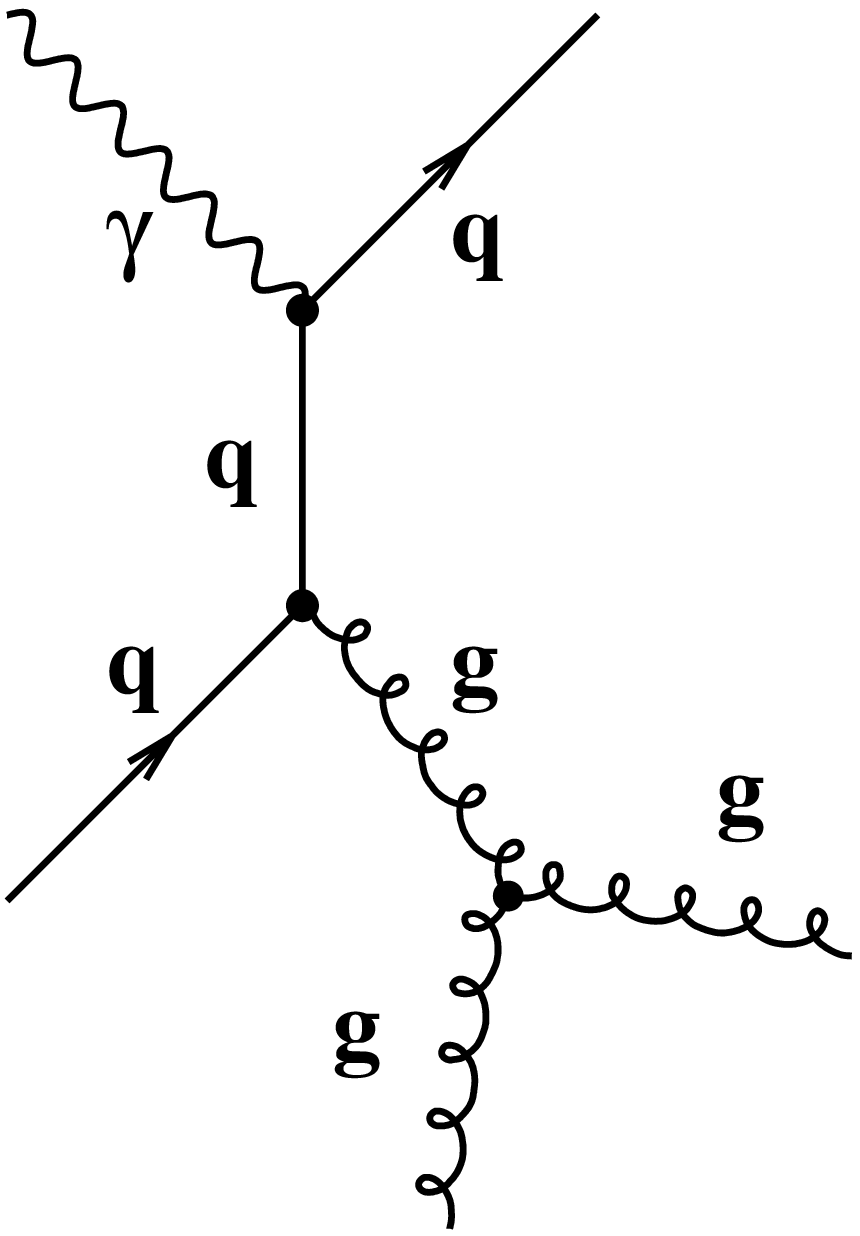,width=5cm}}
\put (3.0,0.0){\epsfig{figure=\figdir 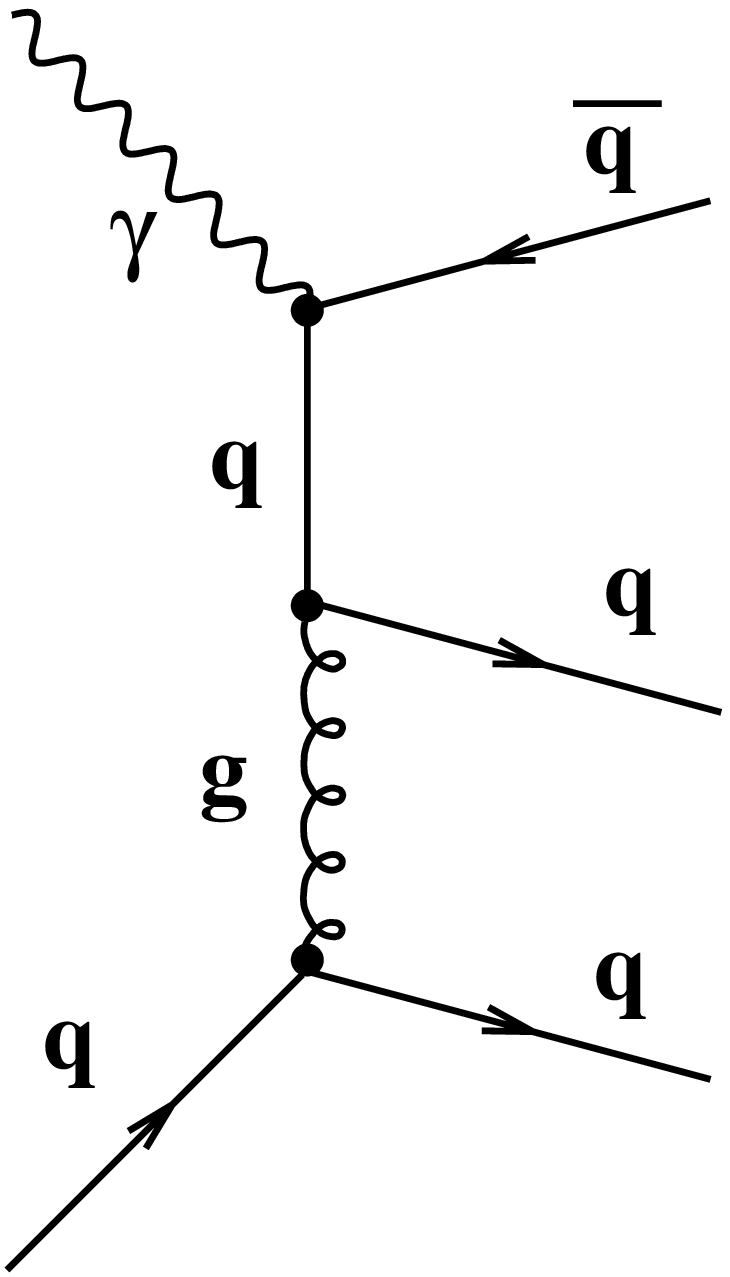,width=4cm}}
\put (10.5,0.5){\epsfig{figure=\figdir 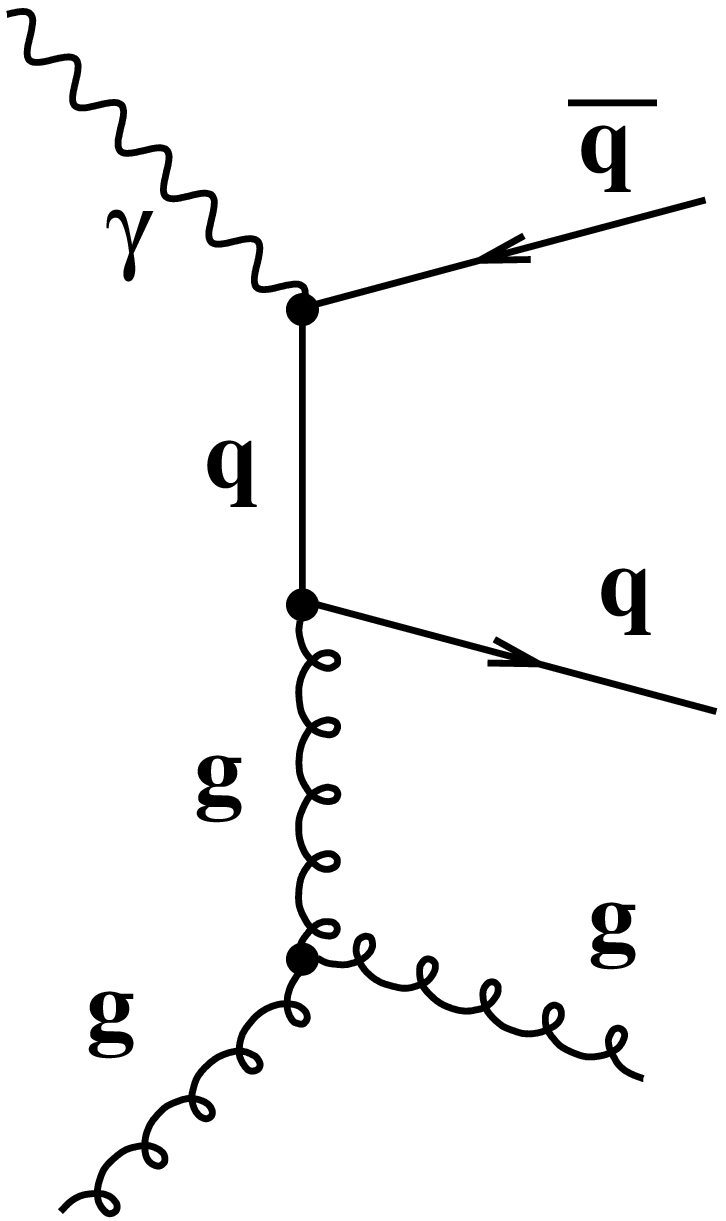,width=4cm}}
\put (4.5,15.0){\bf\small (A)}
\put (12.0,15.0){\bf\small (B)}
\put (4.5,6.7){\bf\small (C)}
\put (12.0,6.7){\bf\small (D)}
\end{picture}
\caption
{\it 
Examples of diagrams for the photoproduction of three-jet events
through direct-photon processes and in NC DIS three-jet events in each
colour configuration:
(A) double-gluon bremsstrahlung from a quark line;
(B) the splitting of a virtual gluon into a pair of final-state
gluons;
(C) the production of a $\qq$ pair through the exchange of a
virtual gluon emitted by an incoming quark;
(D) the production of a $\qq$ pair through the exchange of a 
    virtual gluon arising from the splitting of an incoming gluon.
}
\label{fig1}
\vfill
\end{figure}

A subset of resolved subprocesses with two jets in the final state are
described by diagrams with a TGV; however, such events are difficult
to distinguish from two-jet events without such a contribution. 
Three-jet final states in direct $\gp$ processes also contain
contributions from TGVs and are easier to identify. Since three-jet
production in NC DIS proceeds via the same diagrams as in direct $\gp$,
such processes can also be used to investigate the underlying gauge
symmetry. Examples of diagrams contributing to four colour
configurations are shown in Fig.~\ref{fig1}:
(A) double-gluon bremsstrahlung from a quark line,
(B) the splitting of a virtual gluon into a pair of final-state gluons,
(C) the production of a $\qq$ pair through the exchange of a virtual
gluon emitted by an incoming quark, and 
(D) the production of a $\qq$ pair through the exchange of a virtual
gluon arising from the splitting of an incoming gluon.

Other possible diagrams and interferences correspond to one of
the four configurations. The production rate of all
contributions is proportional to the so-called colour factors, $C_F$,
$C_A$ and $T_F$, which are a physical manifestation of the underlying
group structure. For QCD, these factors represent the relative
strengths of the processes $q\rightarrow qg$, $g\rightarrow gg$ and
$g\rightarrow \qq$. The contributions of the diagrams of
Fig.~\ref{fig1} are proportional to $C_F^2$, $C_FC_A$, $C_FT_F$ and
$T_FC_A$, respectively, independently of the underlying gauge
symmetry.

Three-jet cross sections were previously measured in
$\gp$~\cite{pl:b443:394,*np:b792:1} and in NC
DIS~\cite{epj:c44:183,pl:b515:17}. The shape of the measured cross
sections was well reproduced by perturbative QCD (pQCD)
calculations and a value of $\as$ was extracted~\cite{epj:c44:183}. In
this paper, measurements of angular correlations in three-jet events
in $\gp$ and NC DIS are presented. The comparison between the
measurements and fixed-order $\oass$ and $\oasss$ perturbative
calculations based on different colour configurations provides a
stringent test of pQCD predictions directly beyond LO and gives
insight into the underlying group symmetry. Phase-space regions where
the angular correlations show potential sensitivity to the presence of
the TGV were identified.

\section{Theoretical framework}

The dynamics of a gauge theory such as QCD are completely defined by
the commutation relations between its group generators $T^i$,

$$[T^i,T^j]=i \sum_k f^{ijk} \cdot T^k,$$

where $f^{ijk}$ are the structure constants. The generators $T^i$ can
be represented as matrices. In perturbative calculations, the average
(sum) over all possible colour configurations in the initial (final)
states leads to the appearance of combinatoric factors $C_F$, $C_A$
and $T_F$, which are defined by the relations

$$\sum_{k,\eta} T^k_{\alpha\eta} T^k_{\eta\beta}=\delta_{\alpha\beta} C_F,\
\sum_{j,k} f^{jkm} f^{jkn}=\delta^{mn} C_A,$$

$$\sum_{\alpha,\beta} T^m_{\alpha\beta} T^n_{\beta\alpha}=\delta^{mn} T_F.$$

Measurements of the ratios between the colour factors allow the
experimental determination of the underlying gauge symmetry of the
strong interactions. For SU($N$), the predicted values of the colour
factors are:

$$C_A=N,\ \ C_F=\frac{N^2-1}{2N}\ \ {\rm and}\ \ T_F=1/2,$$
where $N$ is the number of colour charges. In particular, SU(3)
predicts $C_A/C_F=9/4$ and $T_F/C_F=3/8$. In contrast, an Abelian
gluon theory based on U(1)$^3$ would predict $C_A/C_F=0$ and
$T_F/C_F=3$. A non-Abelian theory based on SO(3) predicts $C_A/C_F=1$
and $T_F/C_F=1$.

The $\oass$ calculations of three-jet cross sections for
direct $\gp$ and NC DIS processes can be expressed in terms of $C_A$,
$C_F$ and $T_F$ as~\cite{np:b286:553}:

\begin{equation}
\sigma_{ep \rightarrow 3{\rm jets}} = C_F^2 \cdot \sigma_A + 
           C_F C_A \cdot \sigma_B +
           C_F T_F \cdot \sigma_C + T_F C_A \cdot \sigma_D,\label{one}
\end{equation}
where $\sigma_A$, ..., $\sigma_D$ are the partonic cross sections for
the different contributions (see Fig.~\ref{fig1}).

\section{Definition of the angular correlations}

Angular-correlation observables were devised to distinguish the
contributions from the different colour configurations. They are
defined in terms of the three jets with highest transverse energy in
an event and the beam direction as:
\begin{itemize}
\item {\boldmath $\th$}, the angle between the plane determined by the
  highest-transverse-energy jet and the beam and the plane determined
  by the two jets with lowest transverse energy~\cite{pr:d52:3894};
\item {\boldmath $\a34$}, the angle between the two
  lowest-transverse-energy jets. This variable is based on the angle
  $\alpha_{34}^{\ele}$ for
  $\ele\rightarrow {4\ \rm jets}$~\cite{pl:b255:466};
\item {\boldmath $\pksw$}, the angle defined via the equation \\
  \centerline{$\cos(\pksw)=\cos \left [ \frac{1}{2} \left ( \angle[(\vec p_1 \times \vec p_3),(\vec p_2 \times \vec p_B)] + \angle[(\vec p_1 \times \vec p_B),(\vec p_2 \times \vec p_3)] \right ) \right ]$,}
  where $\vec p_i,\ i=1,...,3$ is the momentum of jet $i$ and $\vec p_B$
  is a unit vector in the direction of the beam; the jets are ordered
  according to decreasing transverse energy. This variable is based
  on the K\"orner-Schierholz-Willrodt angle 
  $\Phi_{\rm KSW}^{\ele}$ for 
  $\ele\rightarrow {4\ \rm jets}$~\cite{np:b185:365};
\item {\boldmath $\etajmax$}, the maximum pseudorapidity of the three
  jets.
\end{itemize}

For three-jet events in $ep$ collisions, the variable $\th$ was
designed~\cite{pr:d52:3894} to be sensitive to the TGV in
quark-induced processes (see Fig.~\ref{fig1}B). In $\ele$ annihilation
into four-jet events, the distribution of $\Phi_{\rm KSW}^{\ele}$ is
sensitive to the differences between $\qq gg$ and $\qq\qq$ final
states whereas that of $\alpha_{34}^{\ele}$ distinguishes between
contributions from double-bremsstrahlung diagrams and diagrams
involving the TGV. 

\section{Experimental set-up}

The data samples used in this analysis were collected with the ZEUS
detector at HERA and correspond to an integrated luminosity of 
$44.9\pm 0.8\ (65.1\pm 1.5)$~\pb1\ for $e^+p$ collisions taken during
1995--97 (1999--2000) and $16.7\pm 0.3$~\pb1\ for $e^-p$ collisions
taken during 1998--99. During 1995--97 (1998--2000), HERA operated with
protons of energy $E_p=820$ ($920$)~GeV and positrons or
electrons\footnote{Here and in the following, the term ``electron''
  denotes generically both the electron ($e^-$) and the positron
  ($e^+$).} of energy $E_e=27.5$~GeV, yielding a centre-of-mass energy
of $\sqrt s=300$ ($318$)~GeV.

A detailed description of the ZEUS detector can be found
elsewhere~\cite{pl:b293:465,zeus:1993:bluebook}. A brief outline of
the components that are most relevant for this analysis is given below.
Charged particles were tracked in the central tracking detector 
(CTD)~\cite{nim:a279:290,*npps:b32:181,*nim:a338:254}, which operated
in a magnetic field of $1.43\Tesla$ provided by a thin
superconducting solenoid. The CTD consisted of 72~cylindrical
drift-chamber layers, organised in nine superlayers covering the
polar-angle\footnote{The ZEUS coordinate system is a right-handed
  Cartesian system, with the $Z$ axis pointing in the proton beam
  direction, referred to as the ``forward direction'', and the $X$
  axis pointing left towards the centre of HERA. The coordinate origin
  is at the nominal interaction point.} region 
\mbox{$15^\circ<\theta<164^\circ$}. The transverse-momentum resolution
for full-length tracks was parameterised as 
$\sigma(p_T)/p_T=0.0058p_T\oplus0.0065\oplus0.0014/p_T$, with $p_T$ in
$\Gev$. The tracking system was used to measure the interaction vertex
with a typical resolution along (transverse to) the beam direction of
0.4~(0.1)~cm and to cross-check the energy scale of the calorimeter.

The high-resolution uranium--scintillator calorimeter
(CAL)~\cite{nim:a309:77,*nim:a309:101,*nim:a321:356,*nim:a336:23}
covered $99.7\%$ of the total solid angle and consisted of three
parts: the forward (FCAL), the barrel (BCAL) and the rear (RCAL)
calorimeters. Each part was subdivided transversely into towers and 
longitudinally into one electromagnetic section (EMC) and either one
(in RCAL) or two (in BCAL and FCAL) hadronic sections (HAC). The
smallest subdivision of the calorimeter was called a cell. Under
test-beam conditions, the CAL single-particle relative energy
resolutions were $\sigma(E)/E=0.18/\sqrt E$ for electrons and
$\sigma(E)/E=0.35/\sqrt E$ for hadrons, with $E$ in GeV.

The luminosity was measured from the rate of the bremsstrahlung
process $ep\rightarrow e\gamma p$. The resulting small-angle
energetic photons were measured by the luminosity 
monitor~\cite{desy-92-066,*zfp:c63:391,*acpp:b32:2025}, a
lead--scintillator calorimeter placed in the HERA tunnel at $Z=-107$ m.

\section{Data selection and jet search}

A three-level trigger system was used to select events
online~\cite{zeus:1993:bluebook,proc:chep:1992:222}. At the third
level, jets were reconstructed using the energies and positions of the
CAL cells. Events with at least one (two) jet(s) with transverse
energy in excess of $10\ (6)$~GeV and pseudorapidity below $2.5$ were
accepted. For trigger-efficiency studies, no jet algorithm was applied
and events with a total transverse energy, excluding the energy in the
eight CAL towers immediately surrounding the forward beampipe, of at
least $25$~GeV were selected in the $\gp$ sample; for the NC DIS
sample, events were selected in which the scattered-electron candidate
was identified using localised energy depositions in the CAL.

In the offline selection, a reconstructed event vertex consistent with
the nominal interaction position was required and cuts based on 
tracking information were applied to reduce the contamination from
beam-induced and cosmic-ray background events. The selection criteria
of the $\gp$ and NC DIS samples were analogous to previous
publications~\cite{pl:b560:7,pl:b649:12}.

The selected $\gp$ sample consisted of events from $ep$
interactions with $\q2<1$ \g2\ and a median $\q2\approx
10^{-3}$~\g2. The event sample was restricted to the kinematic range
$0.2<y<0.85$, where $y$ is the inelasticity.

Events from NC DIS interactions were selected from the 1998--2000
data. Two samples were studied: $\q2>125$~\g2\ and
$500<\q2<5000$~\g2. For both samples, $|\cgh|$ was restricted to be
below $0.65$, where $\gamma_h$, which corresponds to the angle of the
scattered quark in the quark-parton model, is defined as

$$\cgh= \frac{(1-y)x E_p - y E_e}{(1-y)x E_p + y E_e}$$
and $x$ is the Bjorken variable. 

The $\kt$ cluster algorithm~\cite{np:b406:187} was used in the
longitudinally invariant inclusive mode~\cite{pr:d48:3160} to
reconstruct jets in the measured hadronic final state from
the energy deposits in the CAL cells (calorimetric jets). The axis of
the jet was defined according to the Snowmass
convention~\cite{proc:snowmass:1990:134}. 

For $\gp$ events, the jet search was performed in the $\etaphi$ plane
of the laboratory frame. Corrections~\cite{pl:b560:7} to the jet
transverse energy, $\etjet$, were applied to the calorimetric jets as
a function of the jet pseudorapidity, $\etajet$, and $\etjet$ and
averaged over the jet azimuthal angle. Events with at least three jets
of $\etjet>14$~GeV and $\etar$ were retained. Direct $\gp$ events were
further selected by requiring $\xo>0.8$, where $\xo$, the fraction of
the photon momentum participating in the production of the three jets
with highest $\etjet$, is defined as

$$\xo=\frac{1}{2 y E_e}\left ( \etJ e^{-\etaJ}+\etjj e^{-\etajj}+\etjjj
e^{-\etajjj}\right ) .$$
The final $\gp$ data sample contained 1888 events.

For NC DIS events, the $\kt$ jet algorithm was applied after excluding
those cells associated with the scattered-electron candidate and the
search was conducted in the Breit frame. Jet transverse-energy
corrections were computed using the method developed in a previous
analysis~\cite{pl:b649:12}. Events were required to have at least
three jets satisfying $\etjbj>8$~GeV, $E^{\rm jet2,3}_{T,{\rm B}}>5$
GeV and $-2<\etajb<1.5$, where $\etjb$ and $\etajb$ are the jet
transverse energy and pseudorapidity in the Breit frame, respectively.
The final NC DIS data sample with $\q2>125$ ($500<\q2<5000$)~\g2\ 
contained 1095 (492) events.

\section{Monte Carlo simulation}
\label{mc}

Samples of Monte Carlo (MC) events were generated to determine the
response of the detector to jets of hadrons and the correction factors
necessary to obtain the hadron-level jet cross sections. The hadron
level is defined by those hadrons with lifetime $\tau\geq 10$~ps. For
the NC DIS sample, the MC events were also used to correct the
measured cross sections for QED radiative effects and the running of 
$\alpha_{\rm em}$.

The generated events were passed through the {\sc
  Geant}~3.13-based~\cite{tech:cern-dd-ee-84-1} ZEUS detector- and
trigger-simulation programs~\cite{zeus:1993:bluebook}. They were
reconstructed and analysed by the same program chain as the data. The
$\kt$ jet algorithm was applied to the MC simulated events using the
CAL cells in the same way as for the data. The jet algorithm was also
applied to the final-state particles (hadron level) and the partons
available after the parton shower (parton level).

The programs {\sc Pythia}~6.1~\cite{cpc:82:74,*cpc:135:238} and
{\sc Herwig}~6.1~\cite{cpc:67:465,*jhep:0101:010} were used to
generate $\gp$ events for resolved and direct processes. Events were
generated using GRV-HO~\cite{pr:d45:3986,*pr:d46:1973} for the photon
and CTEQ4M~\cite{pr:d55:1280} for the proton parton distribution
functions (PDFs). In both generators, the partonic processes are
simulated using LO matrix elements, with the inclusion of initial- and
final-state parton showers. Fragmentation into hadrons is performed
using the Lund string model~\cite{prep:97:31} as implemented in 
{\sc Jetset}~\cite{cpc:82:74,*cpc:135:238,cpc:39:347,*cpc:43:367} in
the case of {\sc Pythia}, and a cluster model~\cite{np:b238:492} in
the case of {\sc Herwig}. 

Neutral current DIS events including radiative effects were simulated
using the {\sc Heracles}~4.6.1~\cite{cpc:69:155,*spi:www:heracles}
program with the {\sc
  Djangoh}~1.1~\cite{cpc:81:381,*spi:www:djangoh11} interface to the
hadronisation programs. {\sc Heracles} includes corrections for
initial- and final-state radiation, vertex and propagator terms, and
two-boson exchange. The QCD cascade is simulated using the
colour-dipole
model (CDM)~\cite{pl:b165:147,*pl:b175:453,*np:b306:746,*zfp:c43:625}
including the LO QCD diagrams as implemented in
{\sc Ariadne}~4.08~\cite{cpc:71:15,*zfp:c65:285}; additional
samples were generated with the MEPS model of {\sc Lepto}
6.5~\cite{cpc:101:108}. Both MC programs use the Lund string model for
the hadronisation. The CTEQ5D~\cite{epj:c12:375} proton PDFs were used
for these simulations.

\section{Fixed-order calculations}
\label{nlo}

The calculations of direct $\gp$ processes used in this analysis are
based on the program by Klasen, Kleinwort and Kramer
(KKK)~\cite{epjdirectcbg:c1:1}. The number of flavours was set to
five; the renormalisation, $\mu_R$, and factorisation scales, $\mu_F$,
were set to $\mu_R=\mu_F=E_T^{\rm max}$, where $E_T^{\rm max}$ is the
highest $\etjet$ in an event. The calculations were performed using
the ZEUS-S~\cite{pr:d67:012007} parameterisations of the proton PDFs;
$\as$ was calculated at two loops using $\Lambda^{(5)}_{\overline{\rm
    MS}}=226$~MeV, which corresponds to $\asz=0.118$. These
calculations are $\oass$ and represent the lowest-order contribution
to three-jet $\gp$. Full $\oasss$ corrections are not yet
available for three-jet cross sections in $\gp$.

The calculations of NC DIS processes used in this analysis are
based on the program {\sc Nlojet++}~\cite{prl:87:082001}, which
provides $\oass$ and $\oasss$ predictions for three-jet cross
sections. The scales were chosen to be $\mu_R=\mu_F=Q$. Other
parameters were set as for the $\gp$ program.

In general, the programs mentioned above are very flexible and provide
observable-independent computations that allow a complete analytical
cancellation of the soft and collinear singularities encountered in
the calculations of jet cross sections. However, these programs were
written assuming the SU(3) gauge group and the different ingredients
necessary to perform a calculation according to Eq. (1) were not
readily available. The programs were rewritten in order to disentagle
the colour components to make separate predictions for $\sigma_A$, ...,
$\sigma_D$.

The $\kt$ jet algorithm was applied to the partons in the events
generated by KKK and {\sc Nlojet++} in order to compute the jet
cross-section predictions. Thus, these predictions refer to jets of
partons. Since the measurements refer to jets of hadrons, the
calculations were corrected to the hadron level. The multiplicative
correction factors, defined as the ratios between the cross section
for jets of hadrons and that for jets of partons, were estimated using
the MC samples described in Section~\ref{mc}. The normalised cross-section
calculations changed typically by less than $\pm 5\ (10)\%$ for the
predictions in $\gp$ (NC DIS) upon application of the parton-to-hadron
corrections. Therefore, the effect of the parton-to-hadron corrections
on the angular distributions is small. In NC DIS processes, other
effects not accounted for in the calculations, namely $\z0$ exchange,
were also corrected for using the MC samples.

The predictions for jet cross sections are expressed as the
convolution of the PDFs and the matrix elements, which depend on
$\as$. Both the PDFs and $\as$ evolve with the energy scale. In the
calculations performed for this analysis, QCD evolution via the DGLAP
and the renormalisation group equations, respectively, were
used. These evolution equations also depend on the colour
factors. This procedure introduces an additional dependence on the
colour factors with respect to that shown in Eq. (1); this dependence
is suppressed by considering normalised cross sections (see
Section~\ref{cs} for the definition of the cross sections). The
remaining dependence was estimated by comparing to calculations with
fixed $\mu_F$ or $\mu_R$. The values chosen for $\mu_F$ and $\mu_R$
were the mean values of the data distributions, 
$\langle E_T^{\rm max}\rangle_{\rm data}=27.8$~GeV for $\gp$ and
$\sqrt{\langle\q2\rangle_{\rm data}}=31.3\ (36.6)$~GeV for NC DIS with
$\q2>125$ ($500<\q2<5000$)~\g2. 

\begin{figure}[p]
\vfill
\setlength{\unitlength}{1.0cm}
\begin{picture} (18.0,7.0)
\put (0.0,1.5){\epsfig{figure=\figdir 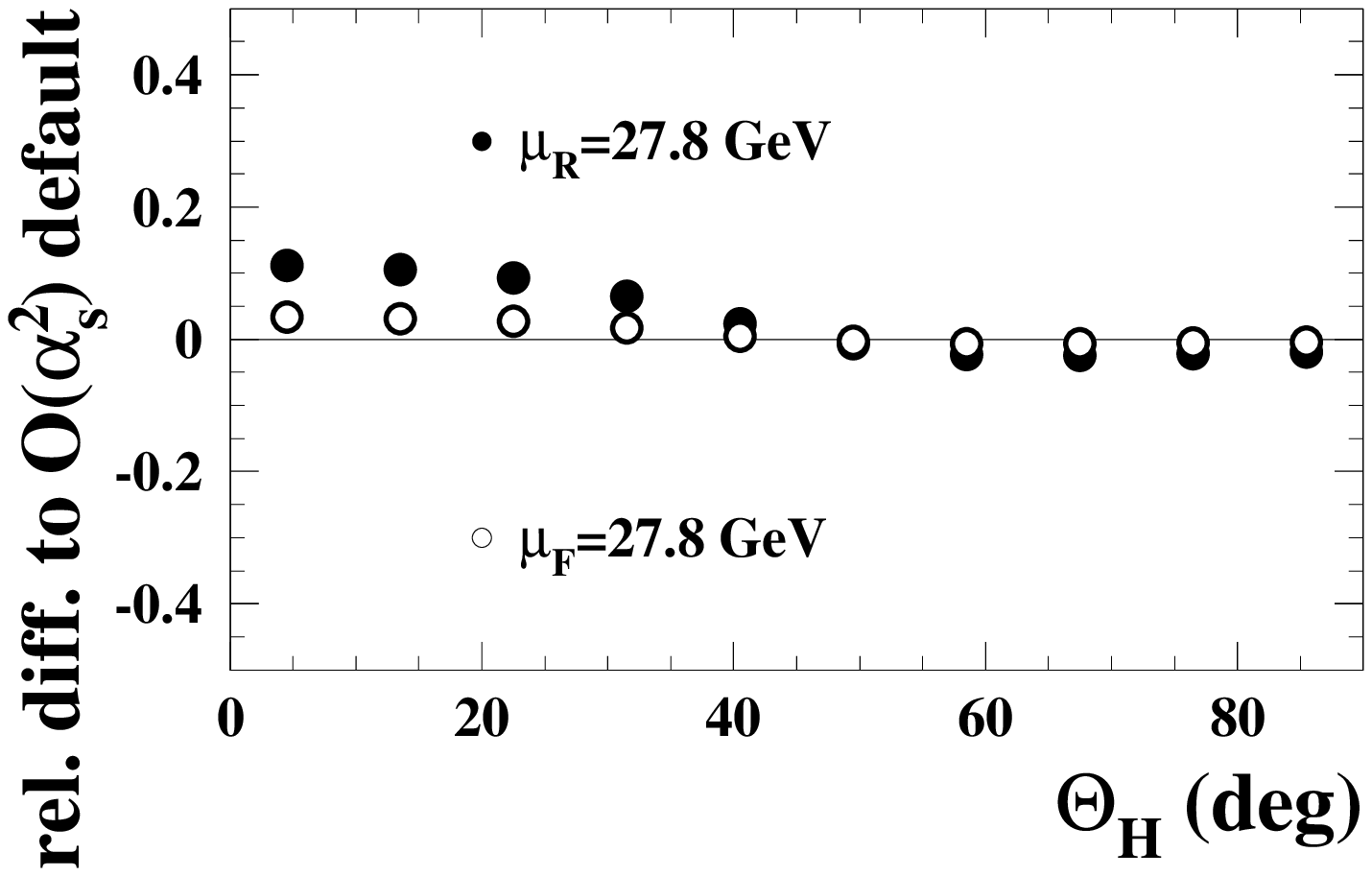,width=10cm}}
\put (7.0,1.5){\epsfig{figure=\figdir 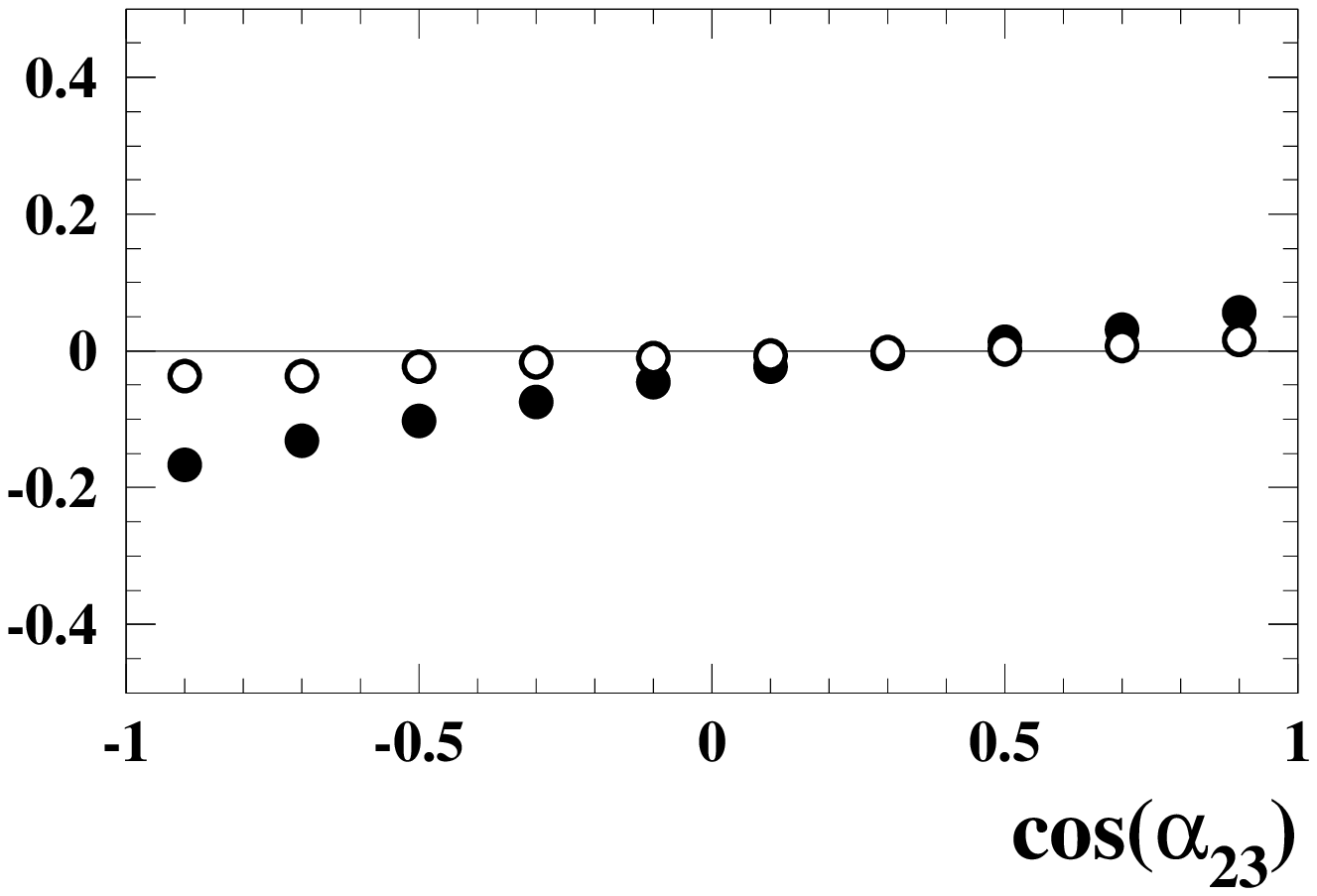,width=10cm}}
\put (0.0,-4.5){\epsfig{figure=\figdir 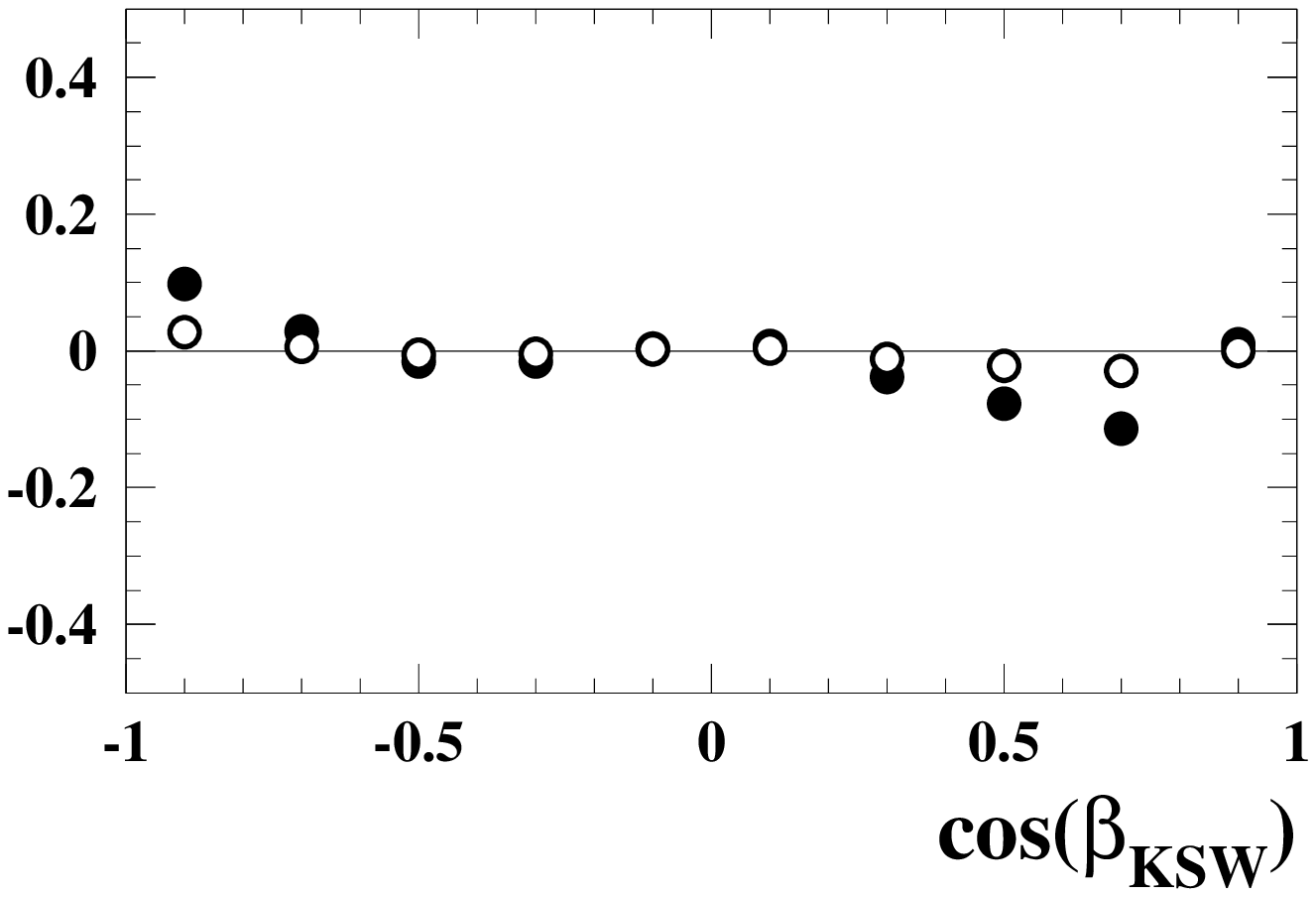,width=10cm}}
\put (7.0,10.0){\bf\small (a)}
\put (14.0,10.0){\bf\small (b)}
\put (7.0,4.0){\bf\small (c)}
\end{picture}
\caption 
{\it 
Relative difference between the $\oass$ calculations with
$\mu_R=27.8$~GeV and the calculations with $\mu_R=E_T^{\rm max}$ (dots)
and between the $\oass$ calculations with $\mu_F=27.8$~GeV and the
calculations with $\mu_F=E_T^{\rm max}$ (open circles) in $\gp$ as
functions of (a) $\th$, (b) $\cos(\a34)$ and (c) $\cos(\pksw)$. These
calculations do not include corrections for hadronisation effects.}
\label{fig7}
\vfill
\end{figure}

\begin{figure}[p]
\vfill
\setlength{\unitlength}{1.0cm}
\begin{picture} (18.0,17.0)
\put (0.0,10.5){\epsfig{figure=\figdir 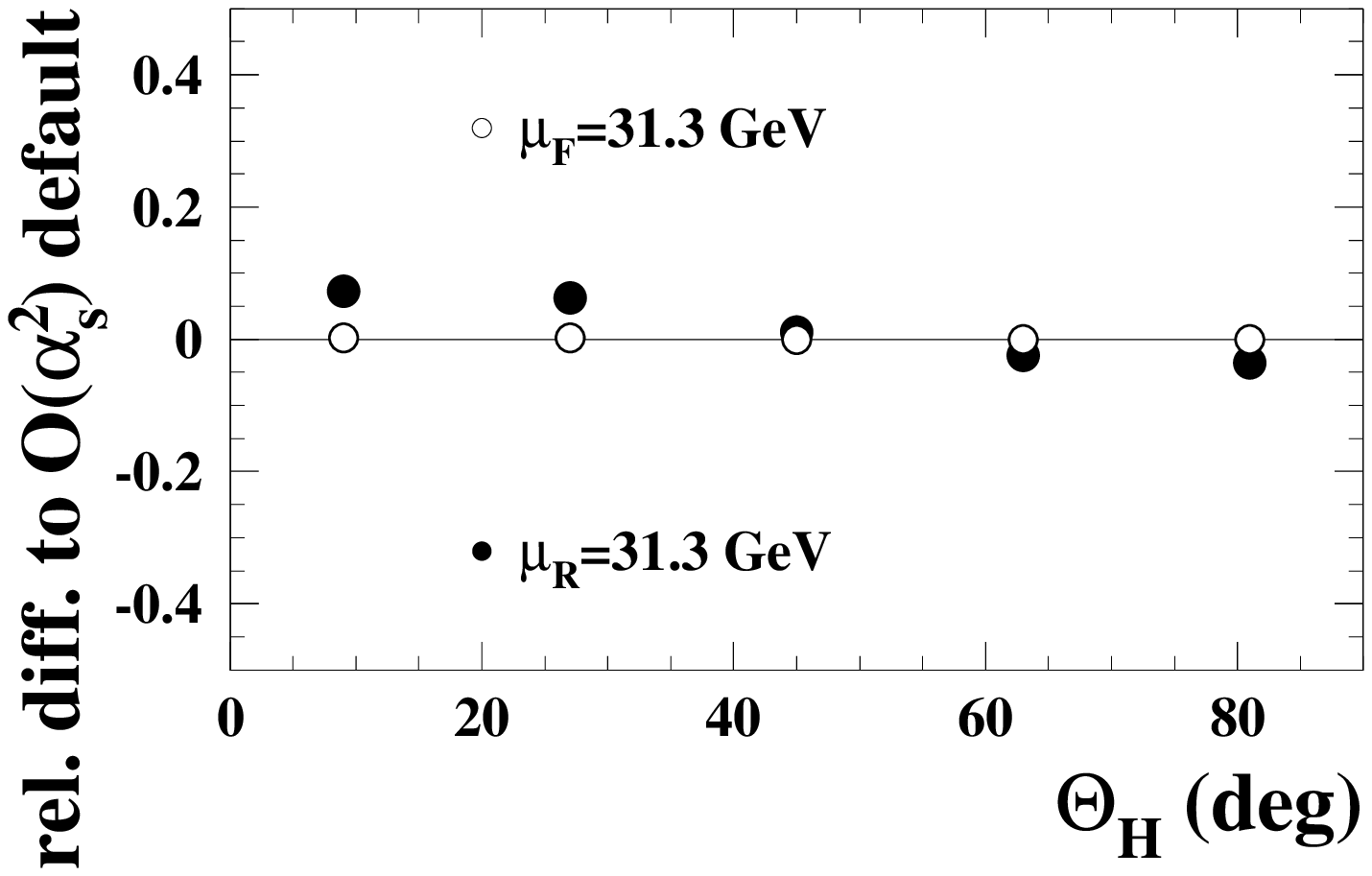,width=10cm}}
\put (7.0,10.5){\epsfig{figure=\figdir 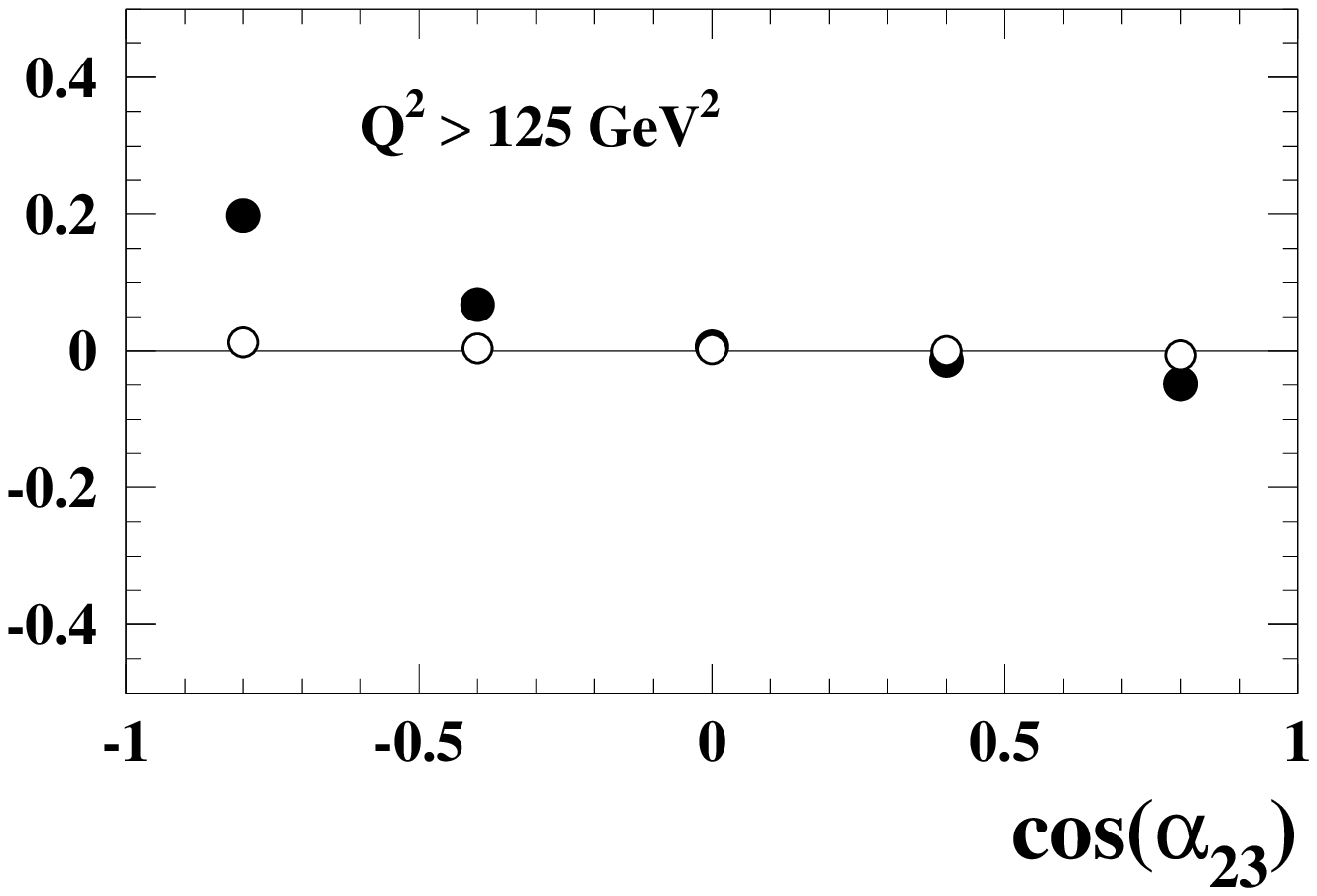,width=10cm}}
\put (0.0,5.5){\epsfig{figure=\figdir 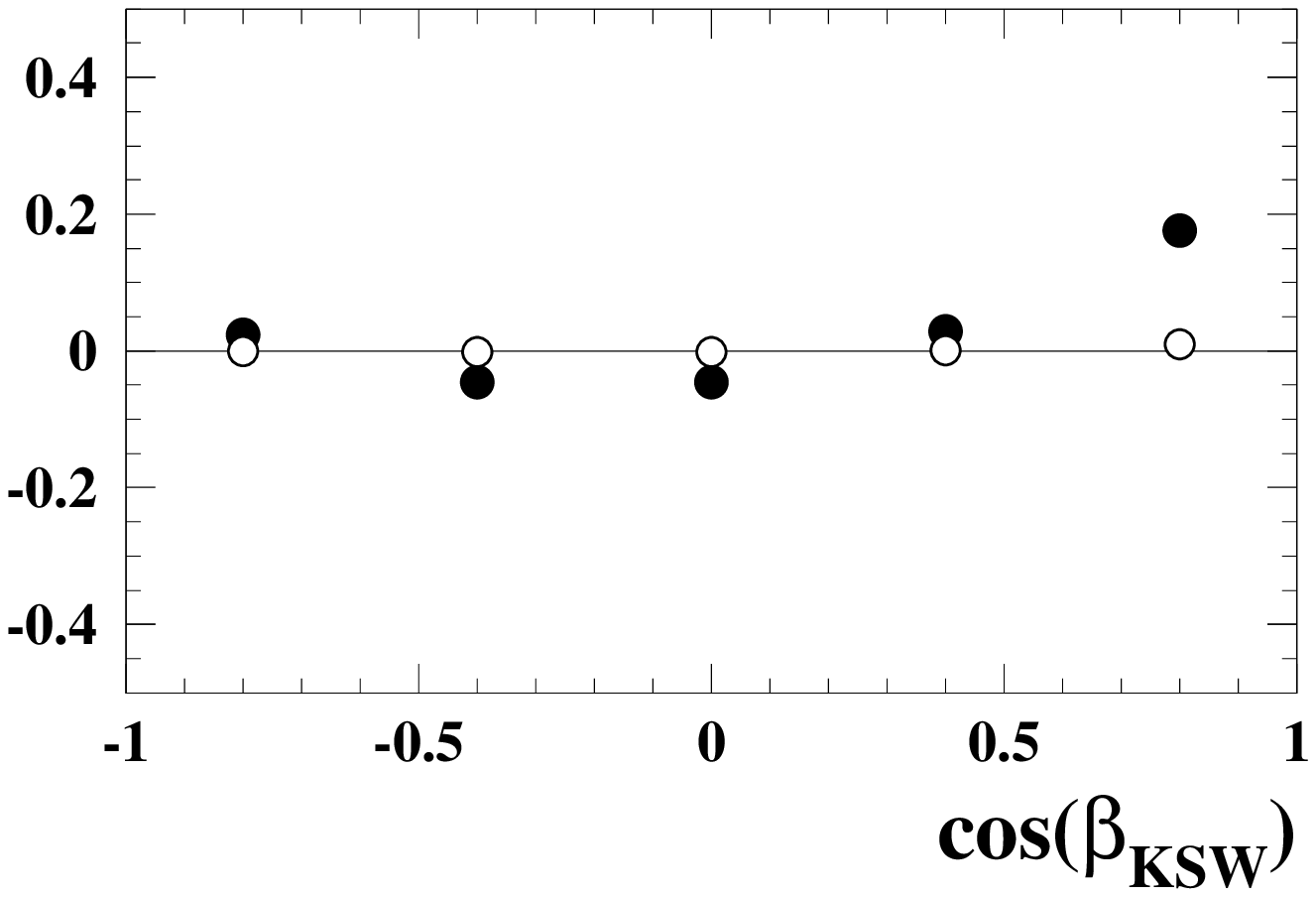,width=10cm}}
\put (7.0,5.5){\epsfig{figure=\figdir 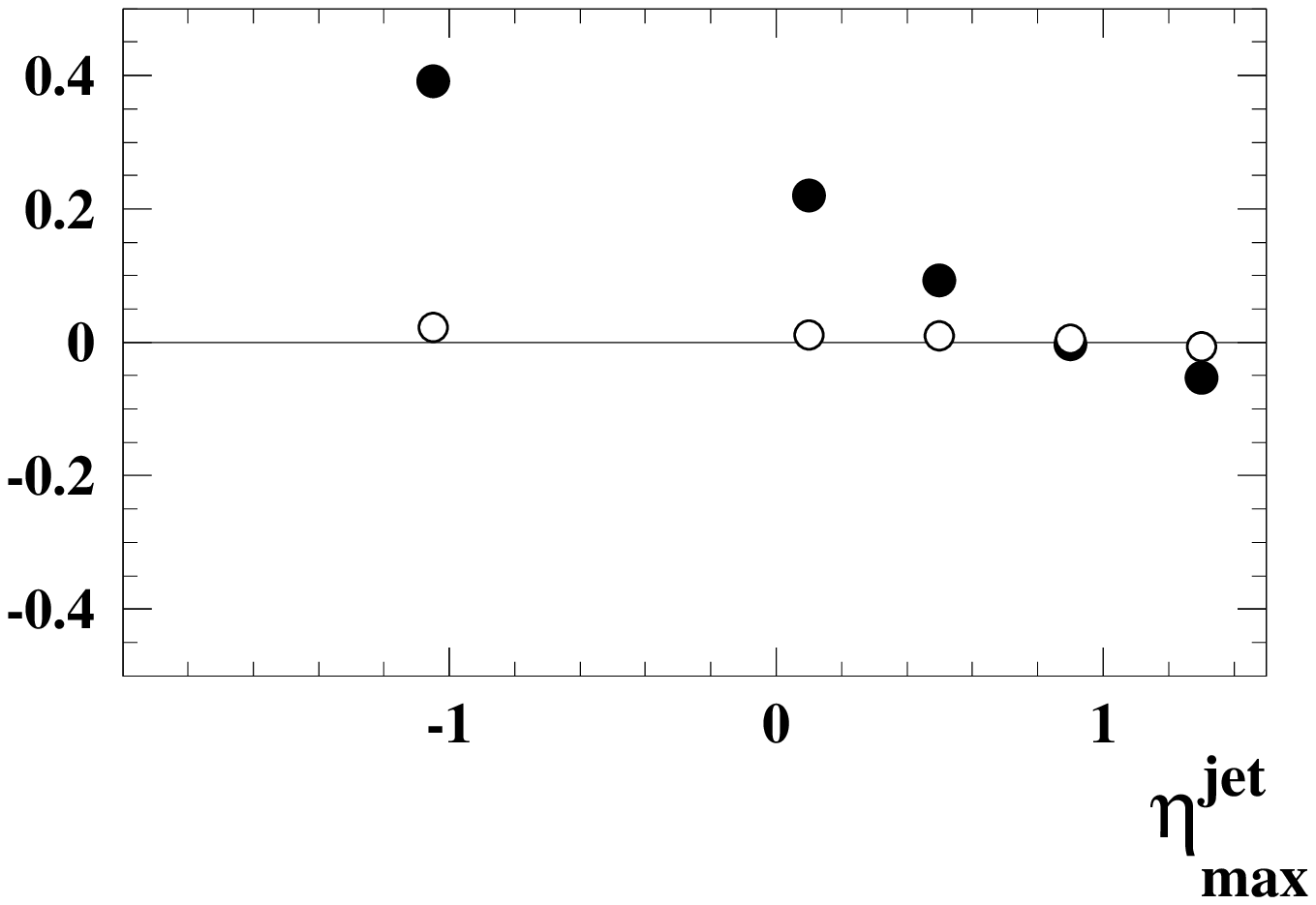,width=10cm}}
\put (0.0,0.5){\epsfig{figure=\figdir 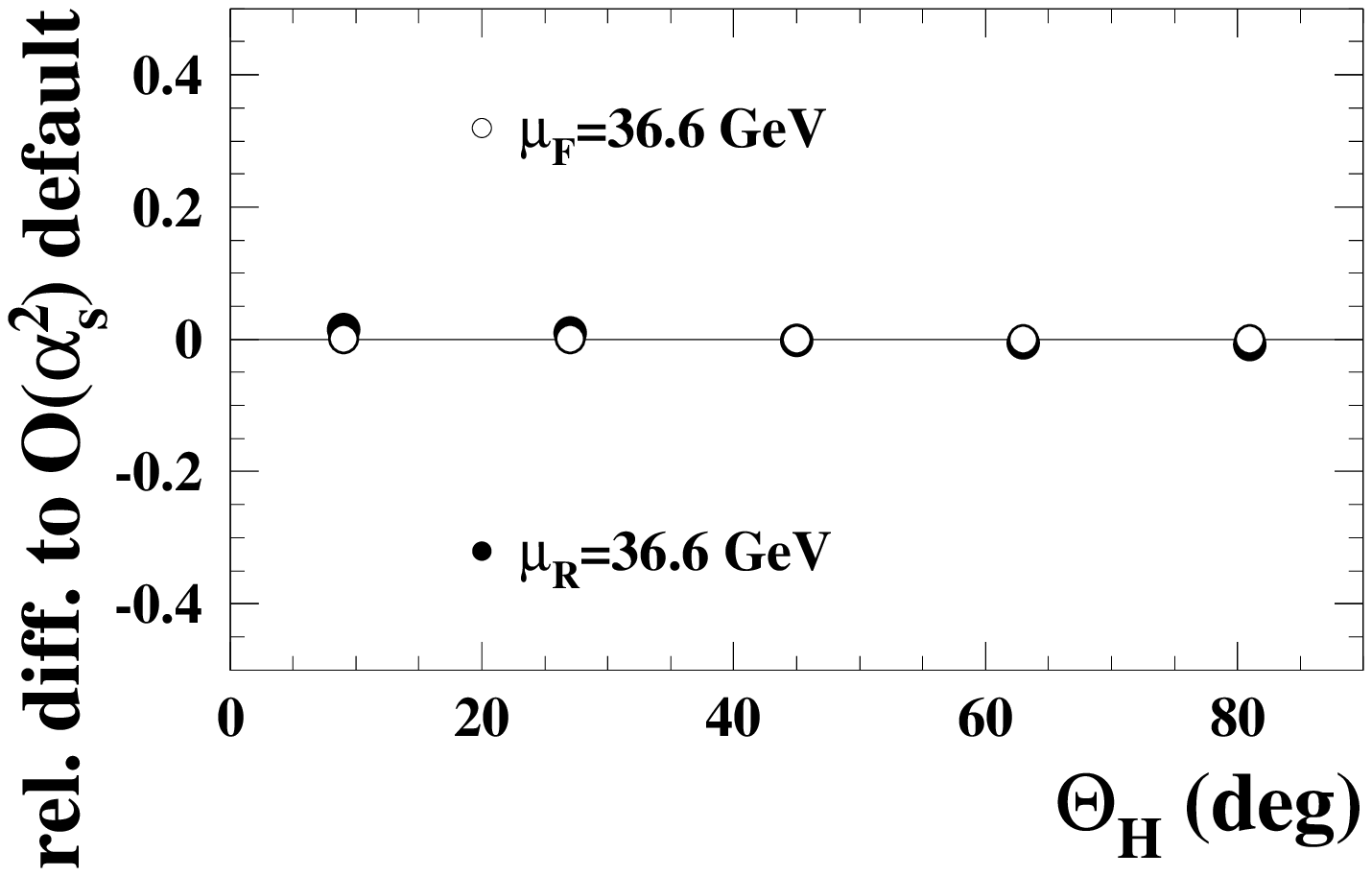,width=10cm}}
\put (7.0,0.5){\epsfig{figure=\figdir 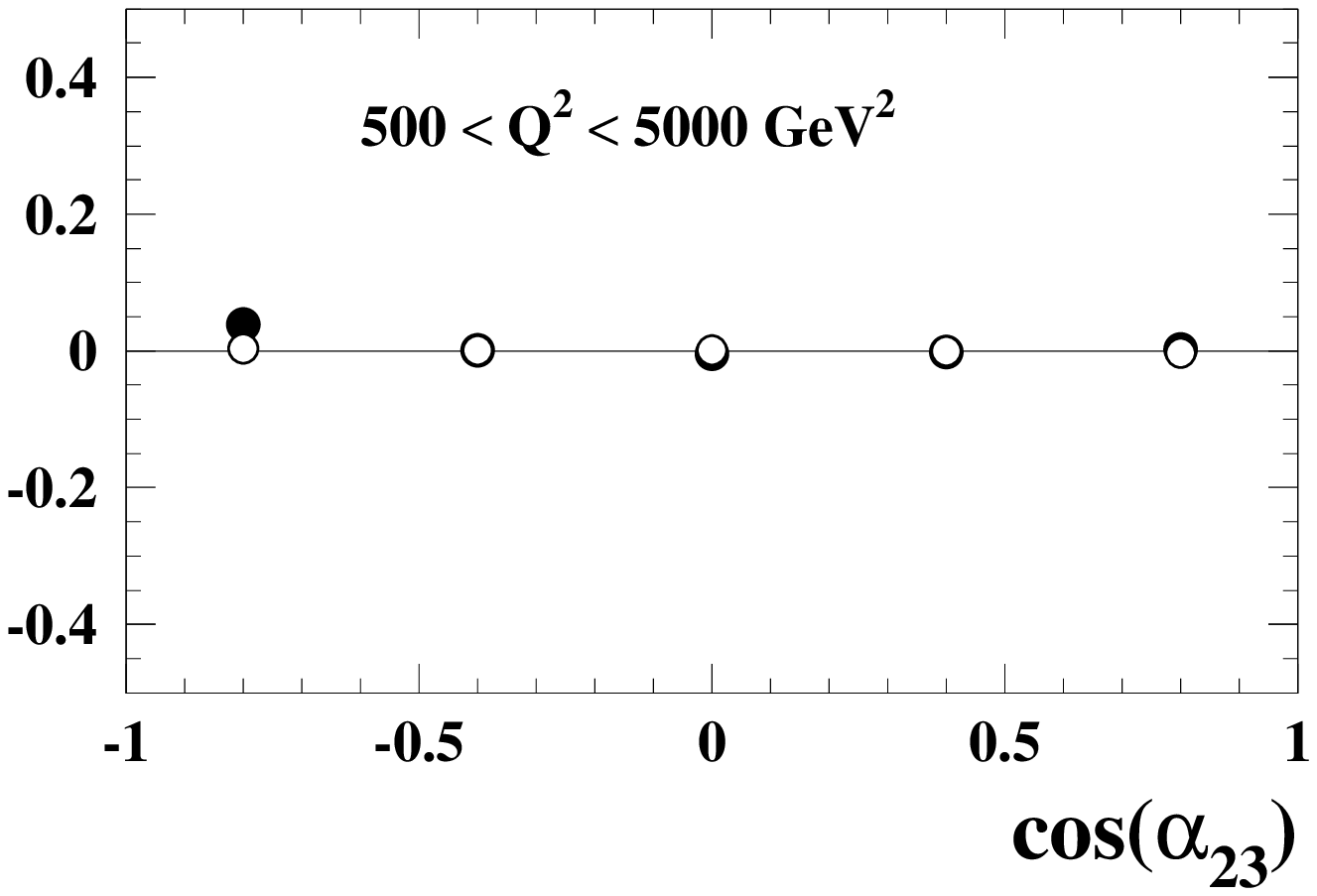,width=10cm}}
\put (0.0,-4.5){\epsfig{figure=\figdir 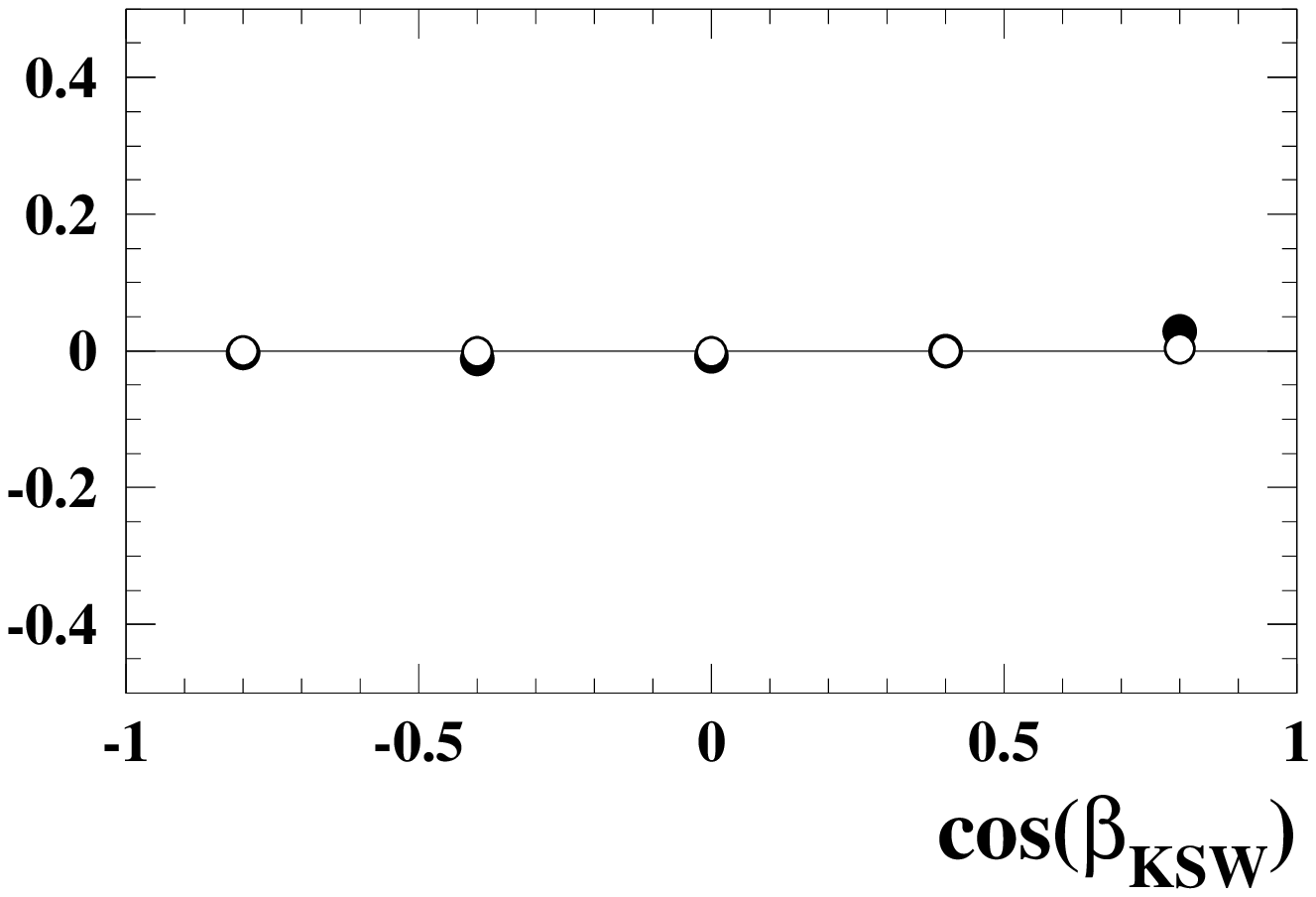,width=10cm}}
\put (7.0,-4.5){\epsfig{figure=\figdir 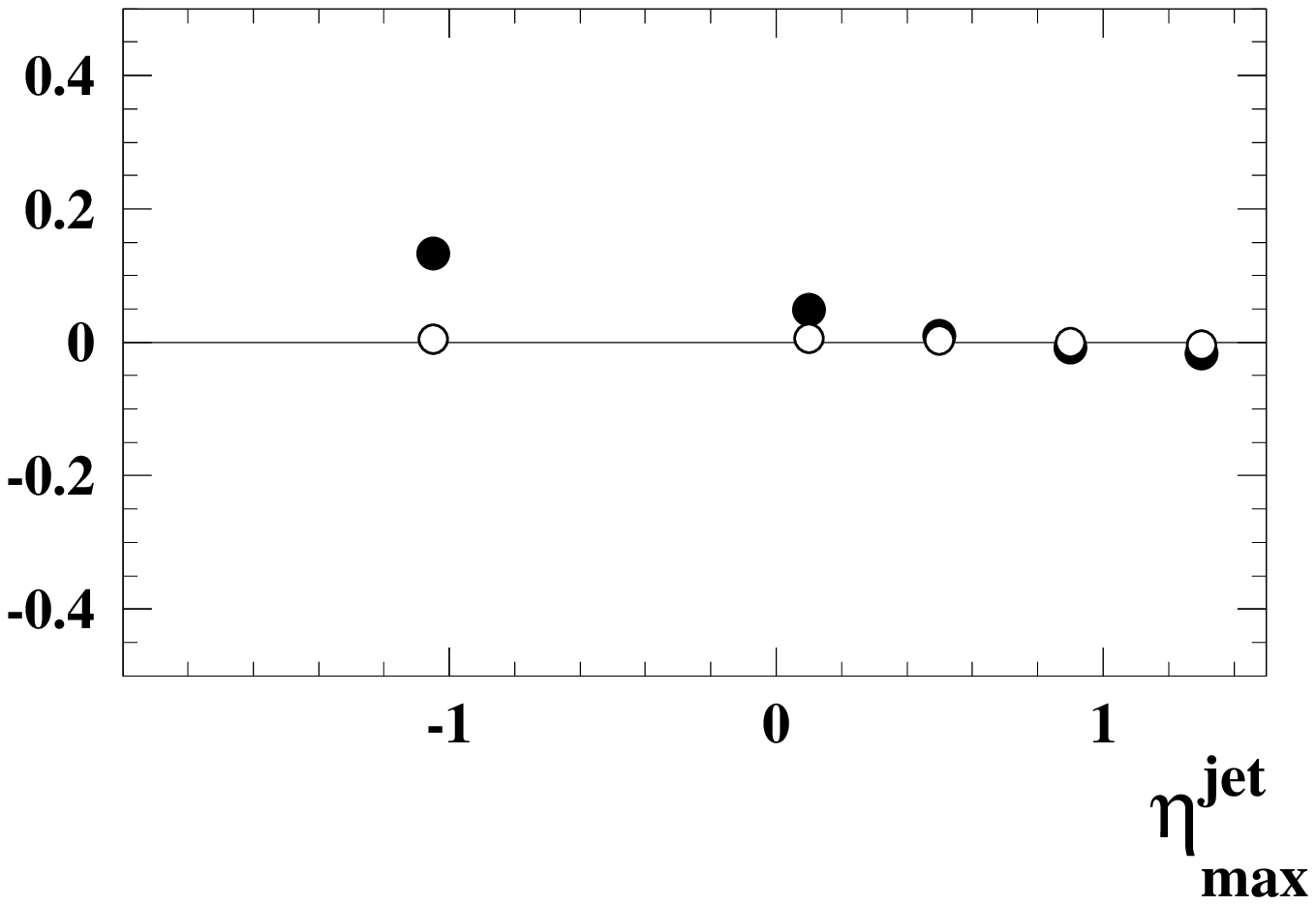,width=10cm}}
\put (7.0,19.0){\bf\small (a)}
\put (14.0,19.0){\bf\small (b)}
\put (7.0,14.0){\bf\small (c)}
\put (14.0,14.0){\bf\small (d)}
\put (7.0,9.0){\bf\small (e)}
\put (14.0,9.0){\bf\small (f)}
\put (7.0,4.0){\bf\small (g)}
\put (14.0,4.0){\bf\small (h)}
\end{picture}
\caption
{\it 
Relative difference between the $\oass$ calculations with
fixed $\mu_R$ and the calculations with $\mu_R=Q$ (dots)
and between the $\oass$ calculations with fixed $\mu_F$ and the
calculations with $\mu_F=Q$ (open circles) in NC DIS as functions of
(a,e) $\th$, (b,f) $\cos(\a34)$, (c,g) $\cos(\pksw)$ and (d,h)
$\etajmax$. These
calculations do not include corrections for hadronisation effects.
}
\label{fig8}
\vfill
\end{figure}

Figure~\ref{fig7} shows the relative difference of the $\oass$ $\gp$
calculations with $\mu_F$ ($\mu_R$) fixed\footnote{When $\mu_F$ was
  fixed, $\mu_R$ was allowed to vary with the scale, and vice-versa.}
to those in which $\mu_F=E_T^{\rm max}$ ($\mu_R=E_T^{\rm max}$) as a
function of the angular variables studied. Figures~\ref{fig8}(a) to
\ref{fig8}(d) show the same relative difference for the $\oass$ {\sc
  Nlojet++} calculations for $\q2>125$~\g2. 

Very small differences are observed for the $\mu_F$
variation. Sizeable differences for the $\mu_R$ variation are seen in
some regions; in particular, a trend is observed for the relative
difference as a function of $\etajmax$: this trend is due to the fact
that the mean values of $\q2$ in each bin of $\etajmax$ increase as
$\etajmax$ decreases.

\begin{figure}[p]
\vfill
\setlength{\unitlength}{1.0cm}
\begin{picture} (18.0,17.0)
\put (0.0,10.5){\epsfig{figure=\figdir 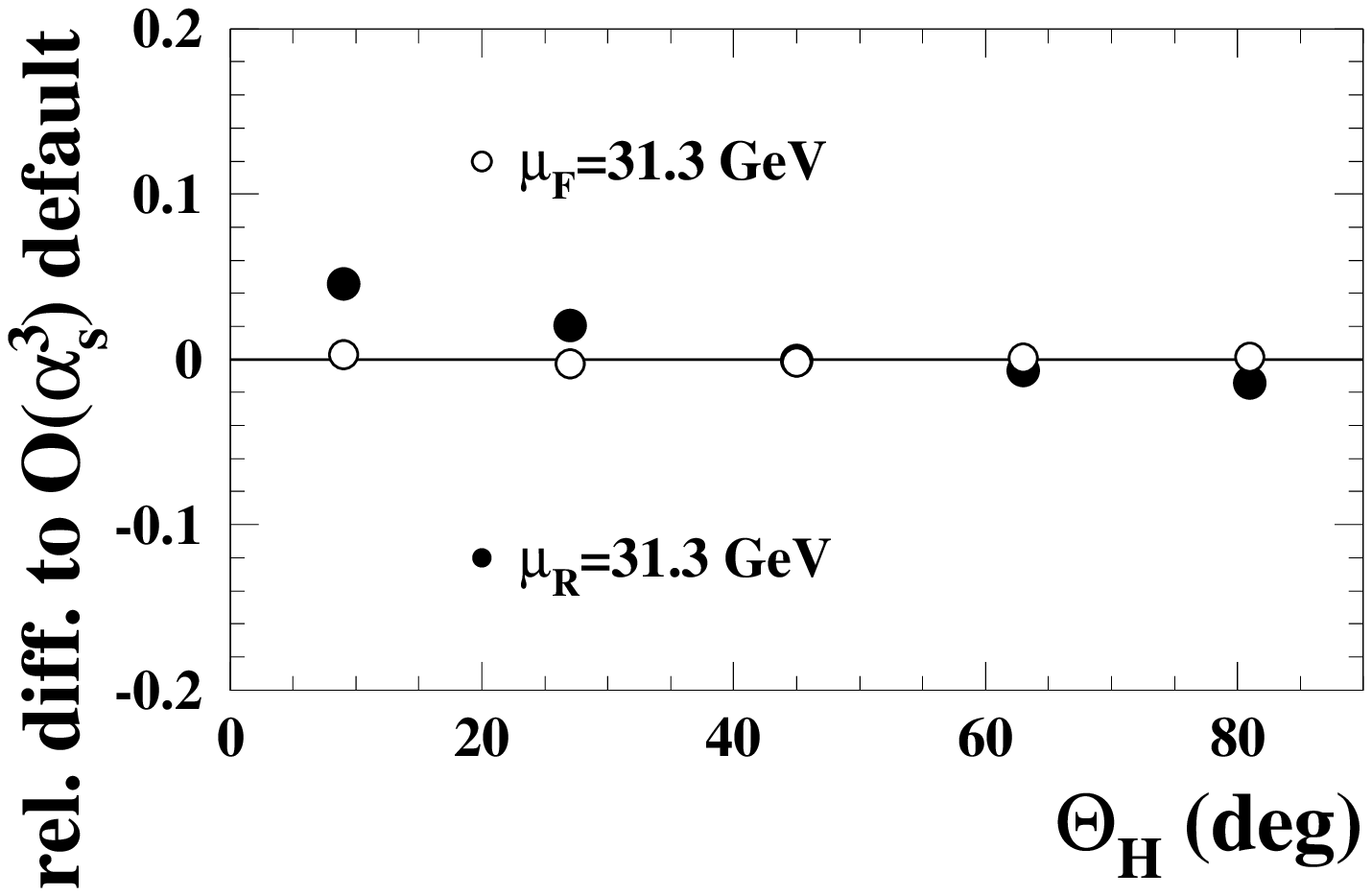,width=10cm}}
\put (7.0,10.5){\epsfig{figure=\figdir 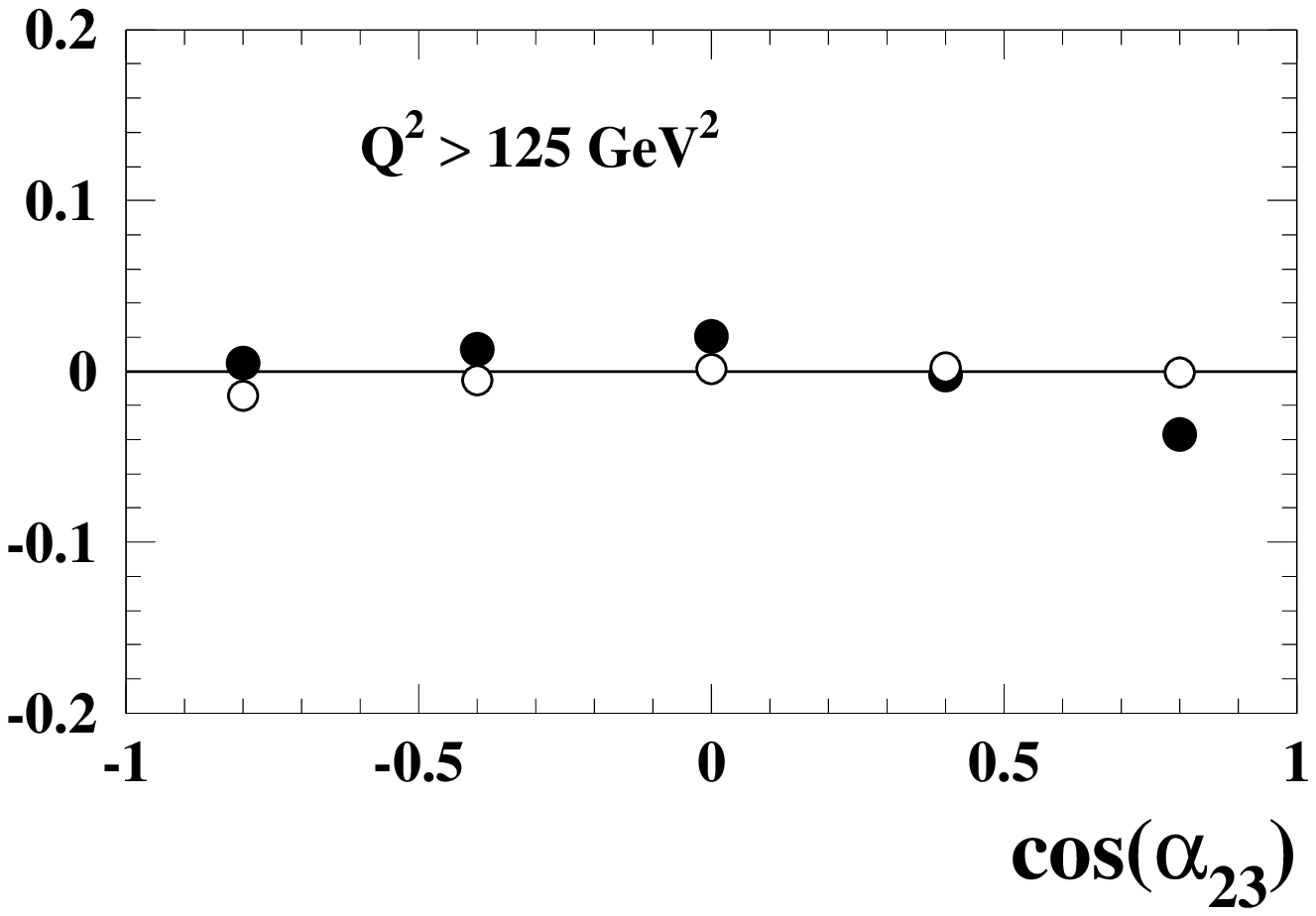,width=10cm}}
\put (0.0,5.5){\epsfig{figure=\figdir 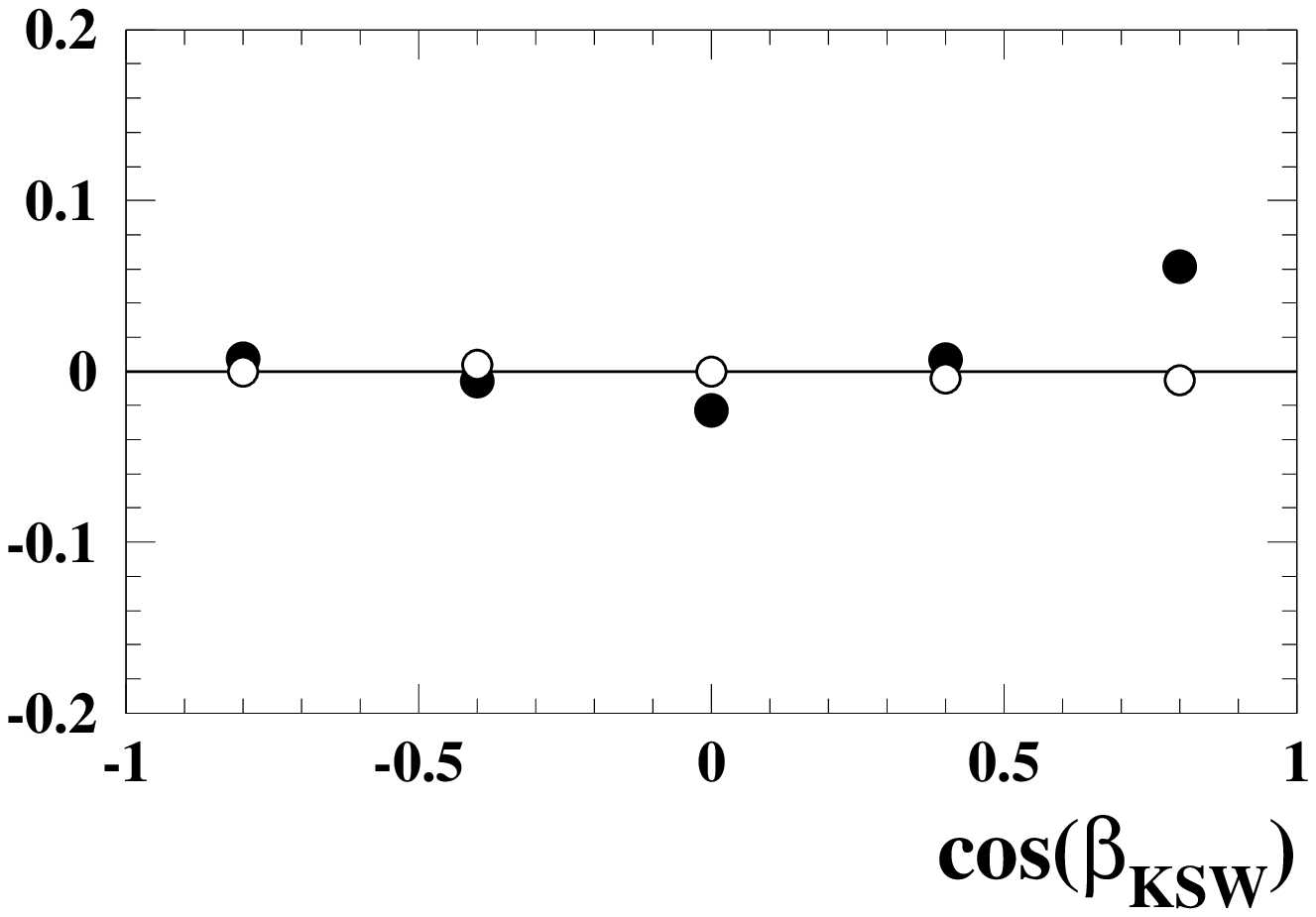,width=10cm}}
\put (7.0,5.5){\epsfig{figure=\figdir 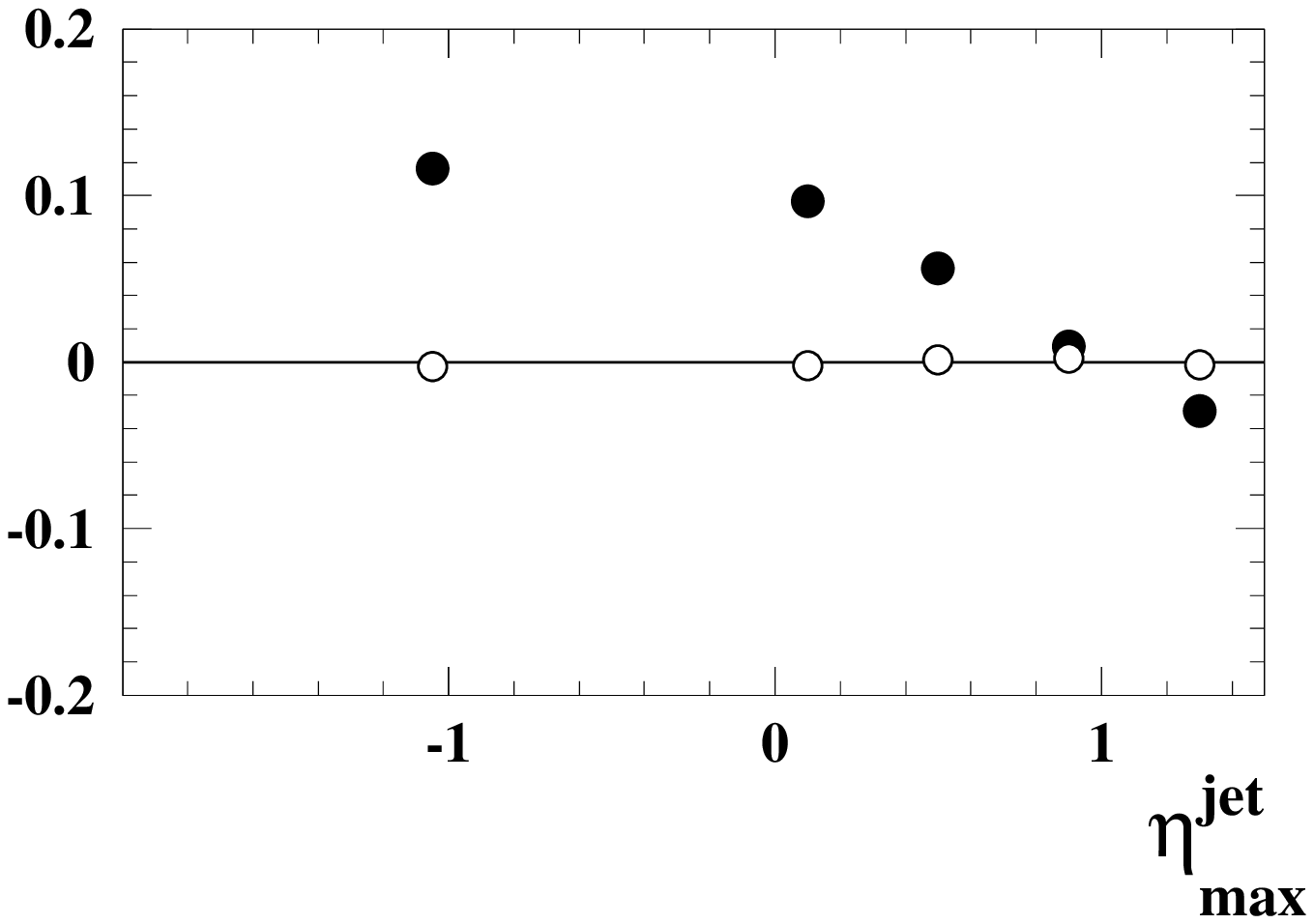,width=10cm}}
\put (0.0,0.5){\epsfig{figure=\figdir 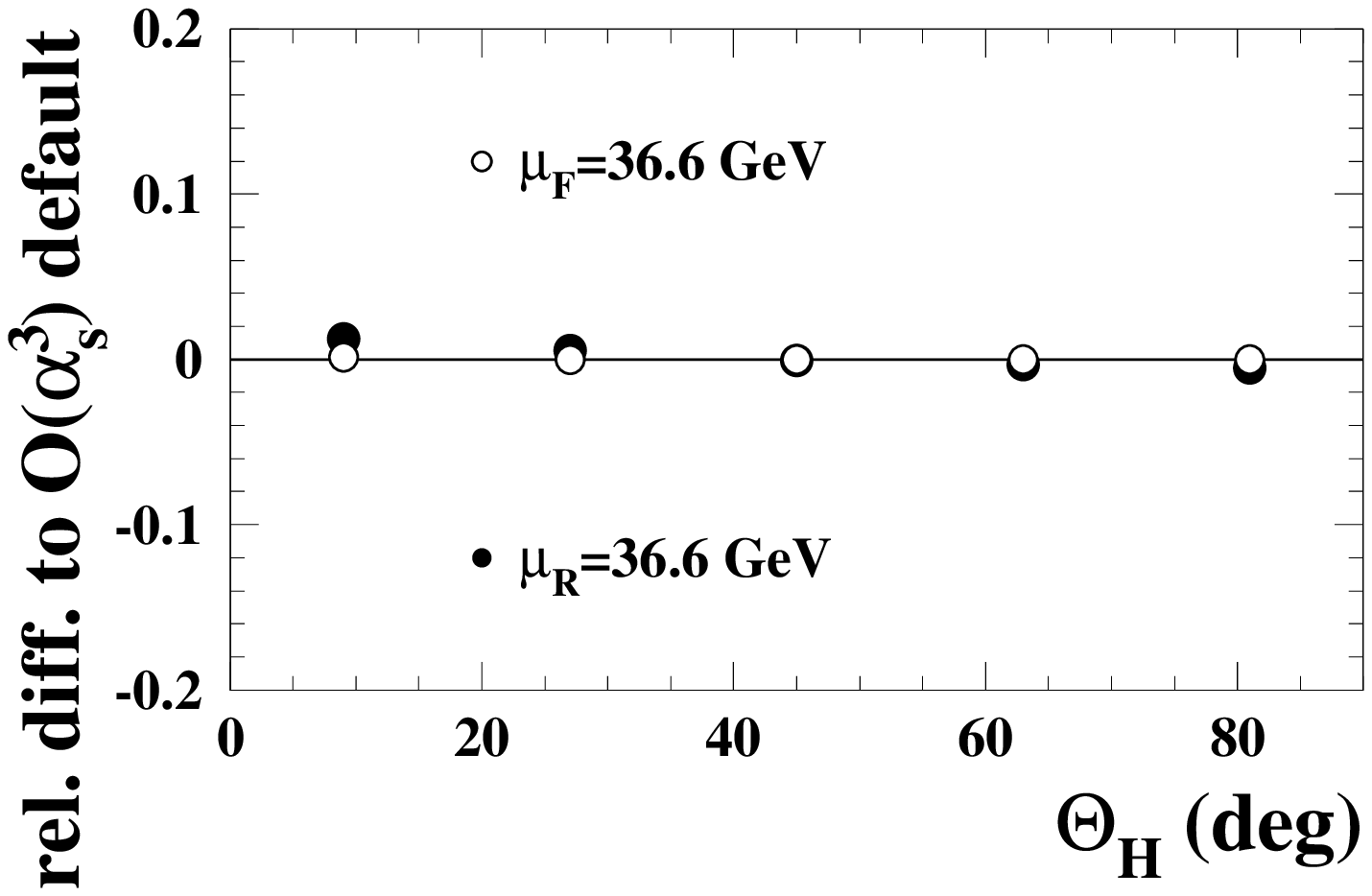,width=10cm}}
\put (7.0,0.5){\epsfig{figure=\figdir 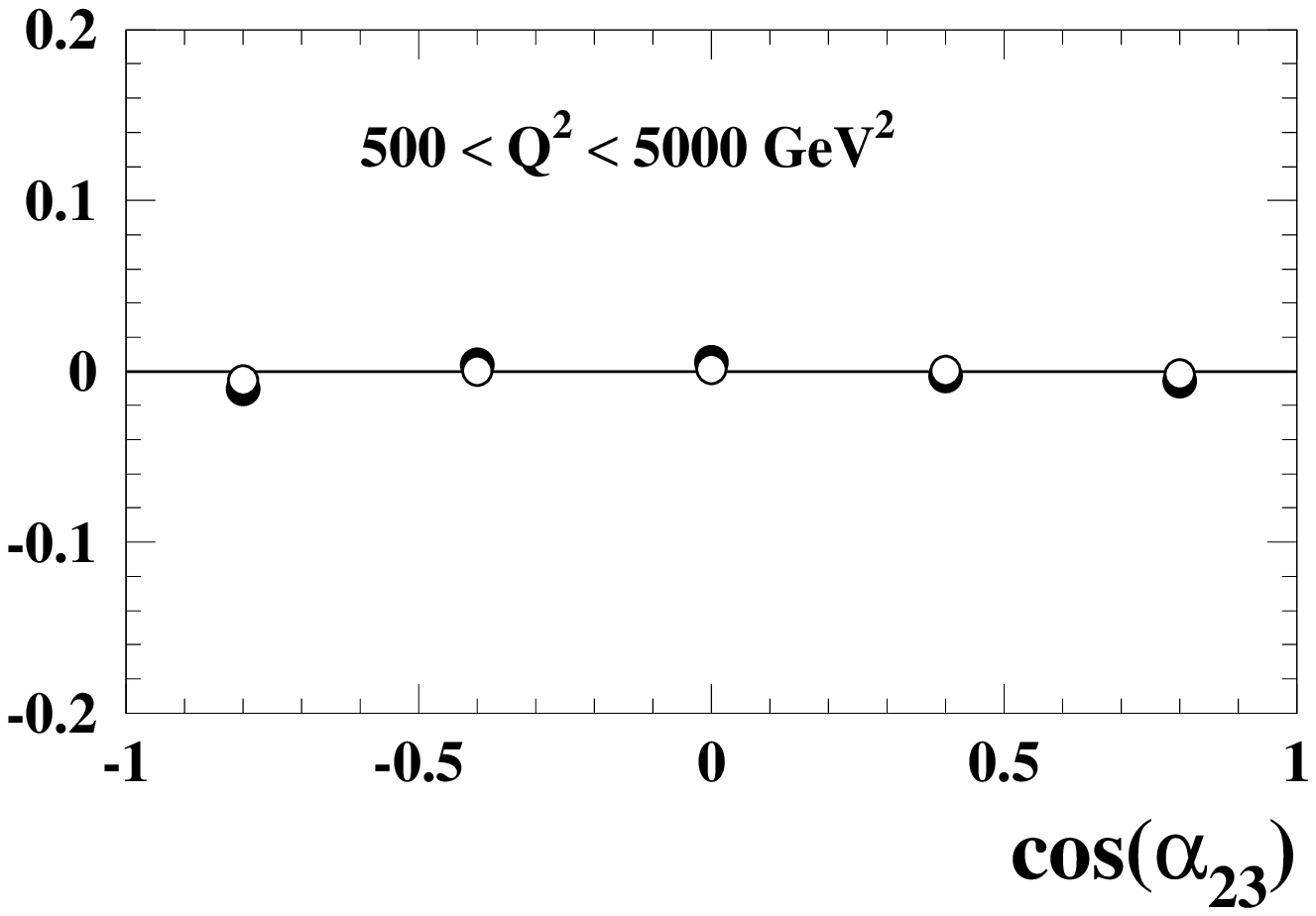,width=10cm}}
\put (0.0,-4.5){\epsfig{figure=\figdir 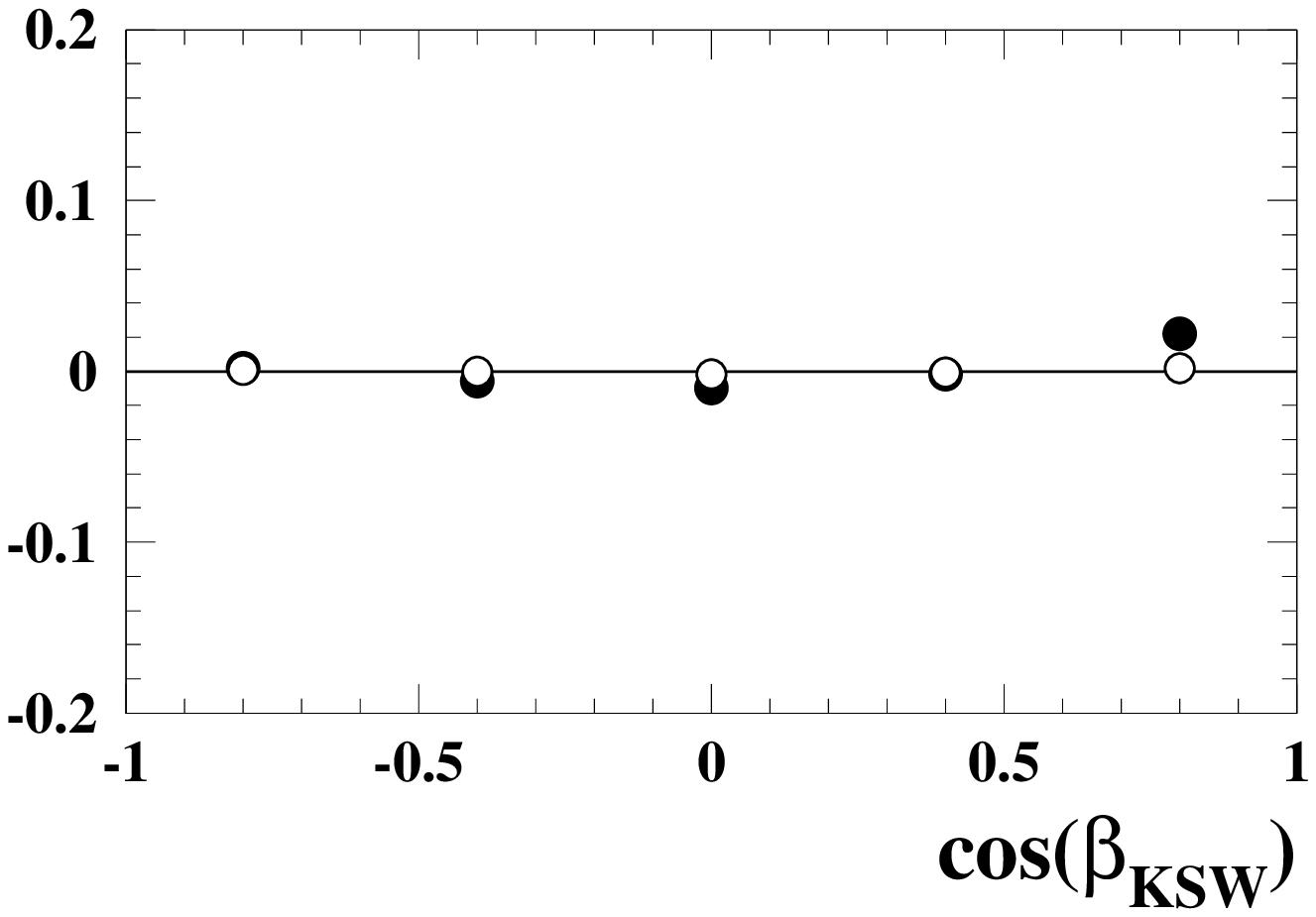,width=10cm}}
\put (7.0,-4.5){\epsfig{figure=\figdir 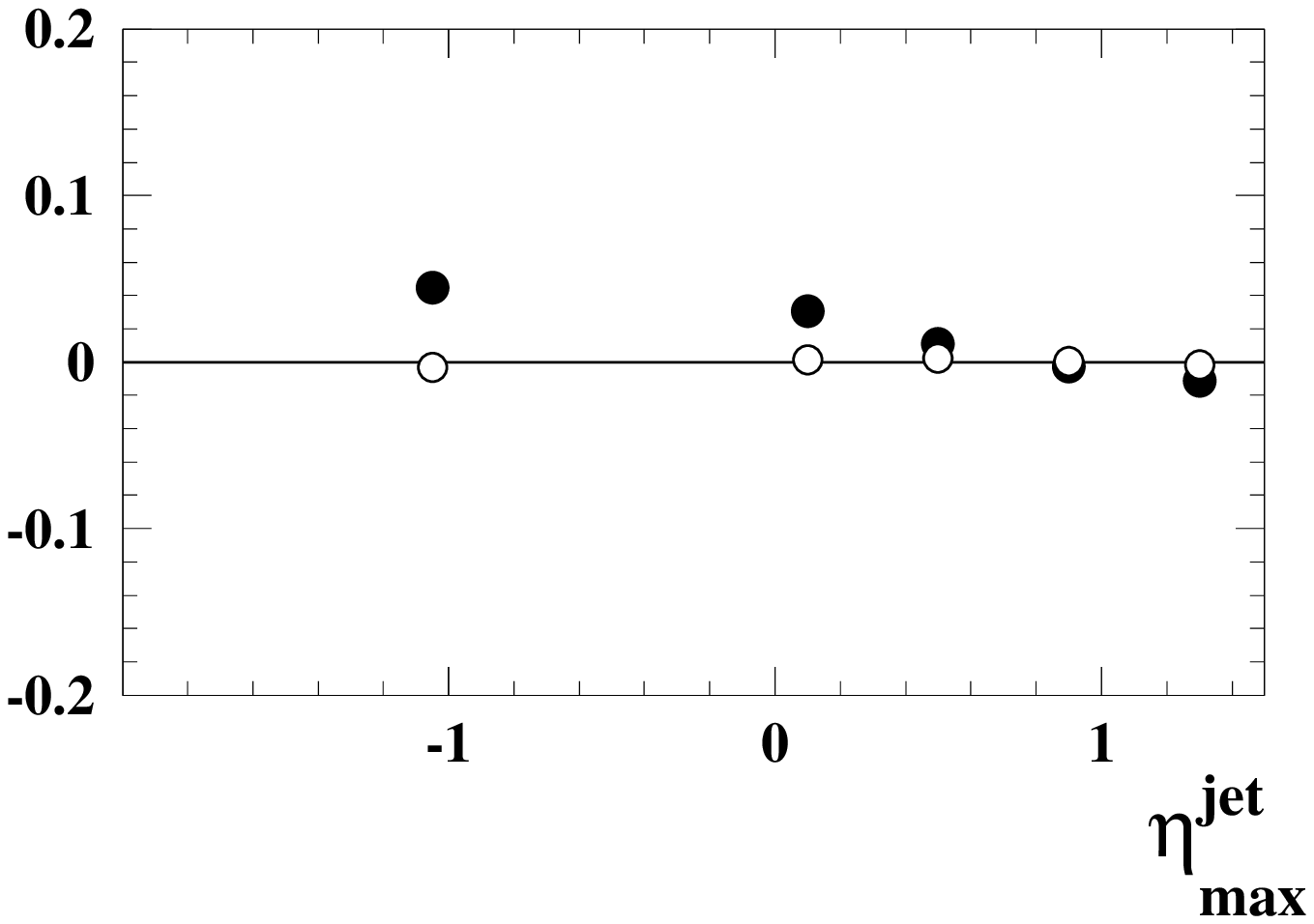,width=10cm}}
\put (7.0,19.0){\bf\small (a)}
\put (14.0,19.0){\bf\small (b)}
\put (7.0,14.0){\bf\small (c)}
\put (14.0,14.0){\bf\small (d)}
\put (7.0,9.0){\bf\small (e)}
\put (14.0,9.0){\bf\small (f)}
\put (7.0,4.0){\bf\small (g)}
\put (14.0,4.0){\bf\small (h)}
\end{picture}
\caption
{\it 
Relative difference between the $\oasss$ calculations with
fixed $\mu_R$ and the calculations with $\mu_R=Q$ (dots)
and between the $\oasss$ calculations with fixed $\mu_F$ and the
calculations with $\mu_F=Q$ (open circles) in NC DIS as functions of
(a,e) $\th$, (b,f) $\cos(\a34)$, (c,g) $\cos(\pksw)$ and (d,h)
$\etajmax$. These
calculations do not include corrections for hadronisation effects.
}
\label{fig9}
\vfill
\end{figure}

These studies demonstrate that the normalised cross sections have
little sensitivity to the evolution of the PDFs. However, there is
still some sensitivity to the running of $\as$. Figures~\ref{fig8}(e)
to \ref{fig8}(h) show the relative difference for $500<\q2<5000$~\g2.
The restriction of the phase space further reduces the dependence on
the running of $\as$; thus, this region is more suitable to extract
the colour factors in NC DIS at $\oass$. At $\oasss$ (see
Fig.~\ref{fig9}), the effect due to the running of $\as$ is already
very small for $\q2>125$~\g2. Therefore, the wider phase-space region
can be kept in an extraction of the colour factors at $\oasss$.

The following theoretical uncertainties were considered (as an example
of the size of the uncertainties, an average value of the effect of
each uncertainty on the normalised cross section as a function of
$\th$ is shown in parentheses for $\gp$, NC DIS with $\q2>125$~\g2\
and NC DIS with $500<\q2<5000$~\g2):
\begin{itemize}
  \item the uncertainty in the modelling of the parton shower was
    estimated by using different models (see Section~\ref{mc}) to
    calculate the parton-to-hadron correction factors ($\pm 2.8\%$,
    $\pm 2.9\%$ and $\pm 5.8\%$);
  \item the uncertainty on the calculations due to higher-order terms
    was estimated by varying $\mu_R$ by a factor of two up and down 
    ($_{-0.8}^{+0.6}\%$, $\pm 1.6\%$ and $\pm 2.2\%$);
  \item the uncertainty on the calculations due to those on the proton
    PDFs was estimated by repeating the calculations using 22
    additional sets from the ZEUS analysis~\cite{pr:d67:012007}; this
    analysis takes into account the statistical and correlated
    systematic experimental uncertainties of each data set used in the
    determination of the proton PDFs ($\pm 0.7\%$, $\pm 0.2\%$ and
    $\pm 0.1\%$);
  \item the uncertainty on the calculations due to that on $\asz$ was
    estimated by repeating the calculations using two additional sets
    of proton PDFs, for which different values of $\asz$ were assumed
    in the fits. The difference between the calculations using these
    various sets was scaled to reflect the uncertainty on the current
    world average of $\as$~\cite{jp:g26:r27} (negligible in all cases);
  \item the uncertainty of the calculations due to the choice of
    $\mu_F$ was estimated by varying $\mu_F$ by a factor of two up and
    down (negligible in all cases).
\end{itemize}

The total theoretical uncertainty was obtained by adding in quadrature
the individual uncertainties listed above. The dominant source of
theoretical uncertainty is that on the modelling of the parton shower.

\section{Definition of the cross sections}
\label{cs}

Normalised differential three-jet cross sections were measured as
functions of $\th$, $\a34$ and $\pksw$ using the selected data samples
in $\gp$ and NC DIS. For NC DIS, the normalised differential three-jet
cross section as a function of $\etajmax$ was also measured. The
normalised differential three-jet cross section in bin $i$ for an
observable $A$ was obtained using

$$\frac{1}{\sigma}\frac{d\sigma_i}{dA}=\frac{1}{\sigma}\frac{N_{{\rm data},i}}{{\cal L}\cdot \Delta A_i}\cdot\frac{N^{\rm had}_{{\rm MC},i}}{N^{\rm det}_{{\rm MC},i}},$$
where $N_{{\rm data},i}$ is the number of data events in bin $i$, 
$N^{\rm had}_{{\rm MC},i}\ (N^{\rm det}_{{\rm MC},i})$ is the number
of MC events at hadron (detector) level, ${\cal L}$ is the integrated
luminosity and $\Delta A_i$ is the bin width. The integrated three-jet
cross section, $\sigma$, was computed using the formula:

$$\sigma=\sum_i\frac{N_{{\rm data},i}}{{\cal L}}\cdot\frac{N^{\rm had}_{{\rm MC},i}}{N^{\rm det}_{{\rm MC},i}},$$
where the sum runs over all bins.

For the $\gp$ sample, due to the different centre-of-mass
energies of the two data sets used in the analysis, the measured
normalised differential three-jet cross sections were combined using

$$\sigma^{\rm
  comb}=\frac{\sigma_{300}\cdot{\cal{L}}_{300}+\sigma_{318}\cdot{\cal{L}}_{318}}{{\cal{L}}_{300}+{\cal{L}}_{318}},$$
where ${\cal{L}}_{\sqrt{s}}$ is the luminosity and $\sigma_{\sqrt{s}}$
is the measured cross section corresponding to $\sqrt s=300$ or
$318$~GeV. This formula was applied for combining the differential and
total cross sections. The same formula was used for computing the
$\oass$ predictions in $\gp$.

\section{Acceptance corrections and experimental uncertainties}
\label{expunc}

The {\sc Pythia} (MEPS) MC samples were used to compute the acceptance
corrections to the angular distributions of the $\gp$ (NC DIS)
data. These correction factors took into account the efficiency of the
trigger, the selection criteria and the purity and efficiency of the
jet reconstruction. The samples of {\sc Herwig} and CDM were used to 
compute the systematic uncertainties coming from the fragmentation and
parton-shower models in $\gp$ and NC DIS, respectively.

\begin{figure}[p]
\vfill
\setlength{\unitlength}{1.0cm}
\begin{picture} (18.0,17.0)
\put (-0.3,8.0){\centerline{\epsfig{figure=\figdir 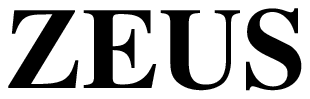,width=10cm}}}
\put (-1.0,7.5){\epsfig{figure=\figdir 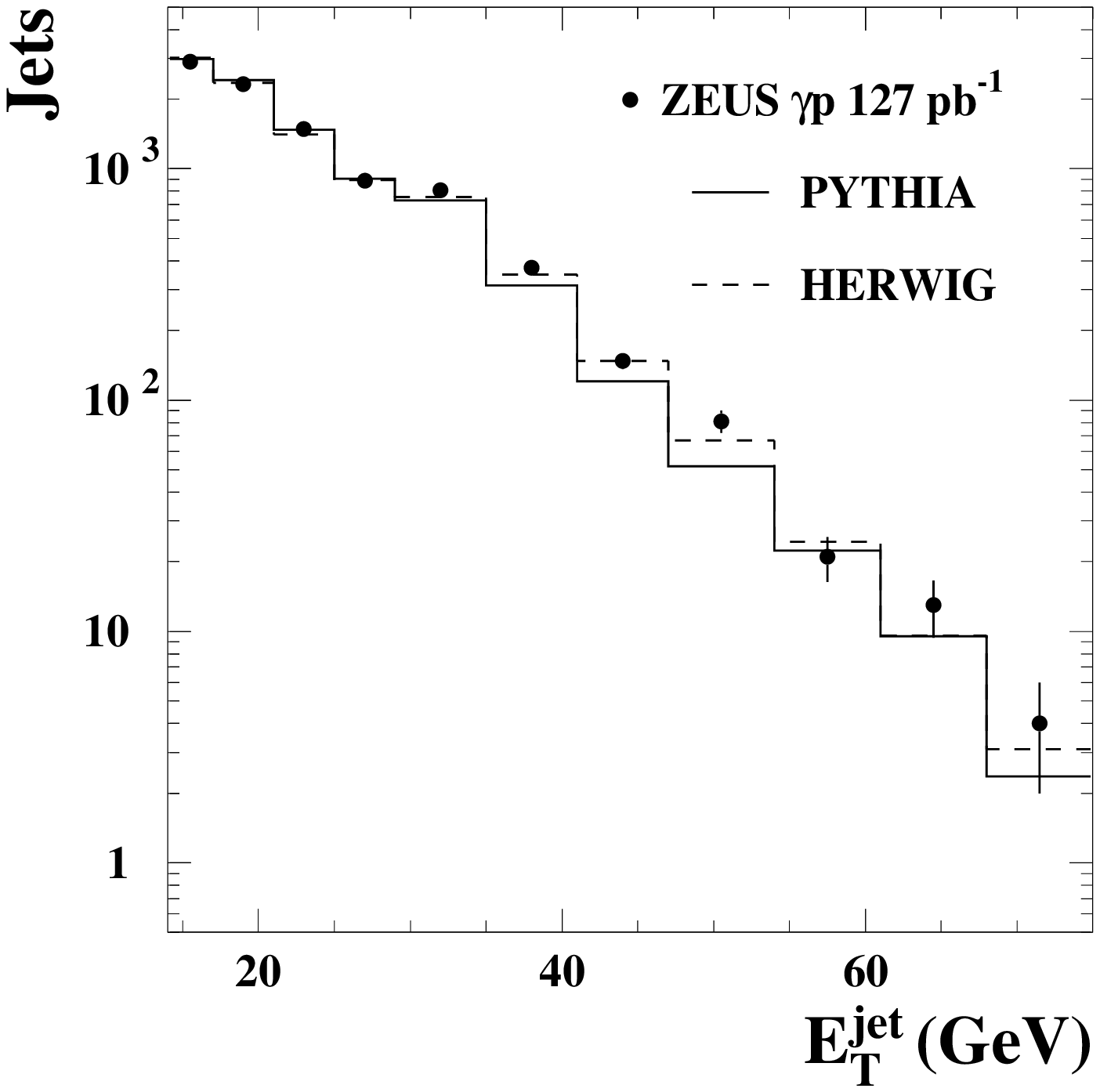,width=10cm}}
\put (6.5,7.5){\epsfig{figure=\figdir 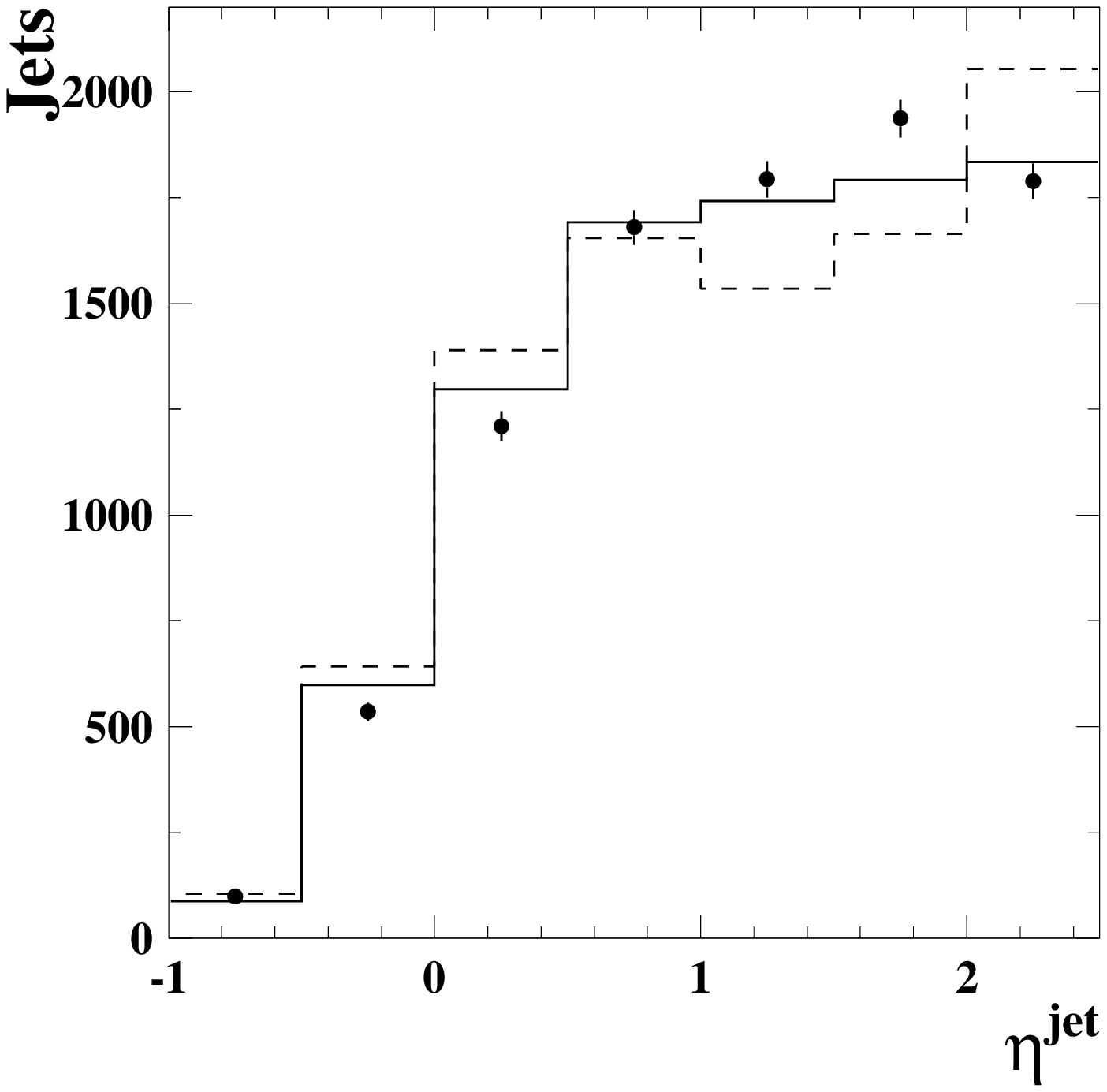,width=10cm}}
\put (-1.0,-0.5){\epsfig{figure=\figdir 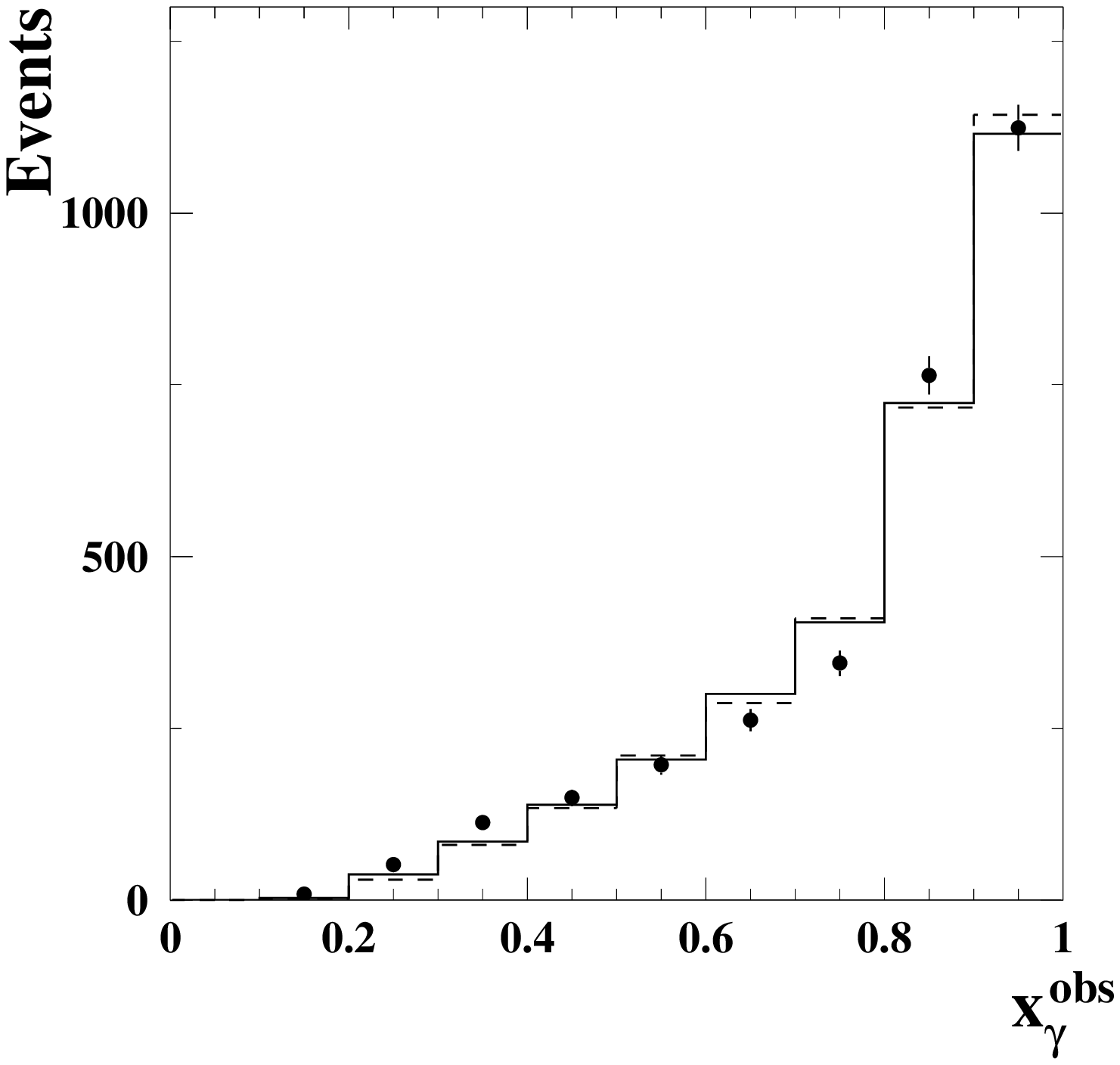,width=10cm}}
\put (6.5,-0.5){\epsfig{figure=\figdir 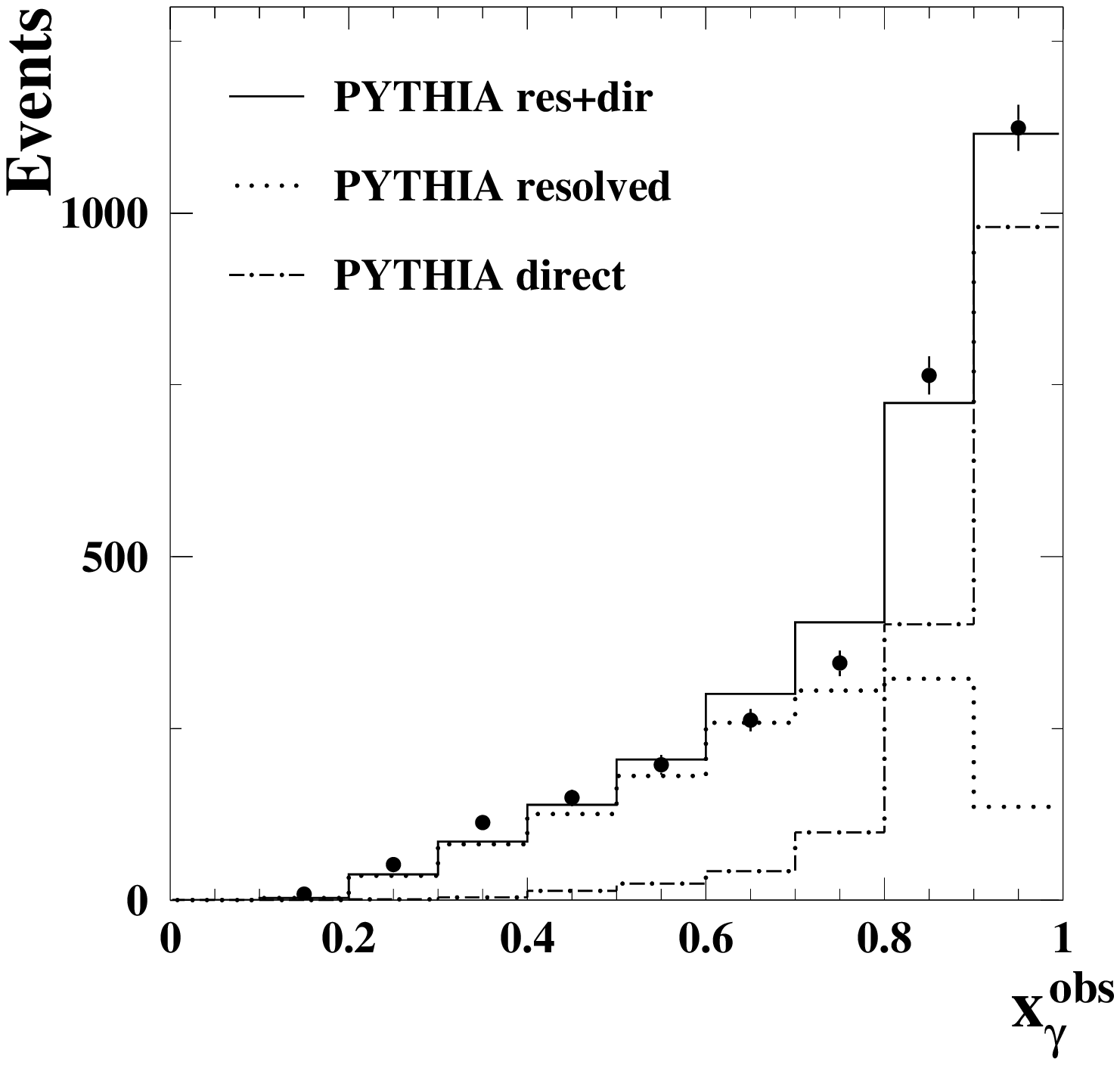,width=10cm}}
\put (6.3,13.0){\bf\small (a)}
\put (8.8,15.0){\bf\small (b)}
\put (6.3,7.0){\bf\small (c)}
\put (13.8,7.0){\bf\small (d)}
\end{picture}
\caption
{\it 
Detector-level data distributions for three-jet photoproduction (dots)
with $\etjet>14$ GeV and $-1<\etajet<2.5$ in the kinematic region
given by $\q2<1$~\gf2\ and $0.2<y<0.85$ as functions of (a) $\etjet$,
(b) $\etajet$ and (c,d) $\xo$. For comparison, the distributions of
the {\sc Pythia} (solid histograms) and {\sc Herwig} (dashed
histograms) MC models for resolved plus direct processes
normalised to the data are included. In (d), the contributions for
resolved (dotted histogram) and direct (dot-dashed histogram)
processes from {\sc Pythia} MC are shown separately.
}
\label{fig2}
\vfill
\end{figure}

The data $\etjet$, $\etajet$ and $\xo$ distributions of the $\gp$
sample, before the $\xo>0.8$ requirement, are shown in Fig.~\ref{fig2}
together with the MC simulations of {\sc Pythia} and {\sc
  Herwig}. Considering that three-jet events in the MC arise only from
the parton-shower approximation, the description of the data is
reasonable. Figure~\ref{fig2}(d) shows the resolved and direct
contributions for the {\sc Pythia} MC separately. It is observed that
the region of $\xo>0.8$ is dominated by direct $\gp$ events. The
remaining contribution in this region from resolved-photon events was
estimated using {\sc Pythia} ({\sc Herwig}) simulated events to be
$\approx 25\ (31)\%$.

\begin{figure}[p]
\vfill
\setlength{\unitlength}{1.0cm}
\begin{picture} (18.0,17.0)
\put (-0.3,8.0){\centerline{\epsfig{figure=\figdir zeus.eps,width=10cm}}}
\put (-1.0,7.5){\epsfig{figure=\figdir 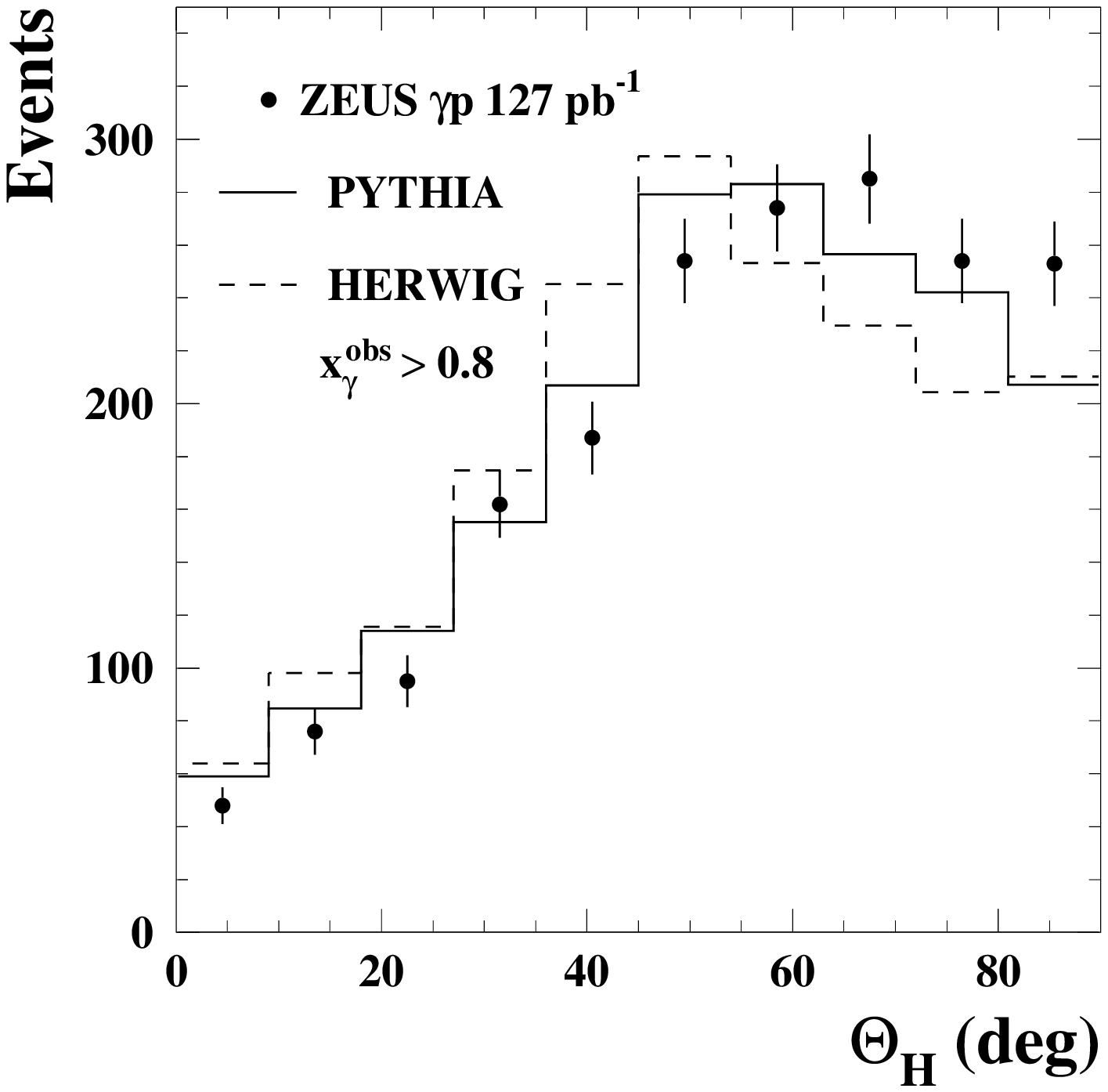,width=10cm}}
\put (6.5,7.5){\epsfig{figure=\figdir 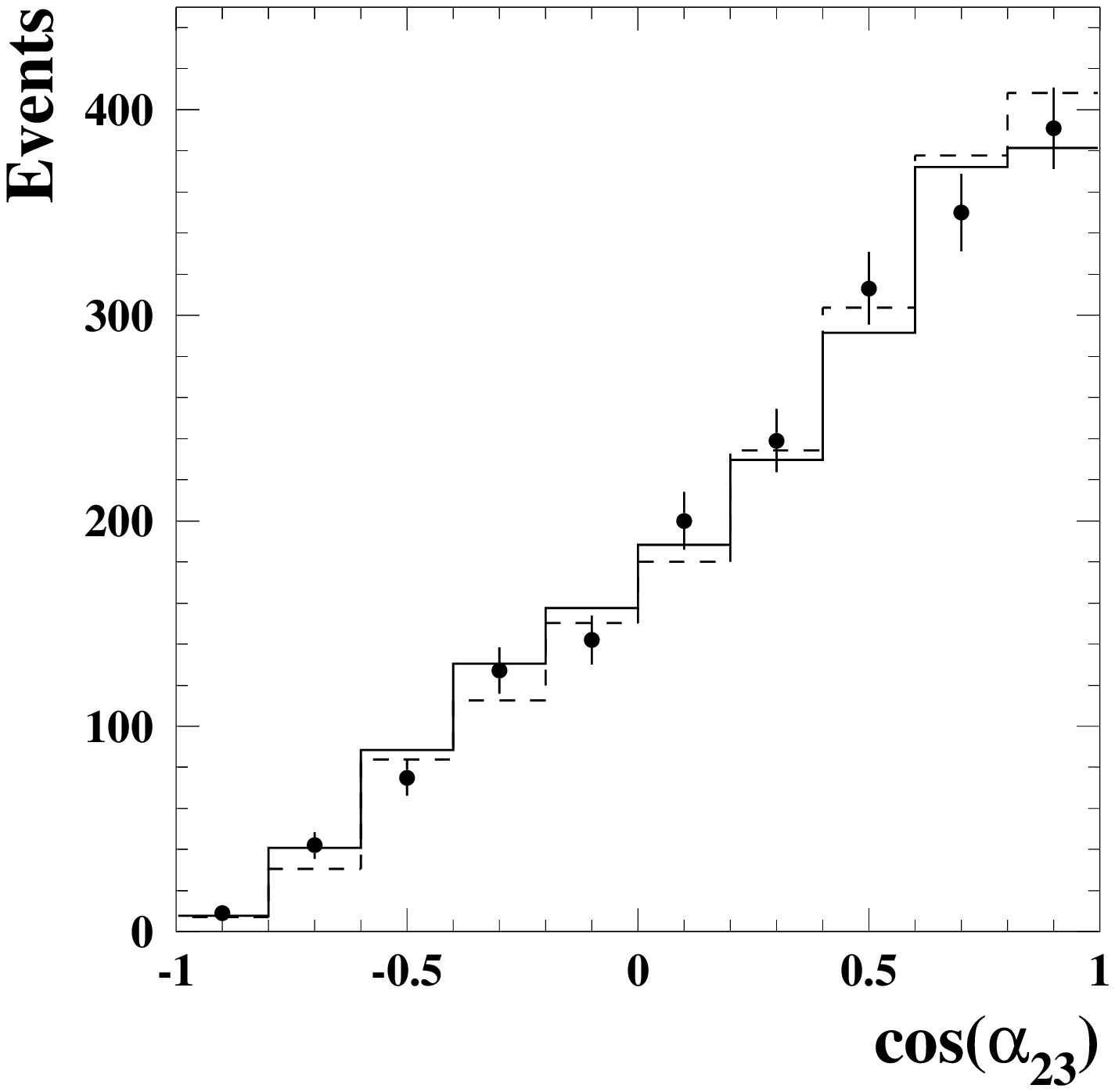,width=10cm}}
\put (-1.0,-0.5){\epsfig{figure=\figdir 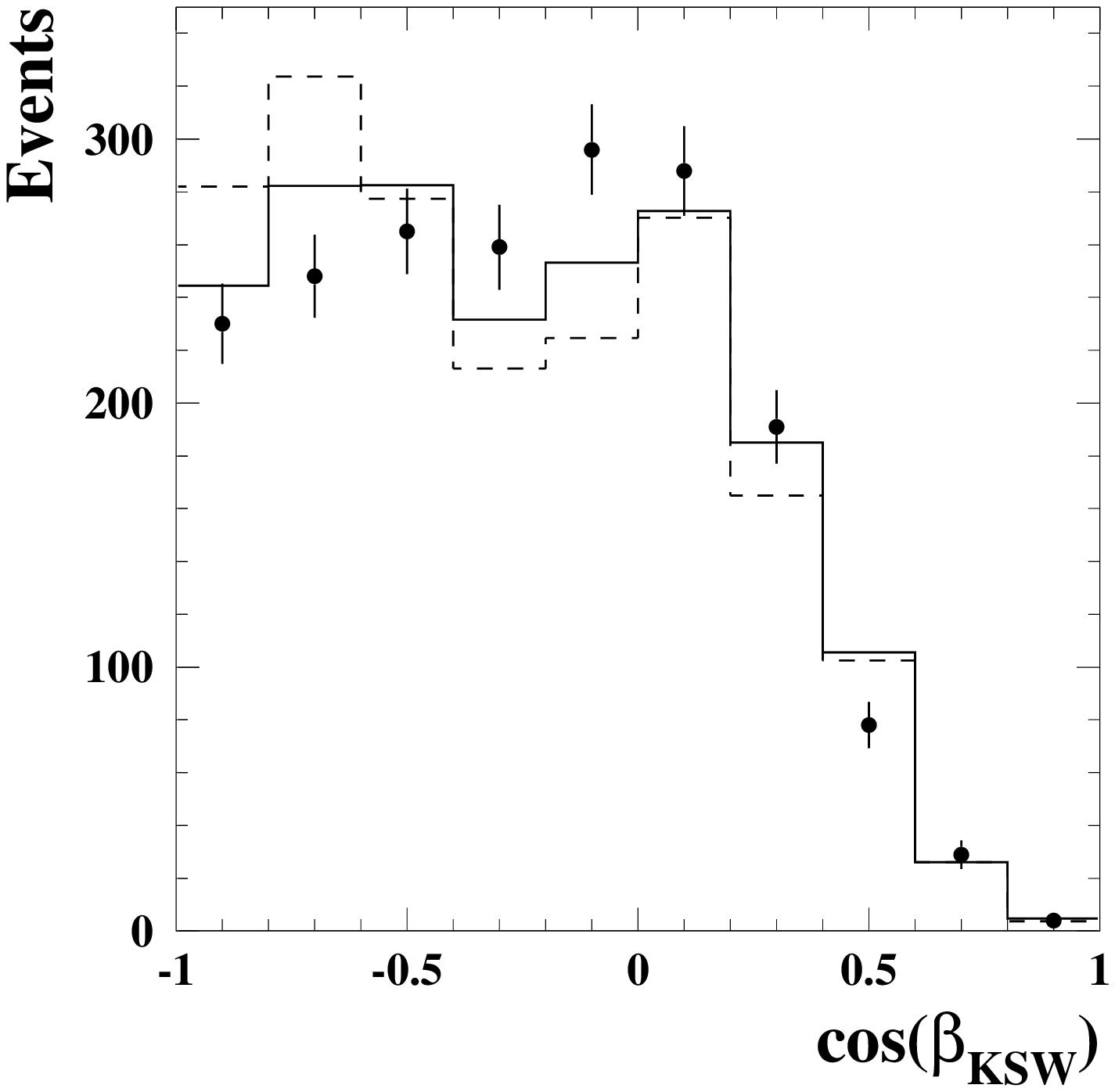,width=10cm}}
\put (6.3,15.0){\bf\small (a)}
\put (8.8,15.0){\bf\small (b)}
\put (6.3,7.0){\bf\small (c)}
\end{picture}
\caption
{\it 
Detector-level data distributions for three-jet photoproduction (dots)
with $\etjet>14$ GeV and $-1<\etajet<2.5$ in the kinematic region
given by $\q2<1$~\gf2, $0.2<y<0.85$ and $\xo>0.8$ as functions of (a)
$\th$, (b) $\cos(\a34)$ and (c) $\cos(\pksw)$. Other details as in the
caption to Fig.~\ref{fig2}.
}
\label{fig3}
\vfill
\end{figure}

\begin{figure}[p]
\vfill
\setlength{\unitlength}{1.0cm}
\begin{picture} (18.0,17.0)
\put (-0.3,8.0){\centerline{\epsfig{figure=\figdir zeus.eps,width=10cm}}}
\put (-1.0,7.5){\epsfig{figure=\figdir 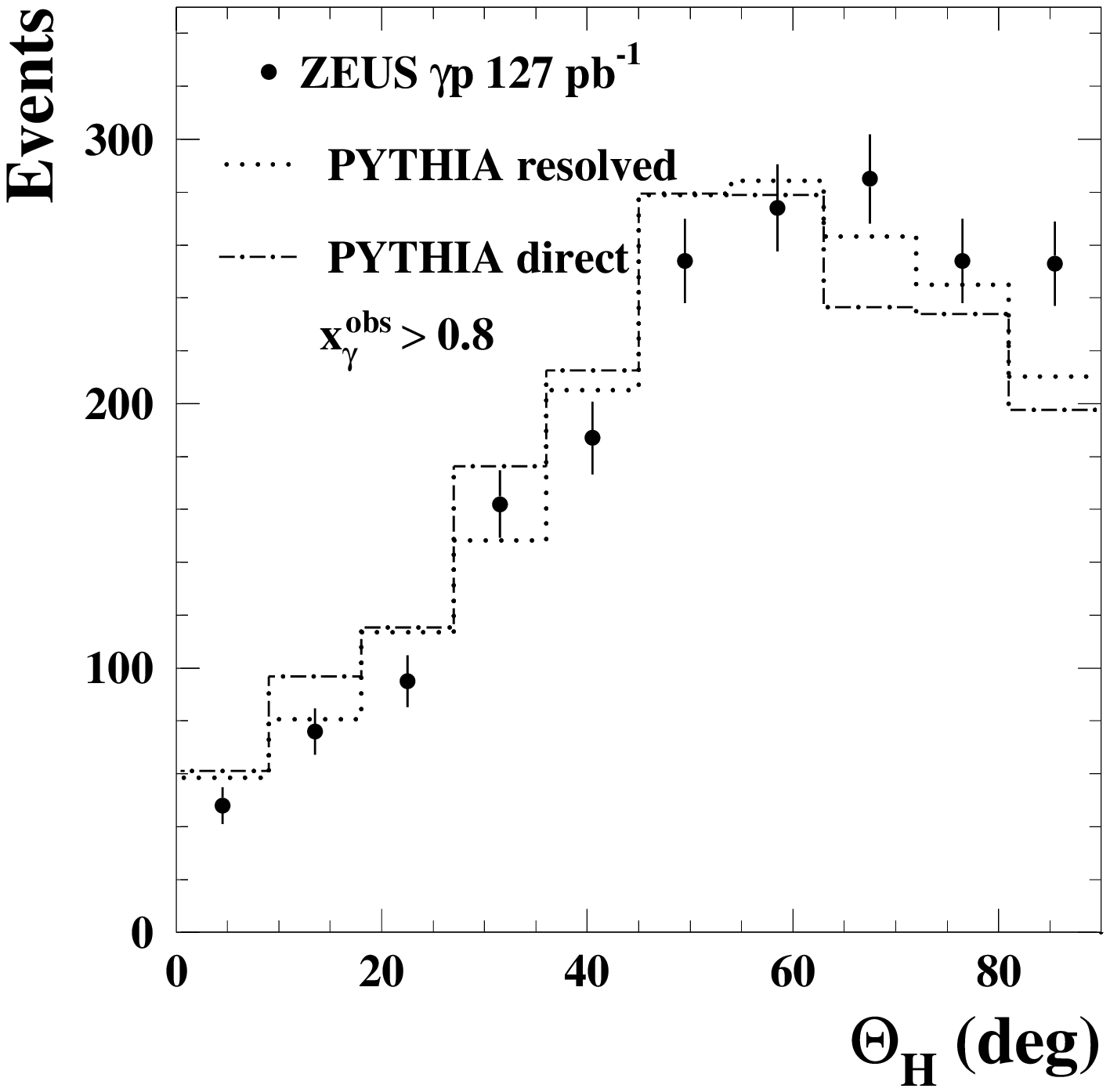,width=10cm}}
\put (6.5,7.5){\epsfig{figure=\figdir 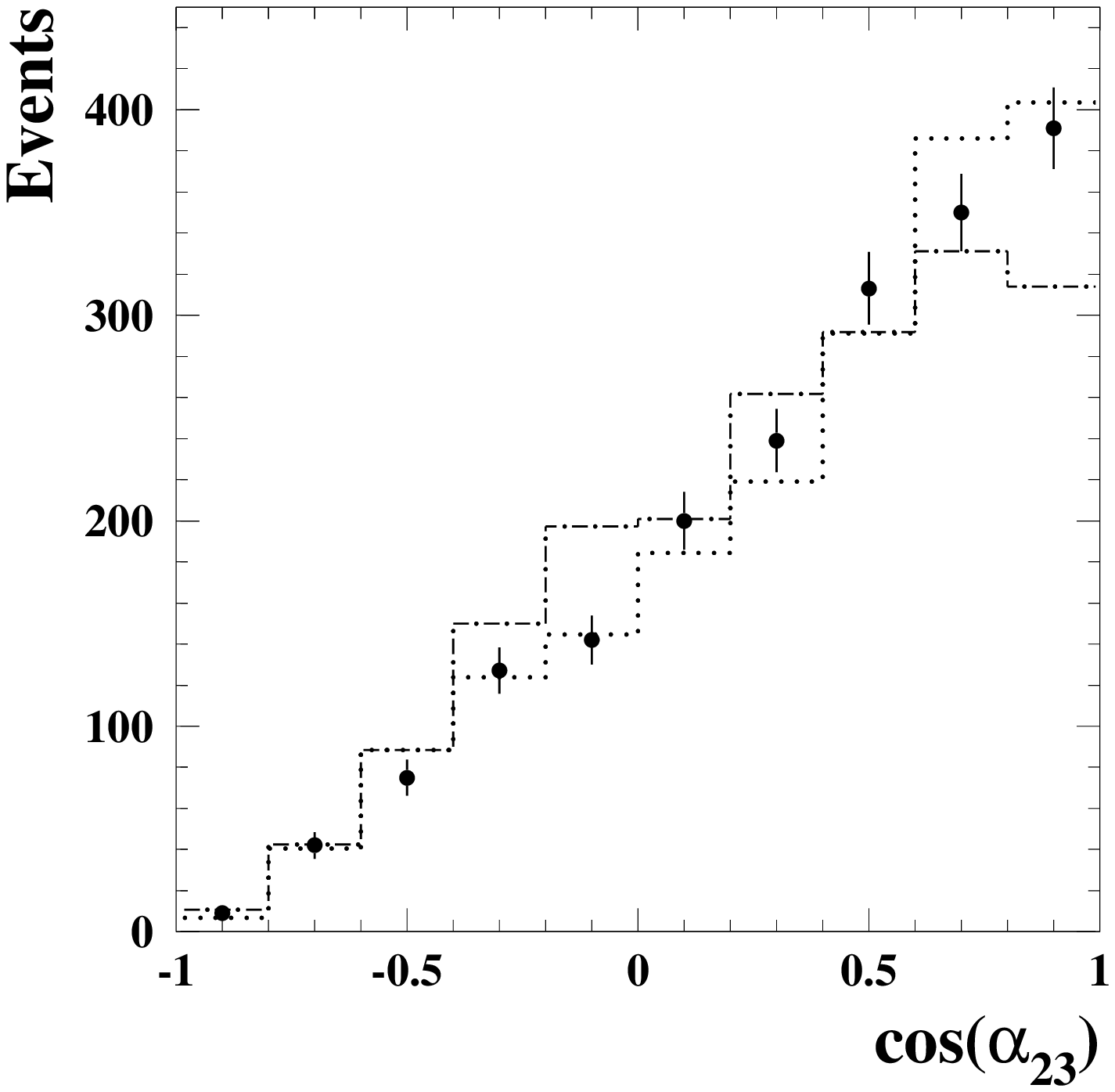,width=10cm}}
\put (-1.0,-0.5){\epsfig{figure=\figdir 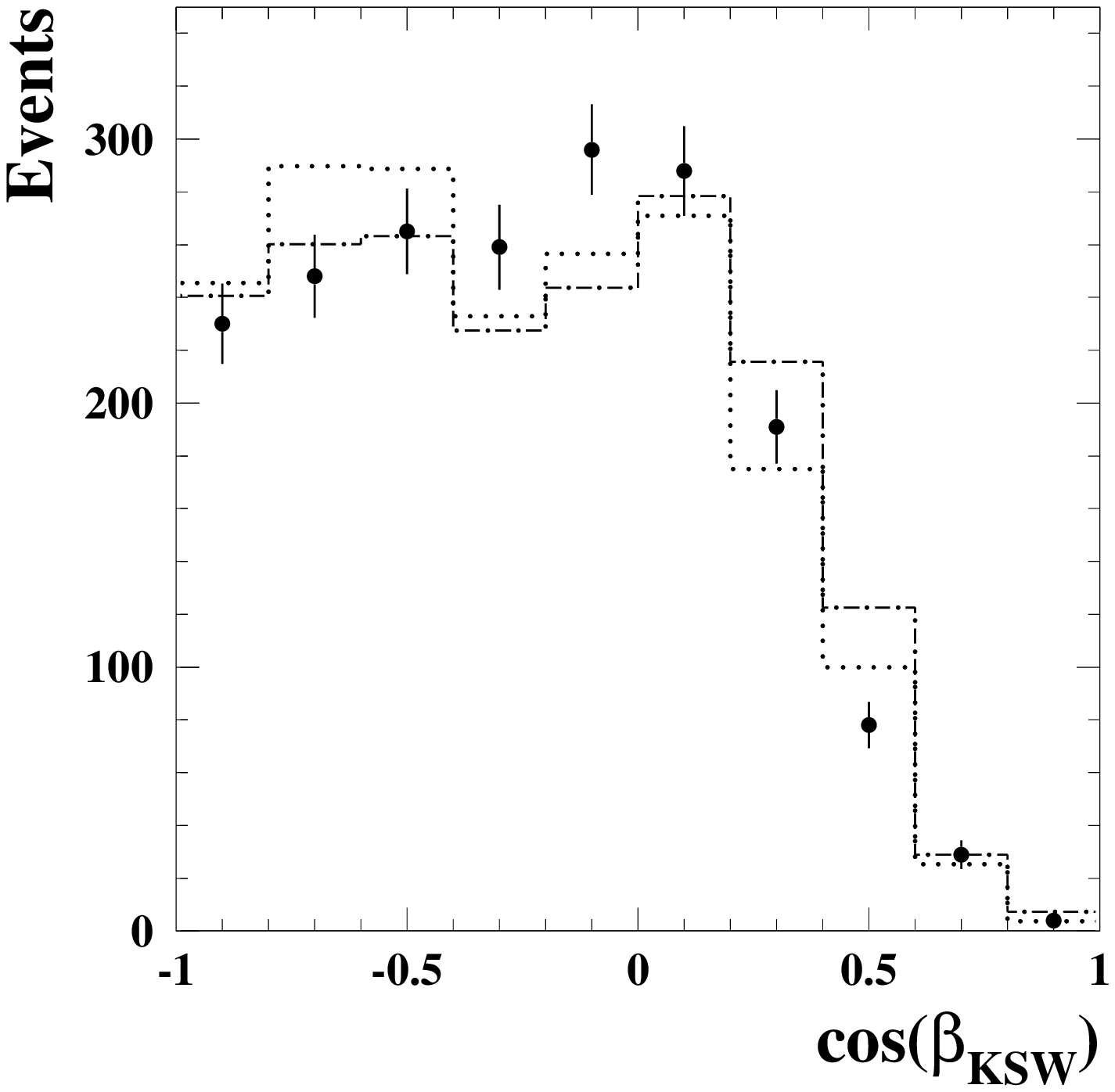,width=10cm}}
\put (6.3,15.0){\bf\small (a)}
\put (8.8,15.0){\bf\small (b)}
\put (6.3,7.0){\bf\small (c)}
\end{picture}
\caption
{\it 
Detector-level data distributions for three-jet photoproduction (dots)
with $\etjet>14$ GeV and $-1<\etajet<2.5$ in the kinematic region
given by $\q2<1$~\gf2, $0.2<y<0.85$ and $\xo>0.8$ as functions of (a)
$\th$, (b) $\cos(\a34)$ and (c) $\cos(\pksw)$.  The predictions for
resolved (dotted histogram) and direct (dot-dashed histogram)
processes from the {\sc Pythia} MC normalised separately to the data
are also shown. Other details as in the caption to Fig.~\ref{fig2}.
}
\label{fig4}
\vfill
\end{figure}

Figure~\ref{fig3} shows the data distributions as functions of $\th$,
$\a34$ and $\pksw$ together with the simulations of {\sc Pythia} and
{\sc Herwig} for $\xo>0.8$. The {\sc Pythia} MC predictions describe
the data distributions well, whereas the description given by {\sc
  Herwig} is somewhat poorer. It was checked that the angular
distributions of the events from resolved processes with $\xo>0.8$
were similar to those from direct processes (see Fig.~\ref{fig4}) and,
therefore, no subtraction of the resolved processes was performed when
comparing to the fixed-order calculations described in
Section~\ref{nlo}.

\begin{figure}[p]
\vfill
\setlength{\unitlength}{1.0cm}
\begin{picture} (18.0,17.0)
\put (-0.3,8.0){\centerline{\epsfig{figure=\figdir zeus.eps,width=10cm}}}
\put (-1.0,7.5){\epsfig{figure=\figdir 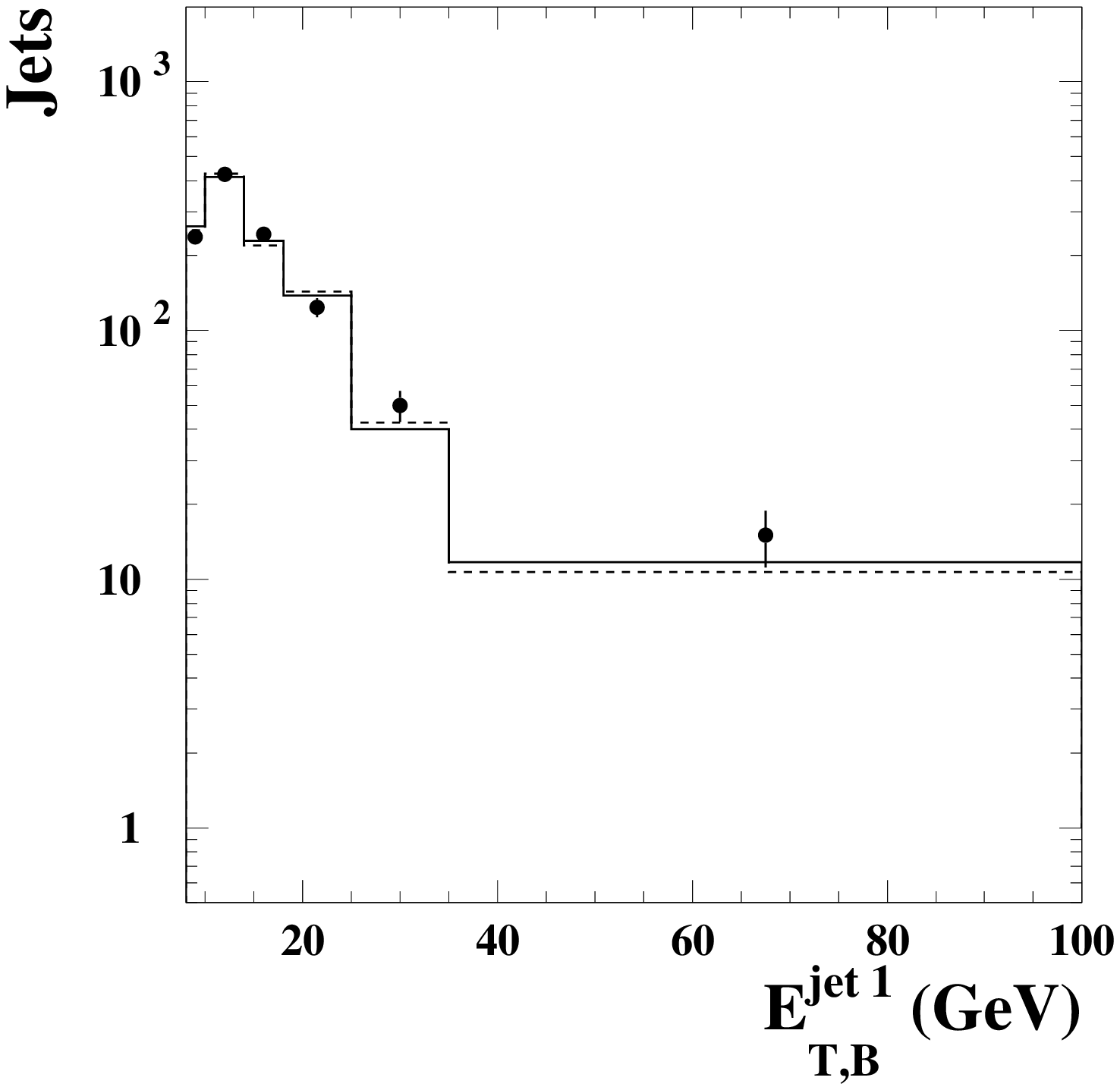,width=10cm}}
\put (-1.0,7.5){\epsfig{figure=\figdir 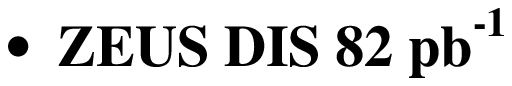,width=10cm}}
\put (6.5,7.5){\epsfig{figure=\figdir 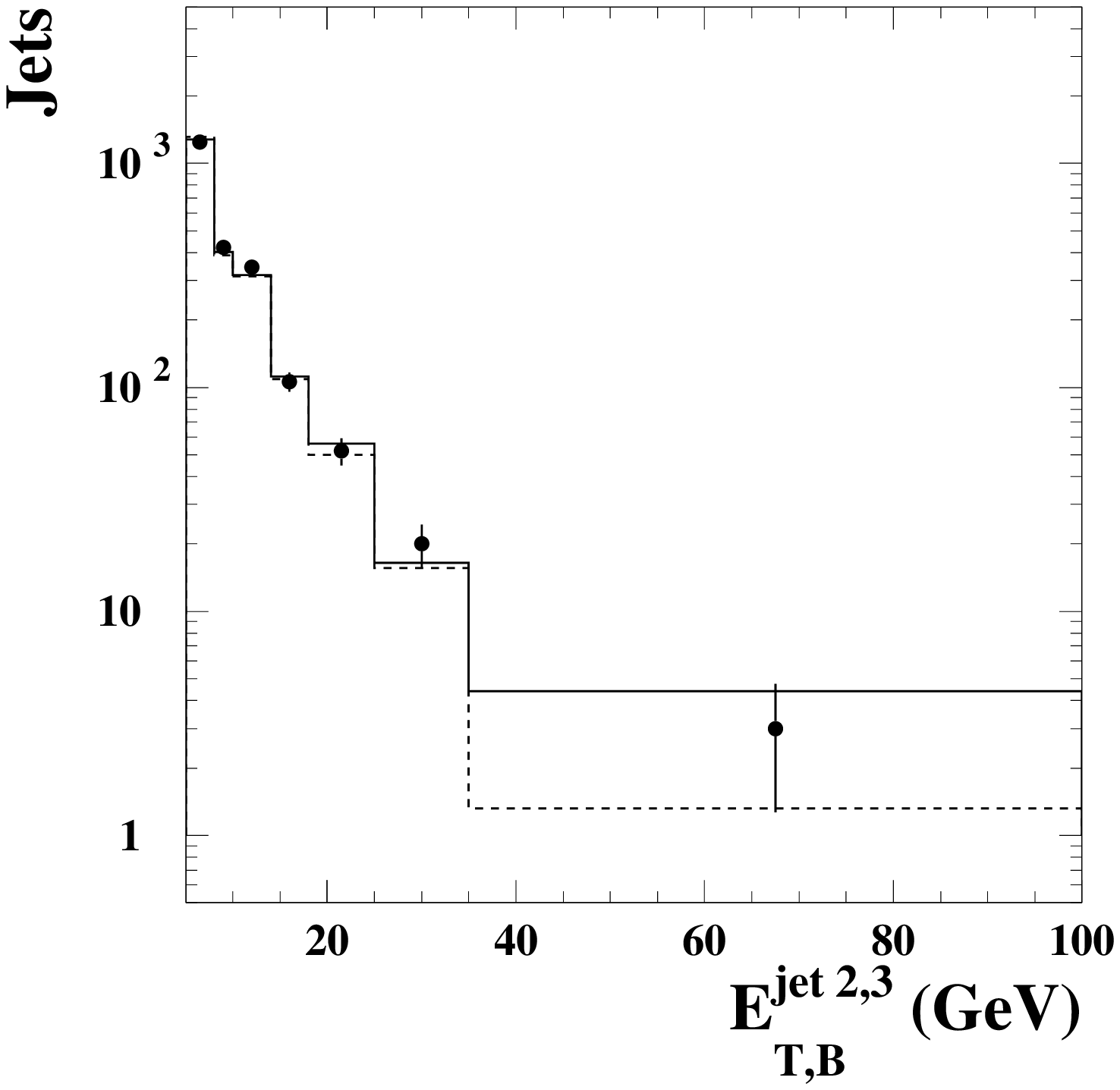,width=10cm}}
\put (-1.0,-0.5){\epsfig{figure=\figdir 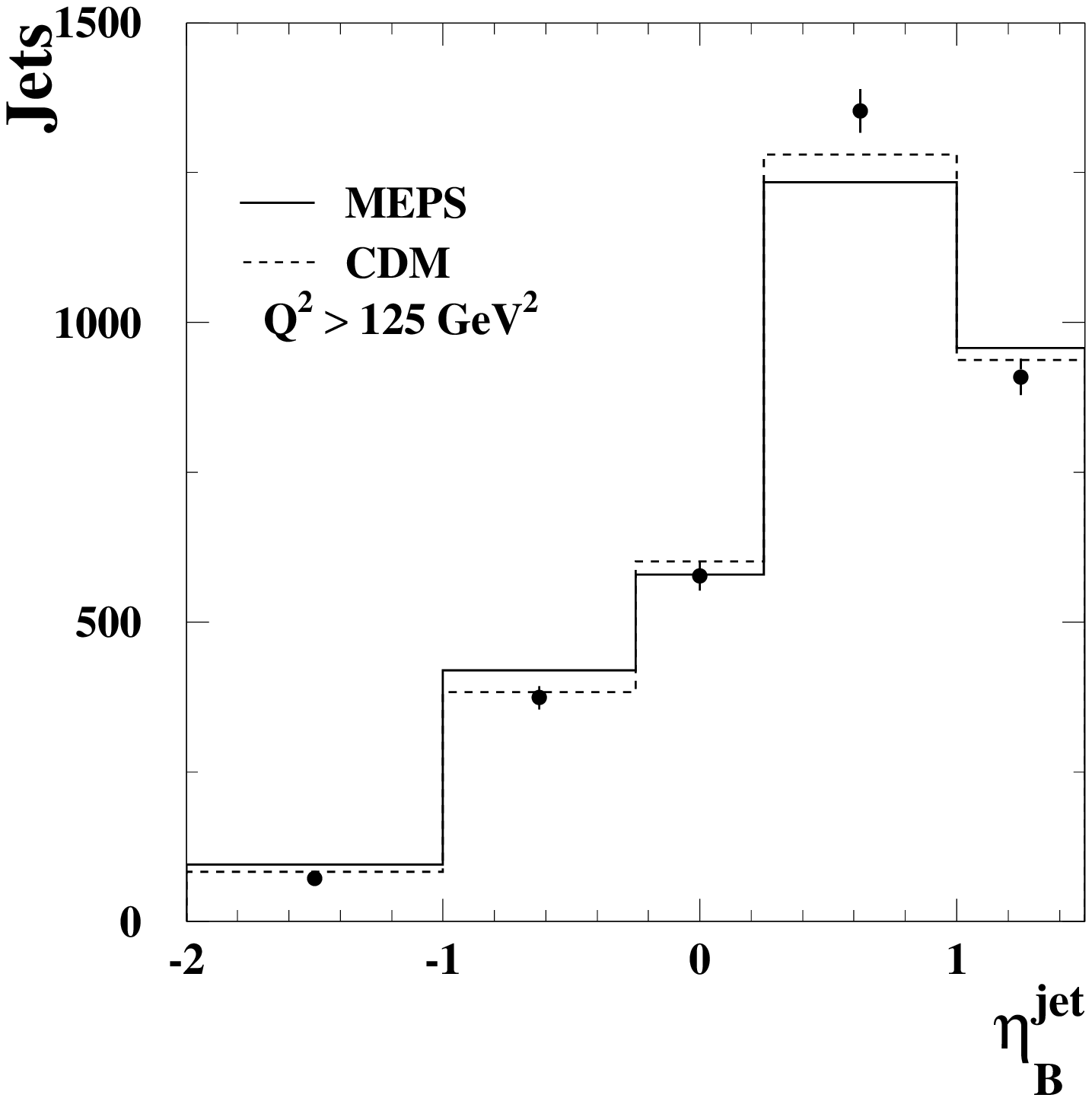,width=10cm}}
\put (6.5,-0.5){\epsfig{figure=\figdir 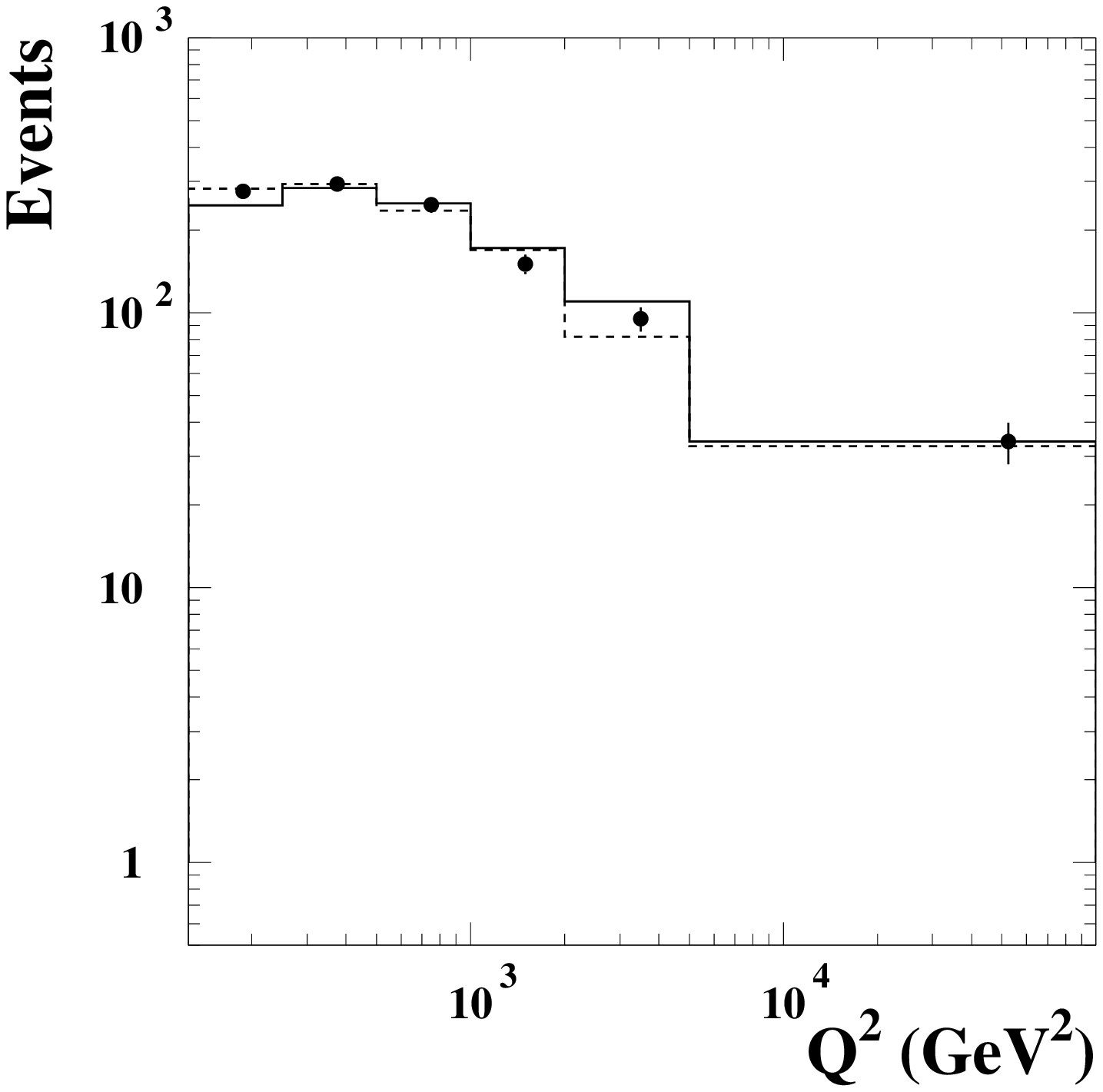,width=10cm}}
\put (6.3,15.0){\bf\small (a)}
\put (13.8,15.0){\bf\small (b)}
\put (6.3,7.0){\bf\small (c)}
\put (13.8,7.0){\bf\small (d)}
\end{picture}
\caption
{\it 
Detector-level data distributions for three-jet production in NC DIS  
(dots) with $\etjbj>8$~GeV, $E^{\rm jet2,3}_{T,{\rm B}}>5$ GeV and 
$-2<\etajb<1.5$ in the kinematic region given by $\q2>125$~\gf2\ and 
$|\cgh|<0.65$ as functions of (a) $\etjbj$, 
(b) $E^{\rm jet2,3}_{T,{\rm B}}$, (c) $\etajb$ and (d) $\q2$. For
comparison, the distributions of the MEPS (solid histograms) and CDM
(dashed histograms) MC models normalised to the data are included.
}
\label{fig23}
\vfill
\end{figure}

\begin{figure}[p]
\vfill
\setlength{\unitlength}{1.0cm}
\begin{picture} (18.0,17.0)
\put (-0.3,8.0){\centerline{\epsfig{figure=\figdir zeus.eps,width=10cm}}}
\put (-1.0,7.5){\epsfig{figure=\figdir 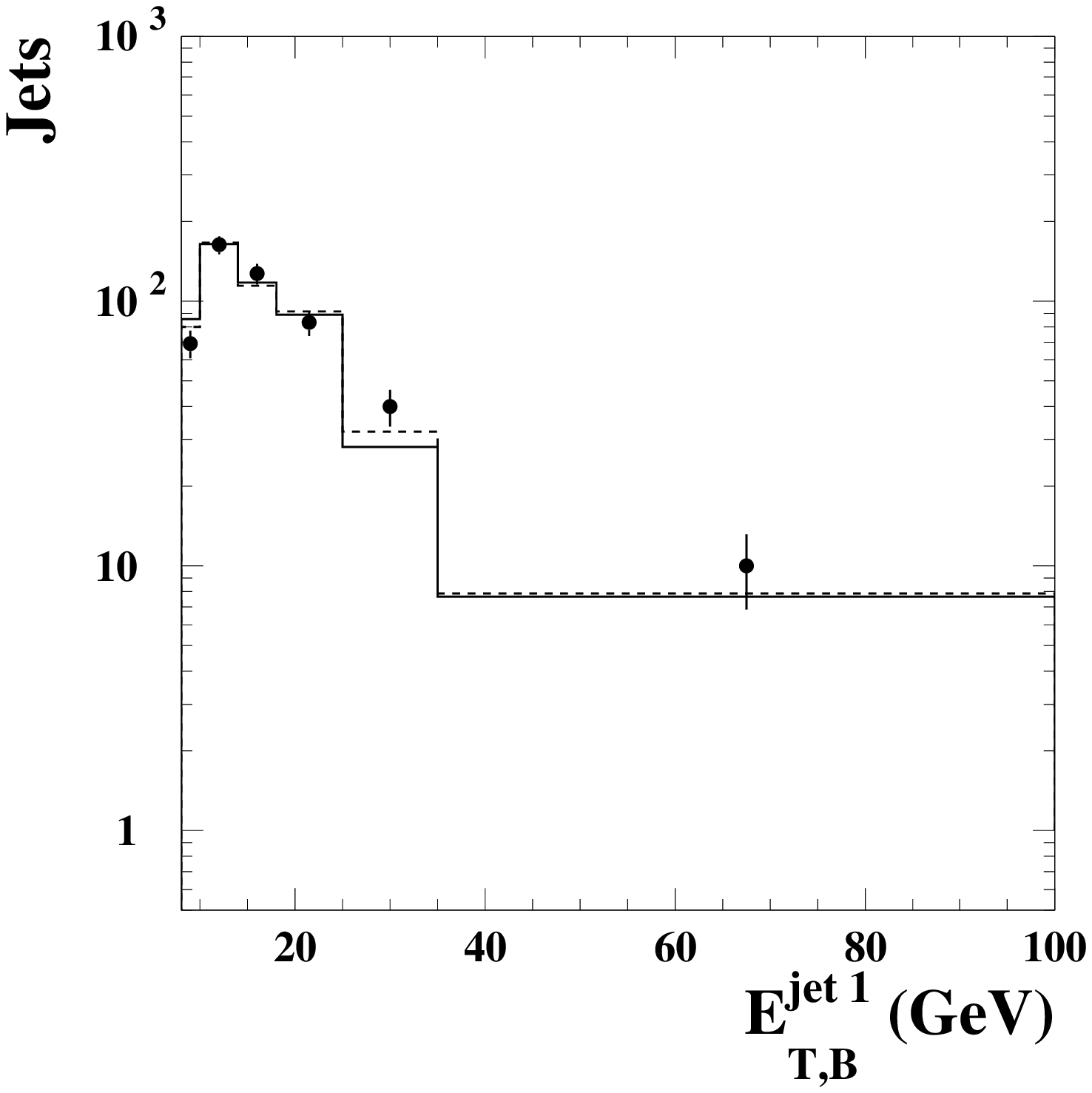,width=10cm}}
\put (-1.0,7.5){\epsfig{figure=\figdir 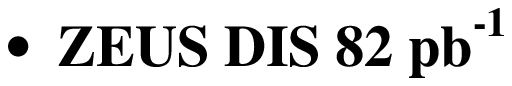,width=10cm}}
\put (6.5,7.5){\epsfig{figure=\figdir 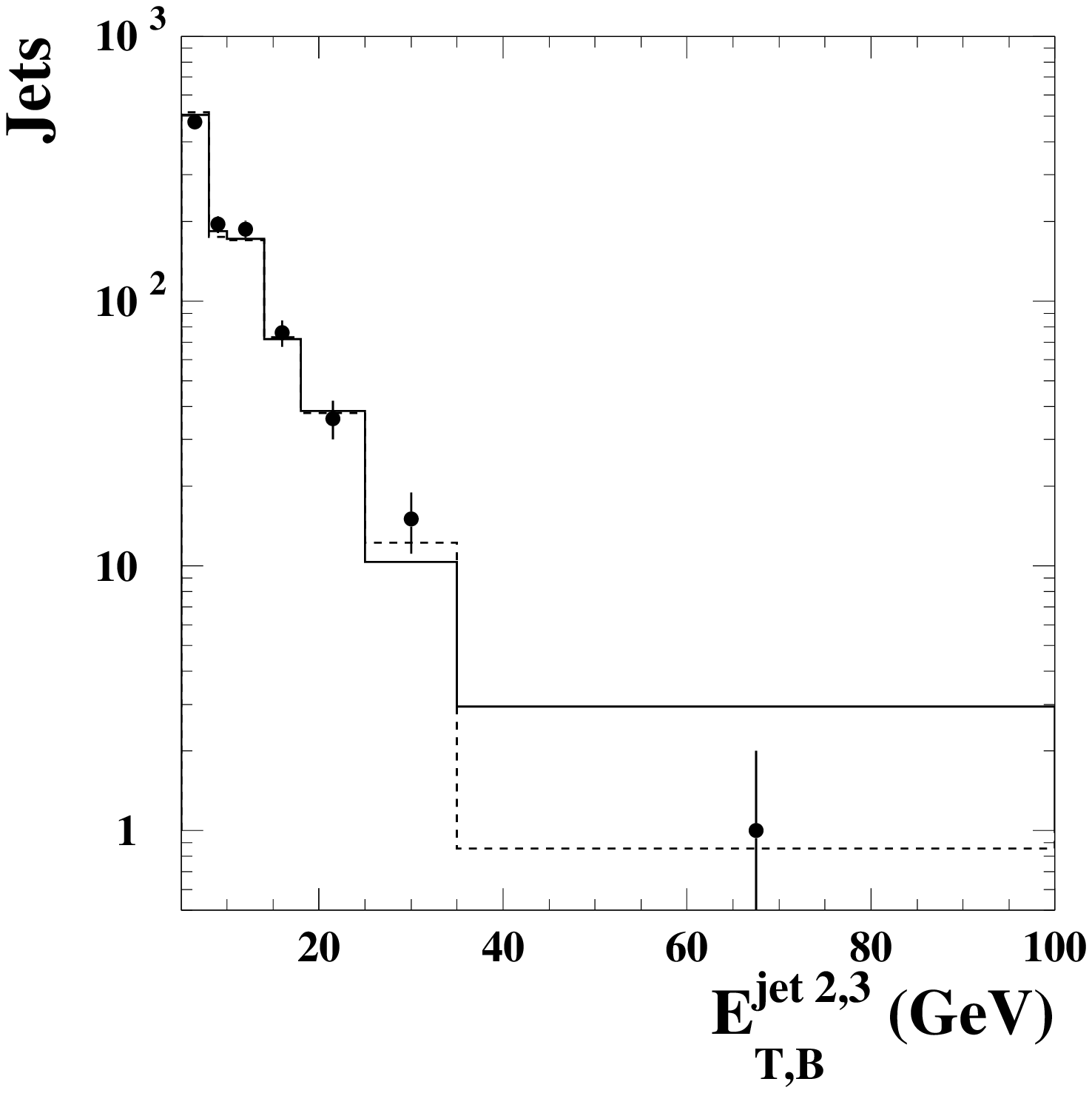,width=10cm}}
\put (-1.0,-0.5){\epsfig{figure=\figdir 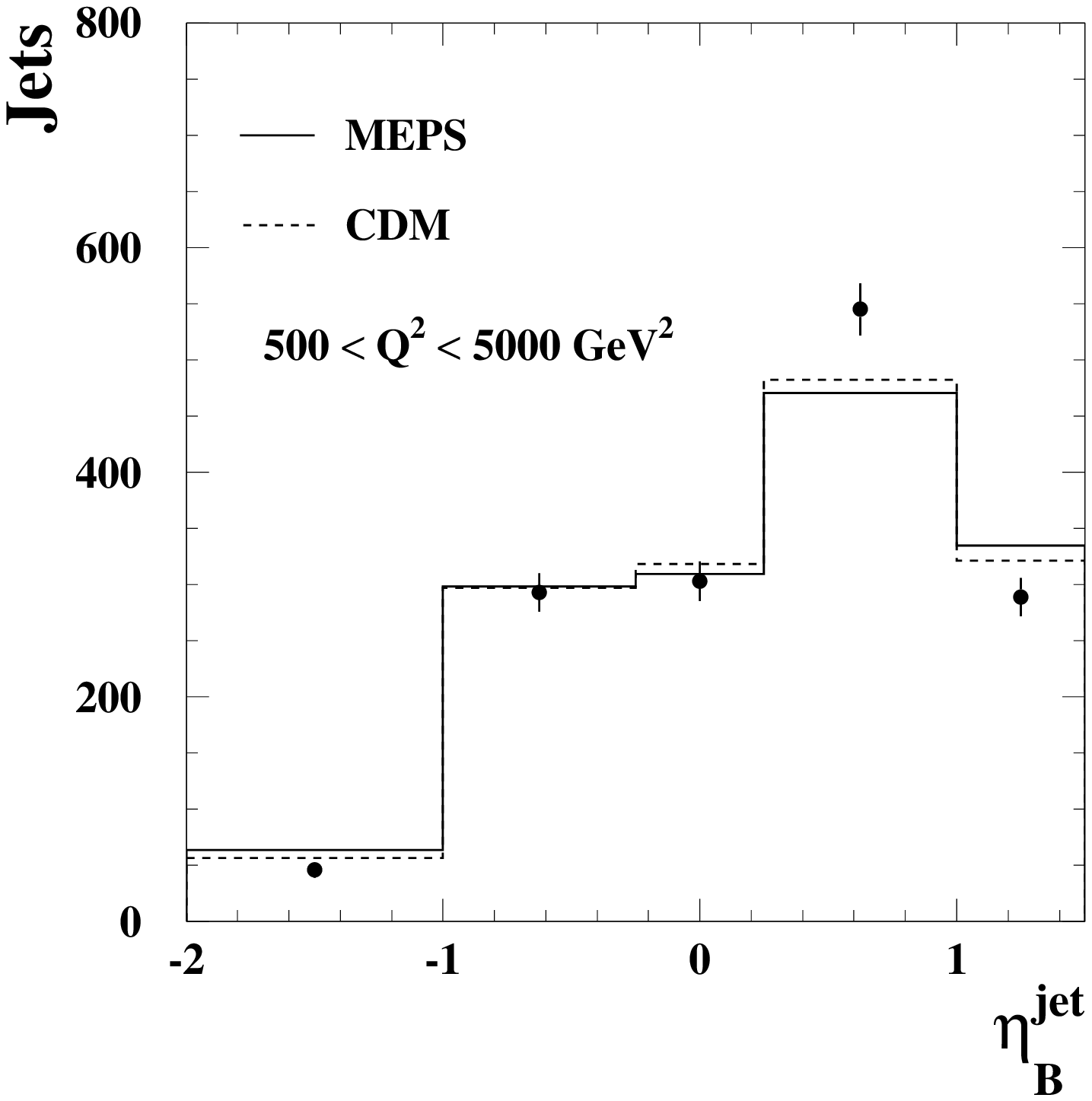,width=10cm}}
\put (6.3,15.0){\bf\small (a)}
\put (13.8,15.0){\bf\small (b)}
\put (6.3,7.0){\bf\small (c)}
\end{picture}
\caption
{\it 
Detector-level data distributions for three-jet production in NC DIS 
(dots) with $\etjbj>8$~GeV, $E^{\rm jet2,3}_{T,{\rm B}}>5$ GeV and 
$-2<\etajb<1.5$ in the kinematic region given by $500<\q2<5000$~\gf2\ and 
$|\cgh|<0.65$ as functions of (a) $\etjbj$, 
(b) $E^{\rm jet2,3}_{T,{\rm B}}$ and (c) $\etajb$. Other details as in
the caption to Fig.~\ref{fig23}.
}
\label{fig24}
\vfill
\end{figure}

\begin{figure}[p]
\vfill
\setlength{\unitlength}{1.0cm}
\begin{picture} (18.0,17.0)
\put (-0.3,8.0){\centerline{\epsfig{figure=\figdir zeus.eps,width=10cm}}}
\put (-1.0,7.5){\epsfig{figure=\figdir 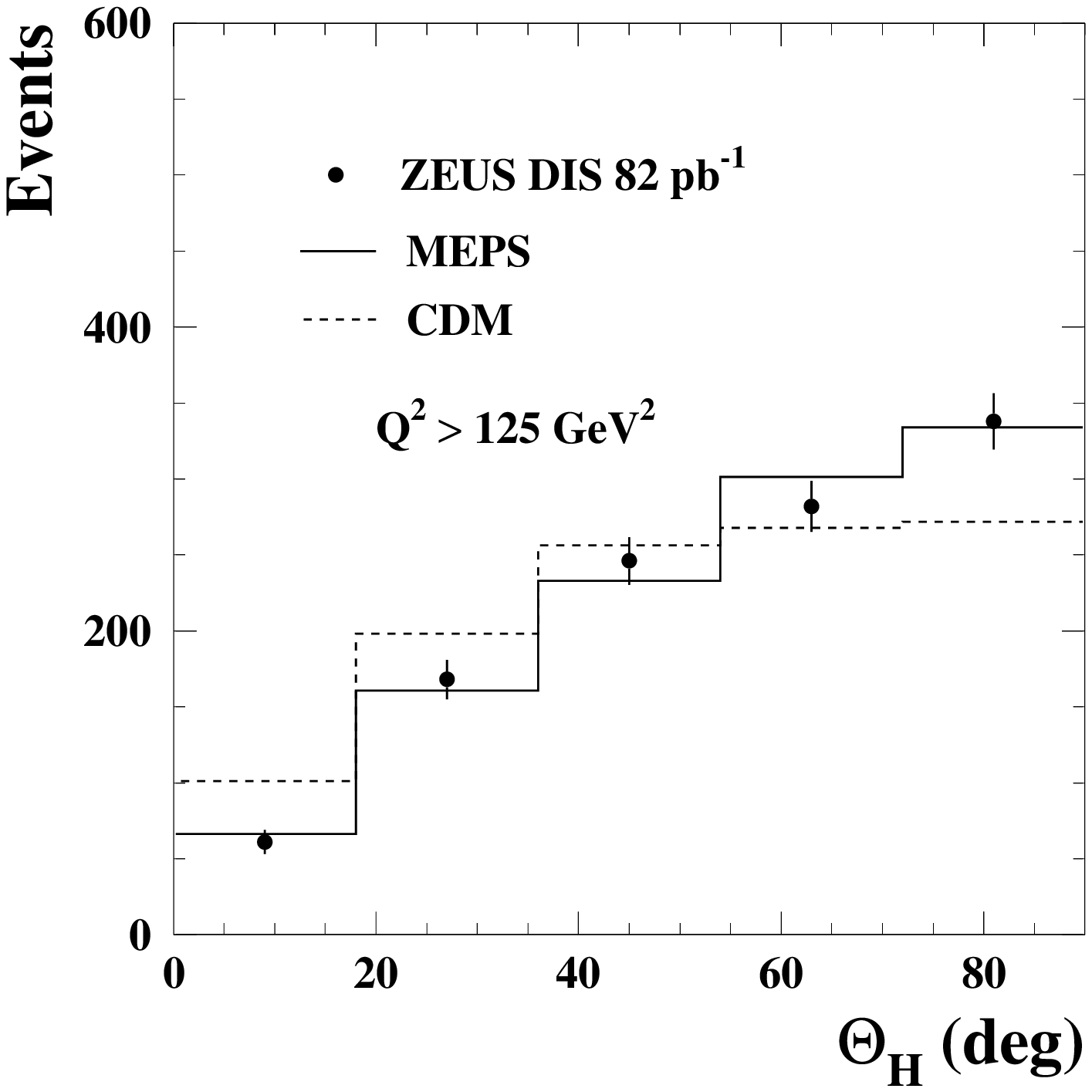,width=10cm}}
\put (6.5,7.5){\epsfig{figure=\figdir 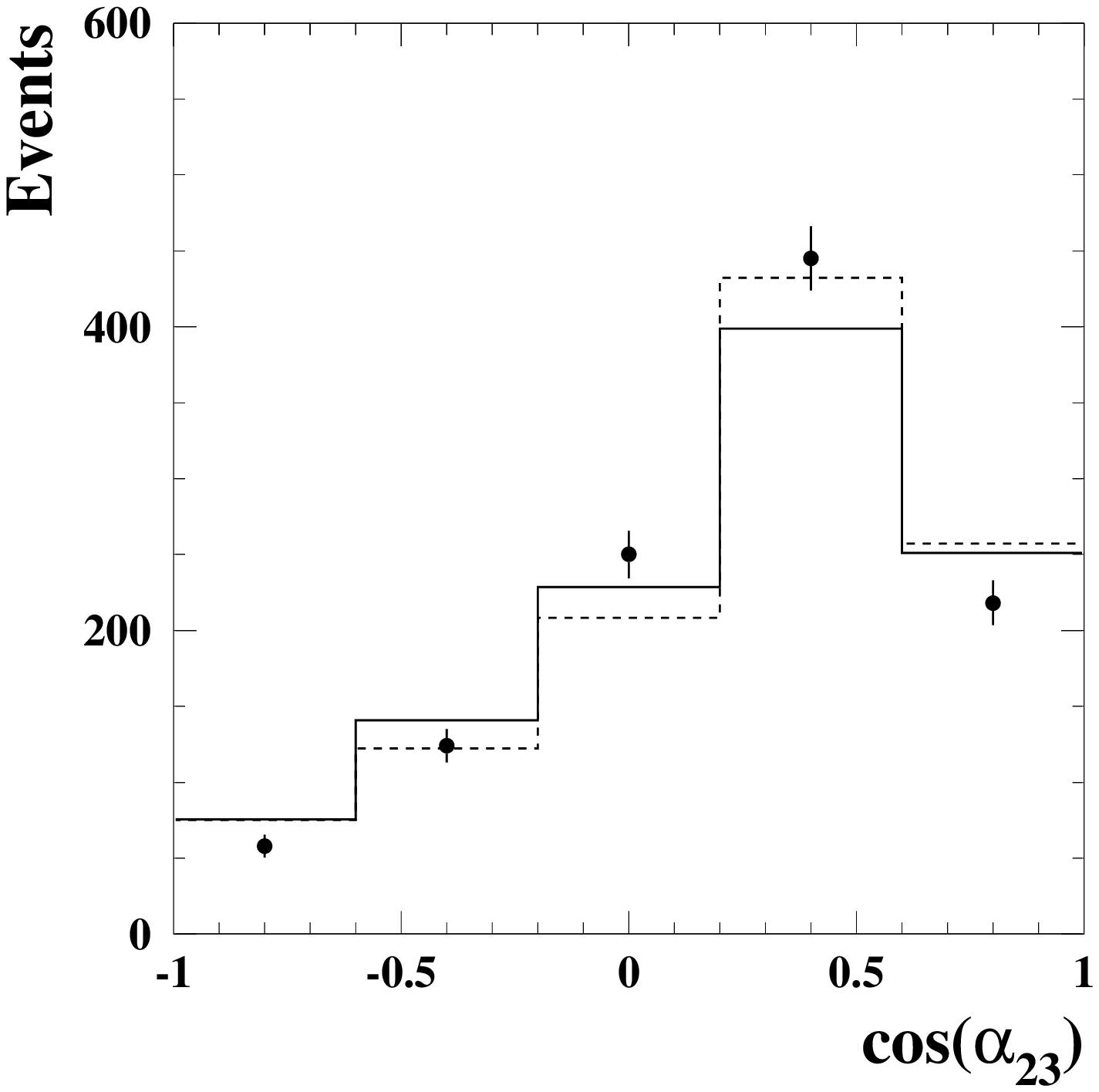,width=10cm}}
\put (-1.0,-0.5){\epsfig{figure=\figdir 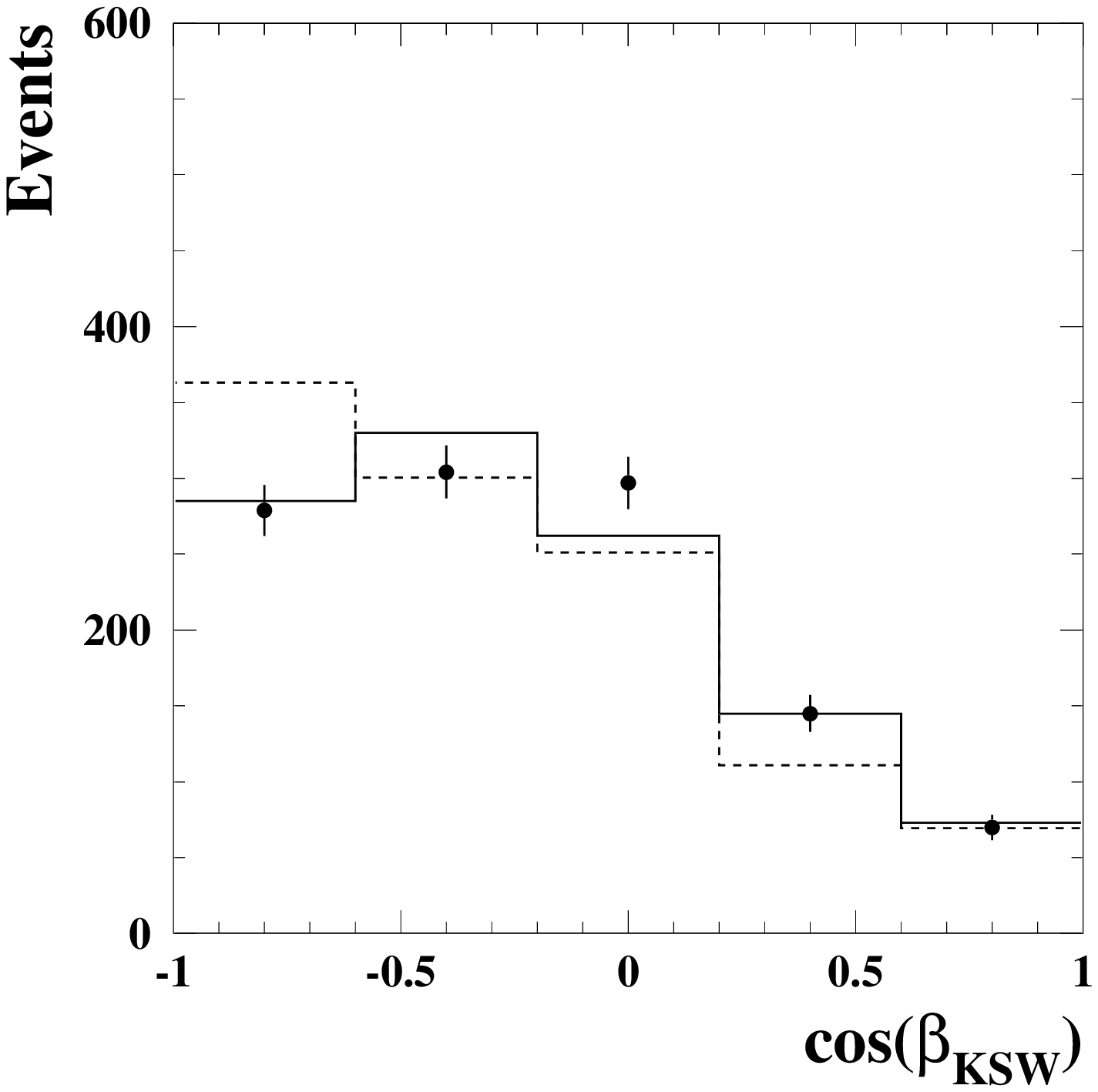,width=10cm}}
\put (6.5,-0.5){\epsfig{figure=\figdir 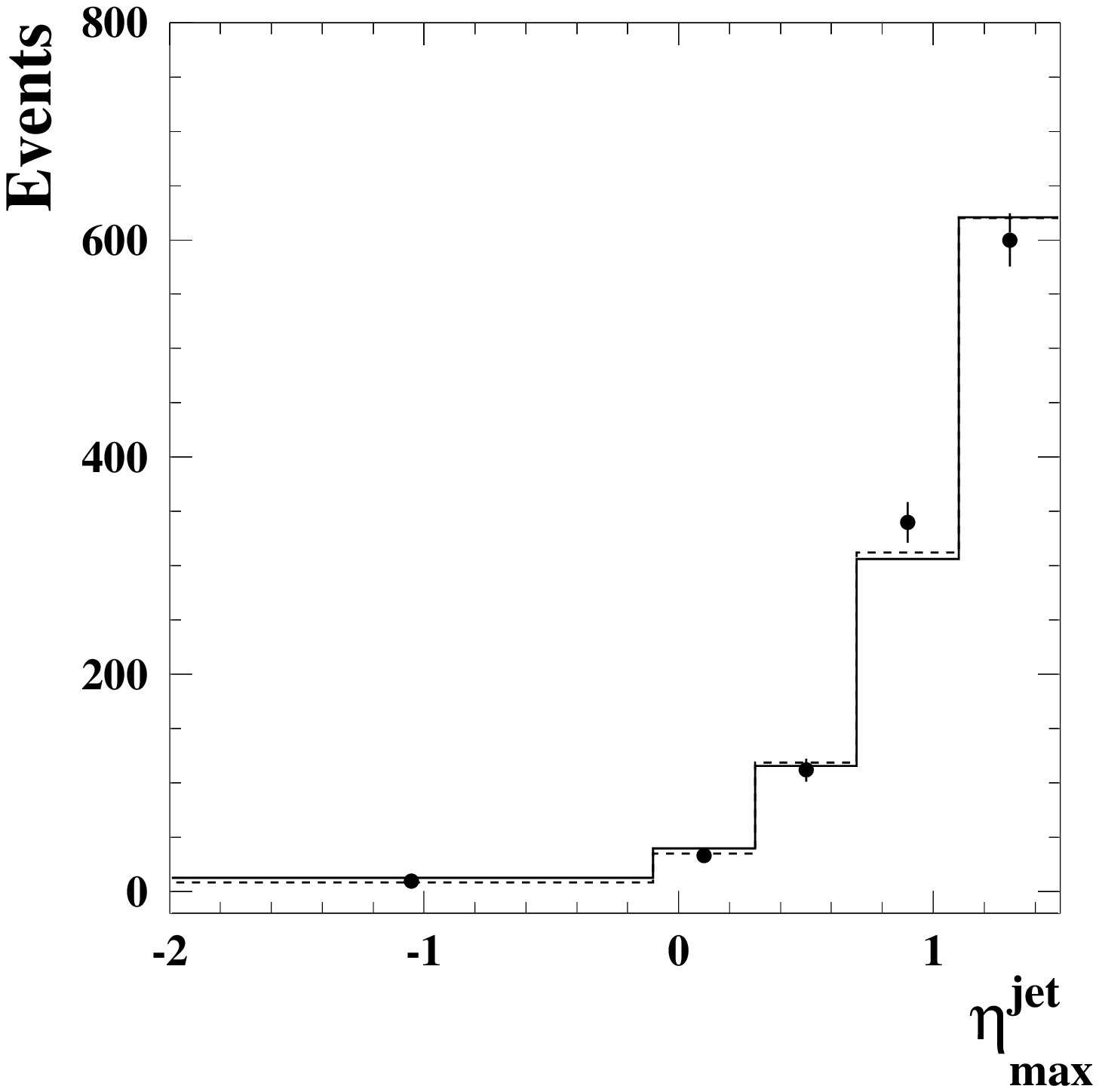,width=10cm}}
\put (6.3,15.0){\bf\small (a)}
\put (13.8,15.0){\bf\small (b)}
\put (6.3,7.0){\bf\small (c)}
\put (13.8,7.0){\bf\small (d)}
\end{picture}
\caption
{\it 
Detector-level data distributions for three-jet production in NC DIS
(dots) with $\etjbj>8$~GeV, $E^{\rm jet2,3}_{T,{\rm B}}>5$ GeV and
$-2<\etajb<1.5$ in the kinematic region given by $\q2>125$~\gf2\ and 
$|\cgh|<0.65$ as functions of (a) $\th$, (b) $\cos(\a34)$, (c)
$\cos(\pksw)$ and (d) $\etajmax$. For comparison, the distributions of
the MEPS (solid histograms) and CDM (dashed histograms) MC
models normalised to the data are included.
}
\label{fig5}
\vfill
\end{figure}

\begin{figure}[p]
\vfill
\setlength{\unitlength}{1.0cm}
\begin{picture} (18.0,17.0)
\put (-0.3,8.0){\centerline{\epsfig{figure=\figdir zeus.eps,width=10cm}}}
\put (-1.0,7.5){\epsfig{figure=\figdir 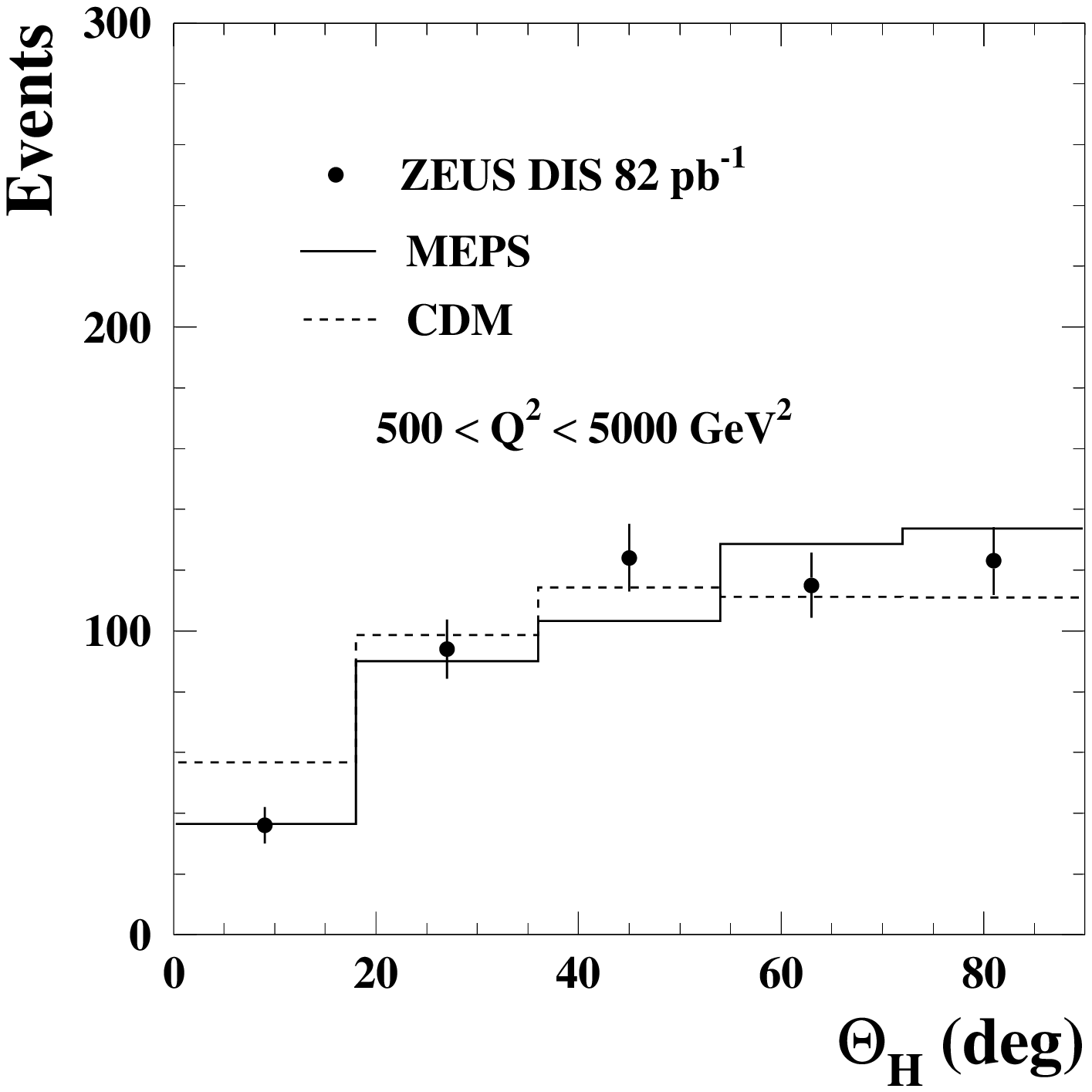,width=10cm}}
\put (6.5,7.5){\epsfig{figure=\figdir 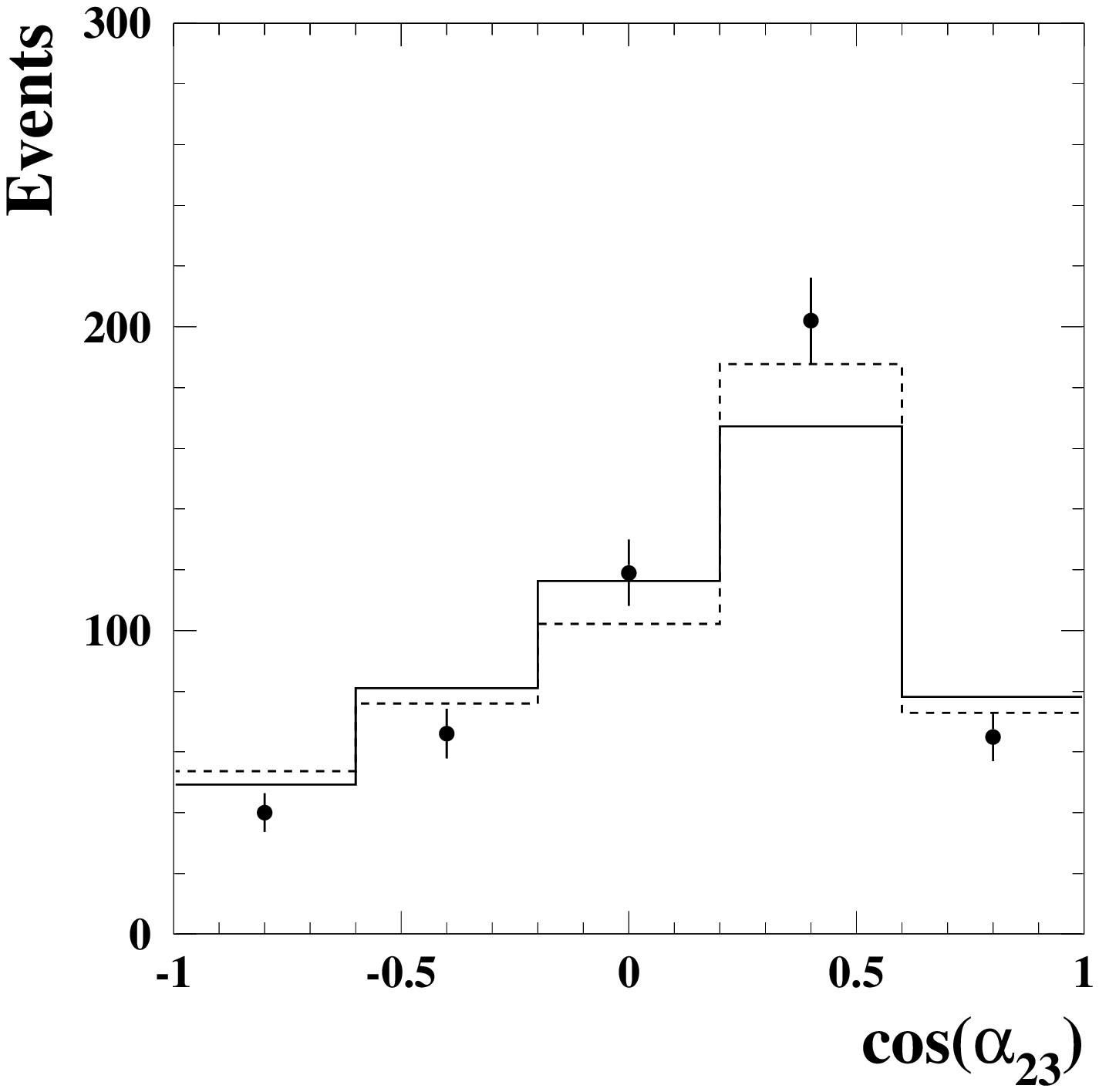,width=10cm}}
\put (-1.0,-0.5){\epsfig{figure=\figdir 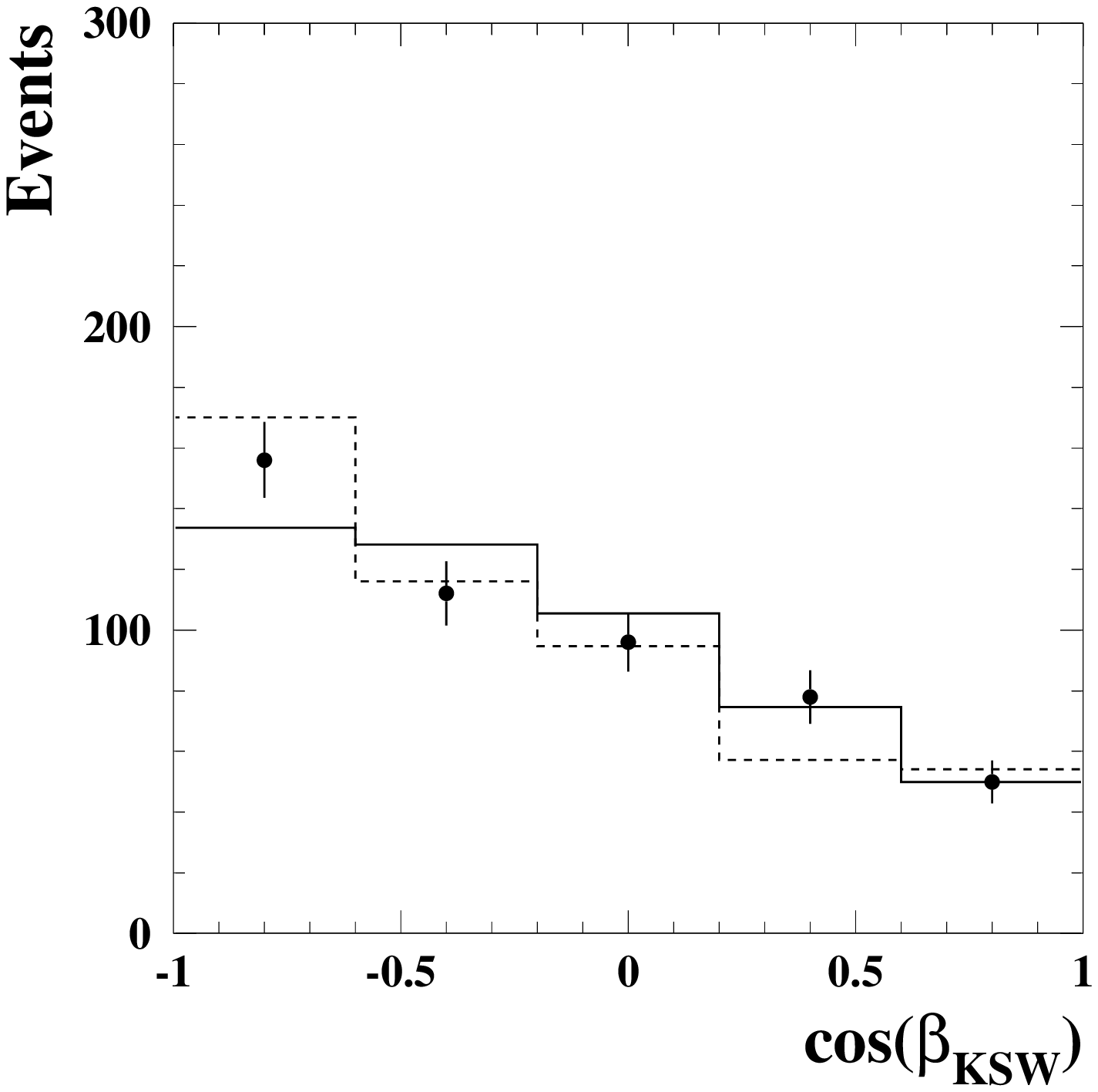,width=10cm}}
\put (6.5,-0.5){\epsfig{figure=\figdir 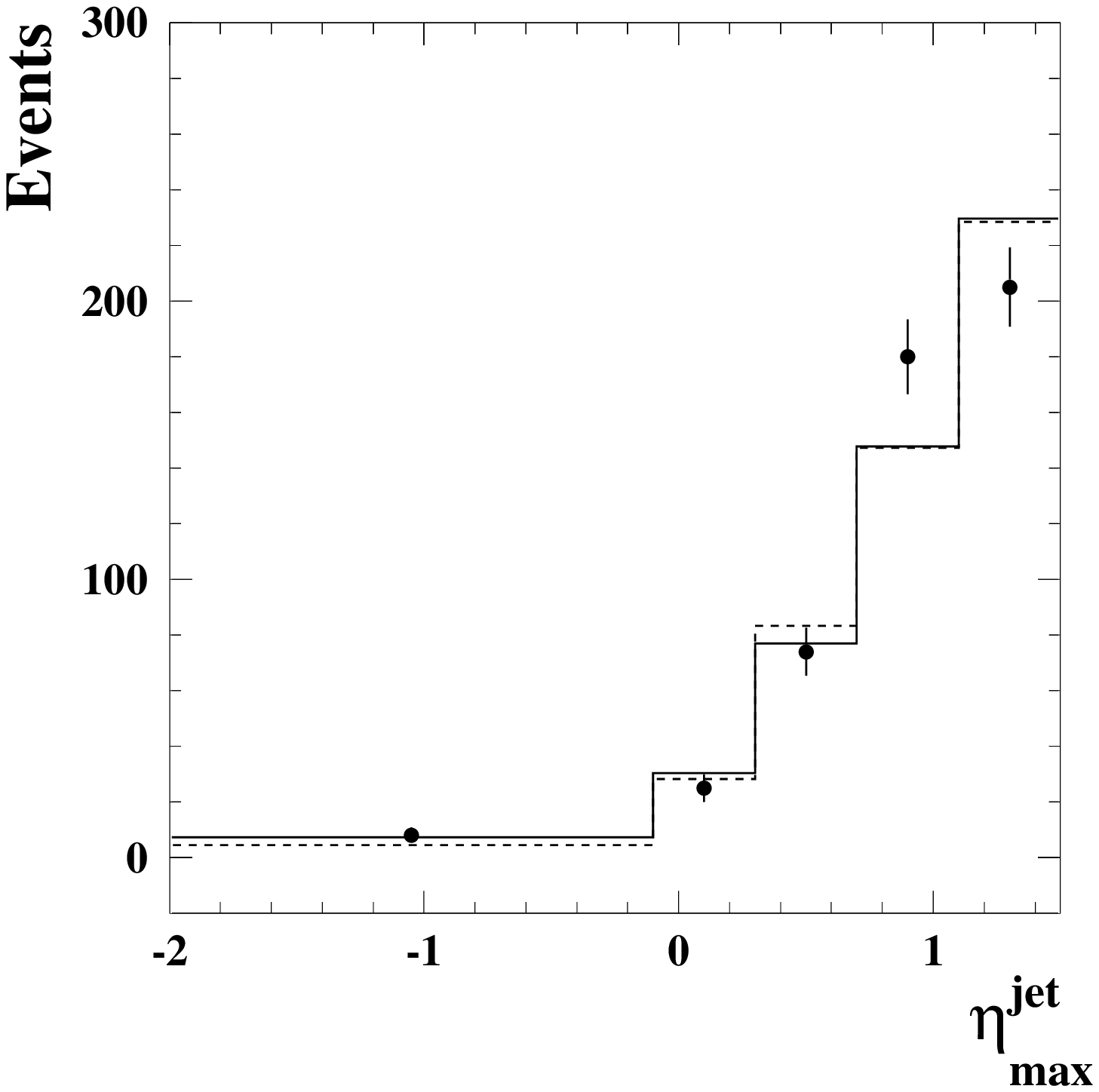,width=10cm}}
\put (6.3,15.0){\bf\small (a)}
\put (13.8,15.0){\bf\small (b)}
\put (6.3,7.0){\bf\small (c)}
\put (13.8,7.0){\bf\small (d)}
\end{picture}
\caption
{\it 
Detector-level data distributions for three-jet production in NC DIS
(dots) with $\etjbj>8$~GeV, $E^{\rm jet2,3}_{T,{\rm B}}>5$ GeV and
$-2<\etajb<1.5$ in the kinematic region given by $500<\q2<5000$~\gf2\
and $|\cgh|<0.65$ as functions of (a) $\th$, (b) $\cos(\a34)$, (c)
$\cos(\pksw)$ and (d) $\etajmax$. Other details as in the caption to
Fig.~\ref{fig5}.
}
\label{fig6}
\vfill
\end{figure}

The data $\etjbj$, $E^{\rm jet2,3}_{T,{\rm B}}$, $\etajb$ and $\q2$
distributions of the NC DIS samples are shown in Fig.~\ref{fig23}
(\ref{fig24}) for $\q2>125$ ($500<\q2<5000$)~\g2\ together with the MC
simulations from the MEPS and CDM models. Both models give a
reasonably good description of the data in both kinematic regions. The
data distributions of $\th$, $\a34$, $\pksw$ and $\etajmax$ are shown
in Fig.~\ref{fig5} (\ref{fig6}) for $\q2>125$
($500<\q2<5000$)~\g2. The MEPS MC predictions describe the data
distributions well, whereas the description given by CDM is somewhat
poorer.

A detailed study of the sources contributing to the experimental
uncertainties was performed~\cite{marcos}.
The following experimental uncertainties were considered for $\gp$ (as
an example of the size of the uncertainties, an average value of the
effect of each uncertainty on the cross section as a function of $\th$
is shown in parentheses):
\begin{itemize}
 \item the effect of the modelling of the parton shower and
   hadronisation was estimated by using {\sc Herwig} instead of {\sc
     Pythia} to evaluate the correction factors ($\pm 6.1\%$);
 \item the effect of the uncertainty on the absolute energy scale of
   the calorimetric jets was estimated by varying $\etjet$ in
   simulated events by its uncertainty of $\pm 1\%$. The method used
   was the same as in earlier
   publications~\cite{proc:calor:2002:767,pl:b560:7,pl:b649:12} 
   ($\pm 1.6\%$);
 \item the effect of the uncertainty on the reconstruction of $y$ was
   estimated by varying its value in simulated events by the
   estimated uncertainty of $\pm 1\%$ ($\pm 1.0\%$);
 \item the effect of the uncertainty on the parameterisations of the
   proton and photon PDFs was estimated by using alternative sets of
   PDFs in the MC simulation to calculate the correction factors
   ($\pm 0.4\%$ and $\pm 2.0\%$, respectively);
  \item the uncertainty in the cross sections due to that in the
    simulation of the trigger ($\pm 0.4\%$).
\end{itemize}

For NC DIS events, the following experimental uncertainties were
considered (as an example of the size of the uncertainties, an average
value of the effect of each uncertainty on the cross section as a
function of $\th$ is shown in parentheses for the $\q2>125$~\g2\ and 
$500<\q2<5000$~\g2\ kinematic regions):
\begin{itemize}
 \item the effect of the modelling of the parton shower
   was estimated by using CDM instead of MEPS to evaluate the
   correction factors ($\pm 5.6\%$ and $\pm 9.1\%$);
  \item the effect of the uncertainty on the absolute energy scale of
   the calorimetric jets was estimated by varying $\etjet$ in
   simulated events by its uncertainty of $\pm 1\%$ for
   $\etjet>10$~GeV and $\pm 3\%$ for lower $\etjet$ values ($\pm 2.3\%$
   and $\pm 1.7\%$);
  \item the uncertainties due to the selection cuts was estimated by
    varying the values of the cuts within the resolution of each
    variable (less than $\pm 1.6\%$ and less than $\pm 4.2\%$ in all
    cases);
  \item the uncertainty on the reconstruction of the boost to the
    Breit frame was estimated by using the direction of the track
    associated with the scattered electron instead of that derived from
    the impact position as determined from the energy depositions in
    the CAL ($\pm 1.6\%$ and $\pm 1.6\%$);
  \item the uncertainty in the absolute energy scale of the electron
    candidate was estimated to be $\pm 1\%$~\cite{epj:c21:443} 
    ($\pm 0.2\%$ and $\pm 0.3\%$);
  \item the uncertainty in the cross sections due to that in the
    simulation of the trigger ($\pm 0.5\%$ and $\pm 0.5\%$).
\end{itemize}

The effect of these uncertainties on the normalised differential
three-jet cross sections is small compared to the statistical
uncertainties for the measurements presented in
Section~\ref{results}. The systematic uncertainties were added in
quadrature to the statistical uncertainties.

\section{Results}
\label{results}

Normalised differential three-jet cross sections were measured in
$\gp$ in the kinematic region $\q2<1$~\g2, $0.2<y<0.85$ and
$\xo>0.8$. The cross sections were determined for jets of hadrons with
$\etjet>14$ GeV and $-1<\etajet<2.5$. In NC DIS, the cross sections
were measured in two kinematic regimes: $\q2>125$~\g2\ and
$500<\q2<5000$~\g2. In both cases, it was required that
$|\cgh|<0.65$. The cross sections correspond to jets of hadrons with
$\etjbj>8$~GeV, $E^{\rm jet2,3}_{T,{\rm B}}>5$ GeV and $-2<\etajb<1.5$.

\subsection{Colour components and the triple-gluon vertex}

\begin{figure}[p]
\vfill
\setlength{\unitlength}{1.0cm}
\begin{picture} (18.0,17.0)
\put (-0.3,8.0){\centerline{\epsfig{figure=\figdir zeus.eps,width=10cm}}}
\put (-1.0,7.5){\epsfig{figure=\figdir 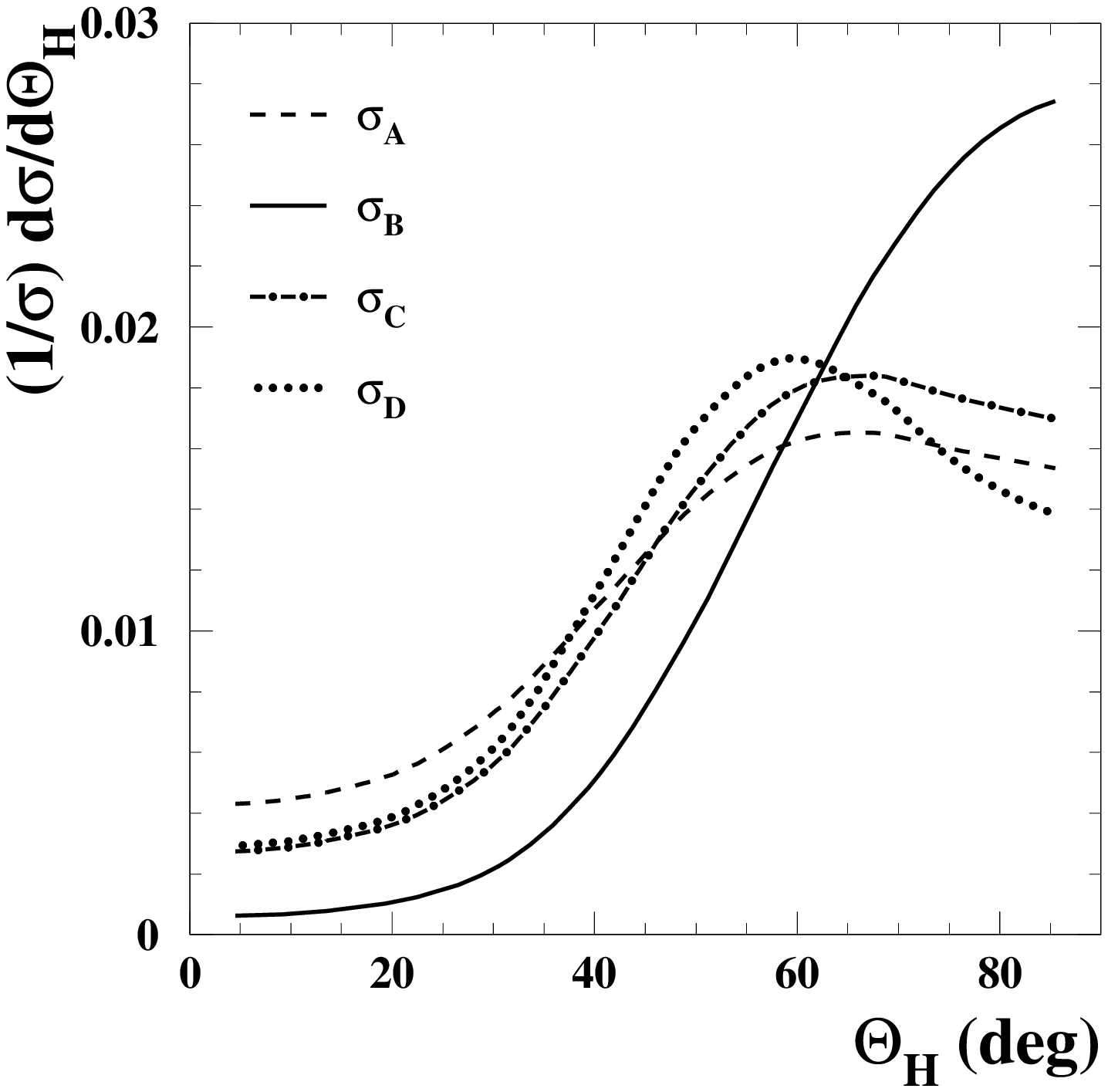,width=10cm}}
\put (6.5,7.5){\epsfig{figure=\figdir 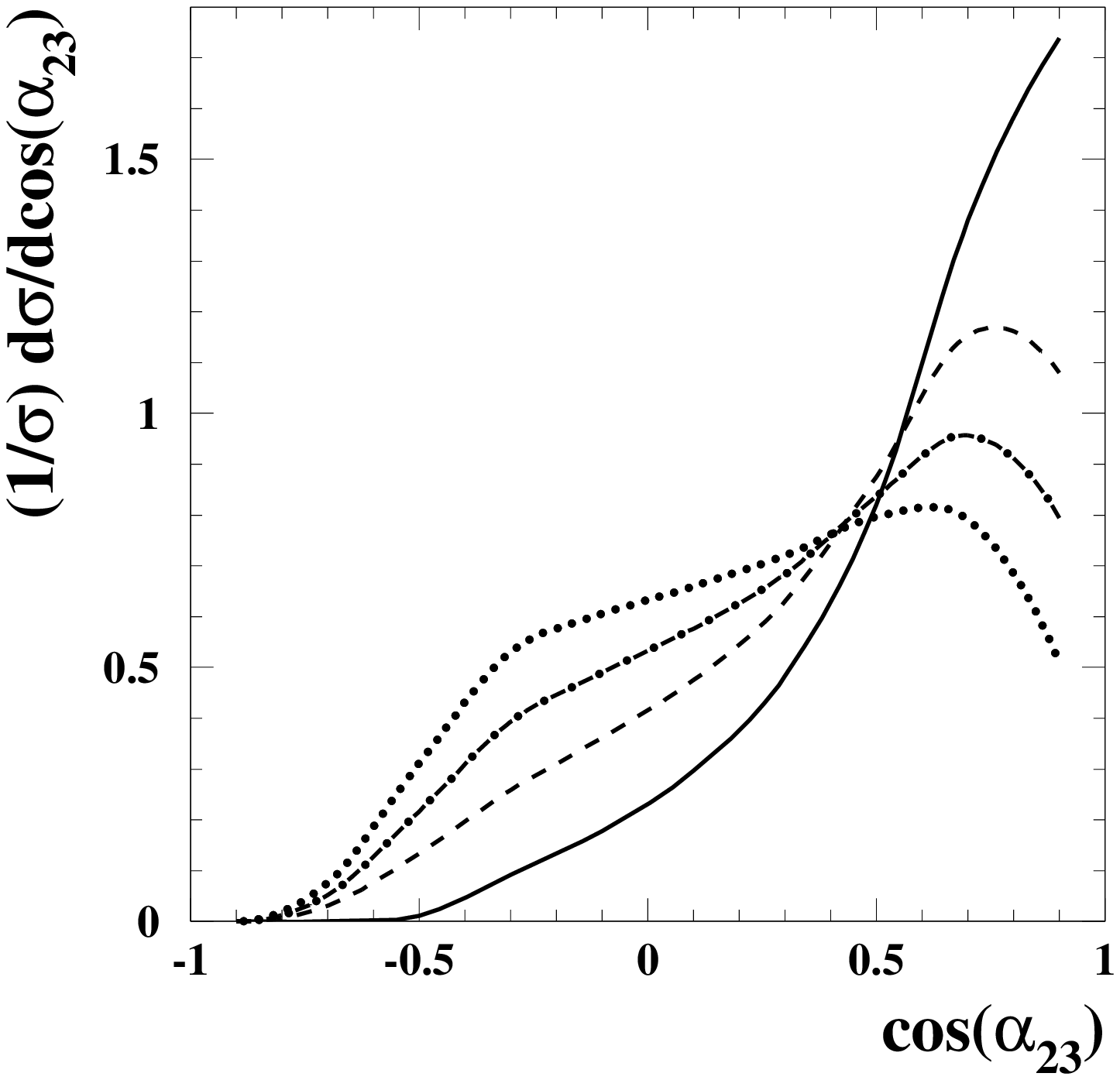,width=10cm}}
\put (-1.0,-0.5){\epsfig{figure=\figdir 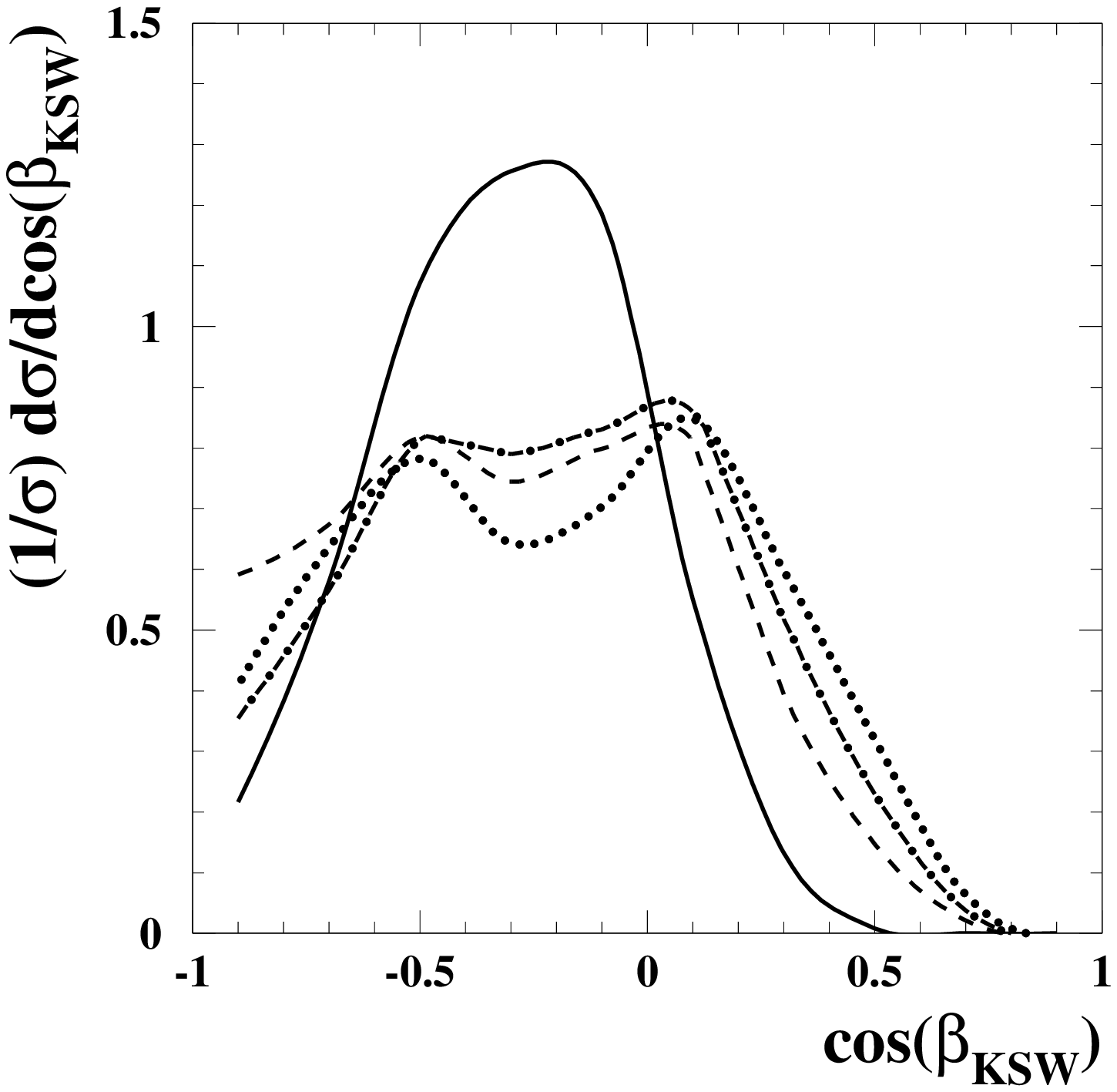,width=10cm}}
\put (5.3,15.0){\bf\small (a)}
\put (8.8,15.0){\bf\small (b)}
\put (6.3,7.0){\bf\small (c)}
\end{picture}
\caption
{\it 
Predicted normalised differential $ep$ cross sections for three-jet
direct-photon processes at $\oass$ integrated over $\etjet>14$ GeV and
$\etar$ in the kinematic region defined by $\q2<1$~\gf2 and
$0.2<y<0.85$ as functions of (a) $\th$, (b) $\cos(\a34)$ and (c)
$\cos(\pksw)$. In each figure, the predictions for the colour components
are shown: $\sigma_A$ (dashed lines), $\sigma_B$ (solid lines),
$\sigma_C$ (dot-dashed lines) and $\sigma_D$ (dotted lines). These
calculations do not include corrections for hadronisation effects.
}
\label{fig14}
\vfill
\end{figure}

\begin{figure}[p]
\vfill
\setlength{\unitlength}{1.0cm}
\begin{picture} (18.0,17.0)
\put (-0.3,8.0){\centerline{\epsfig{figure=\figdir zeus.eps,width=10cm}}}
\put (-1.0,7.5){\epsfig{figure=\figdir 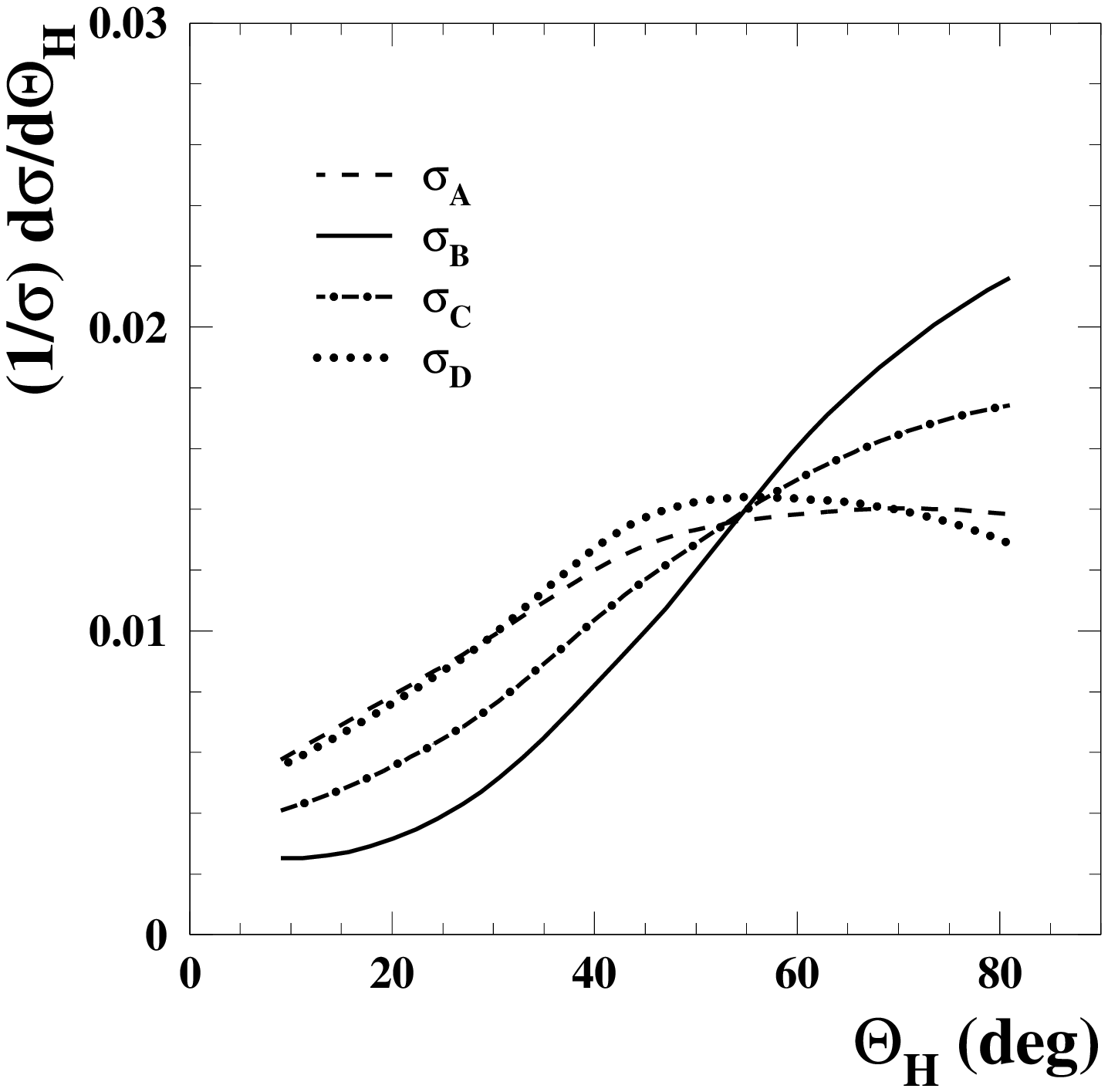,width=10cm}}
\put (6.5,7.5){\epsfig{figure=\figdir 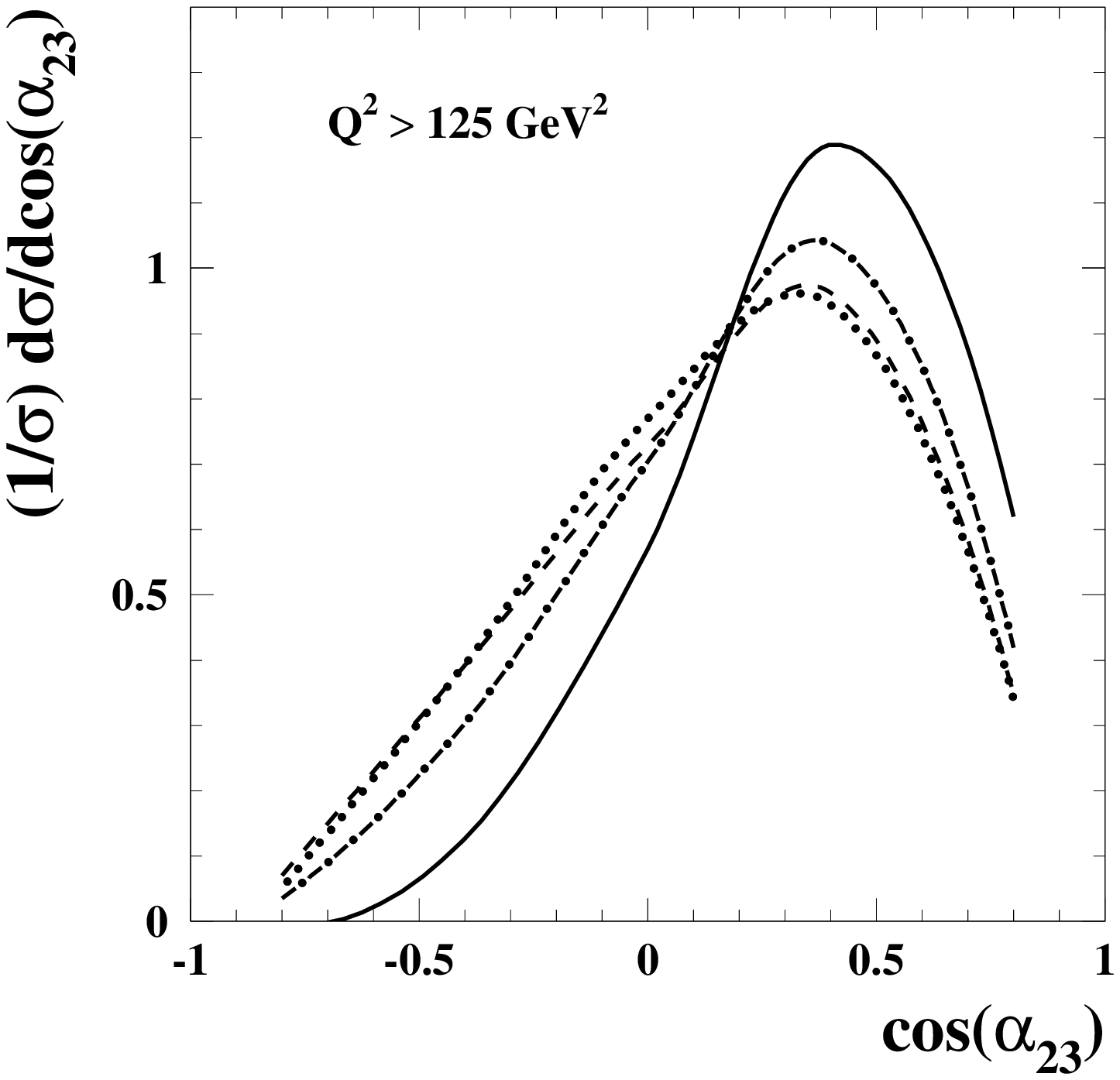,width=10cm}}
\put (-1.0,-0.5){\epsfig{figure=\figdir 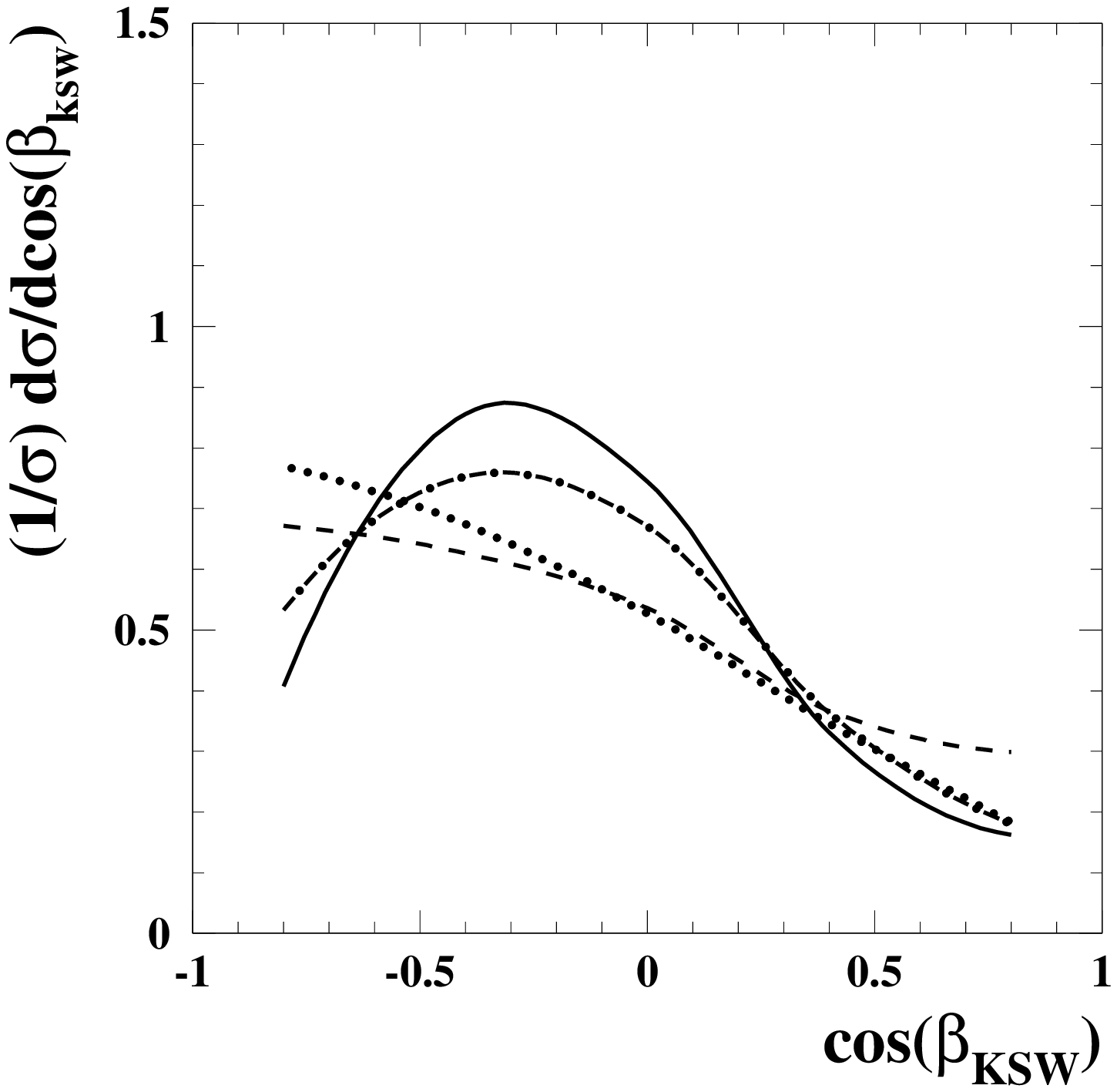,width=10cm}}
\put (6.5,-0.5){\epsfig{figure=\figdir 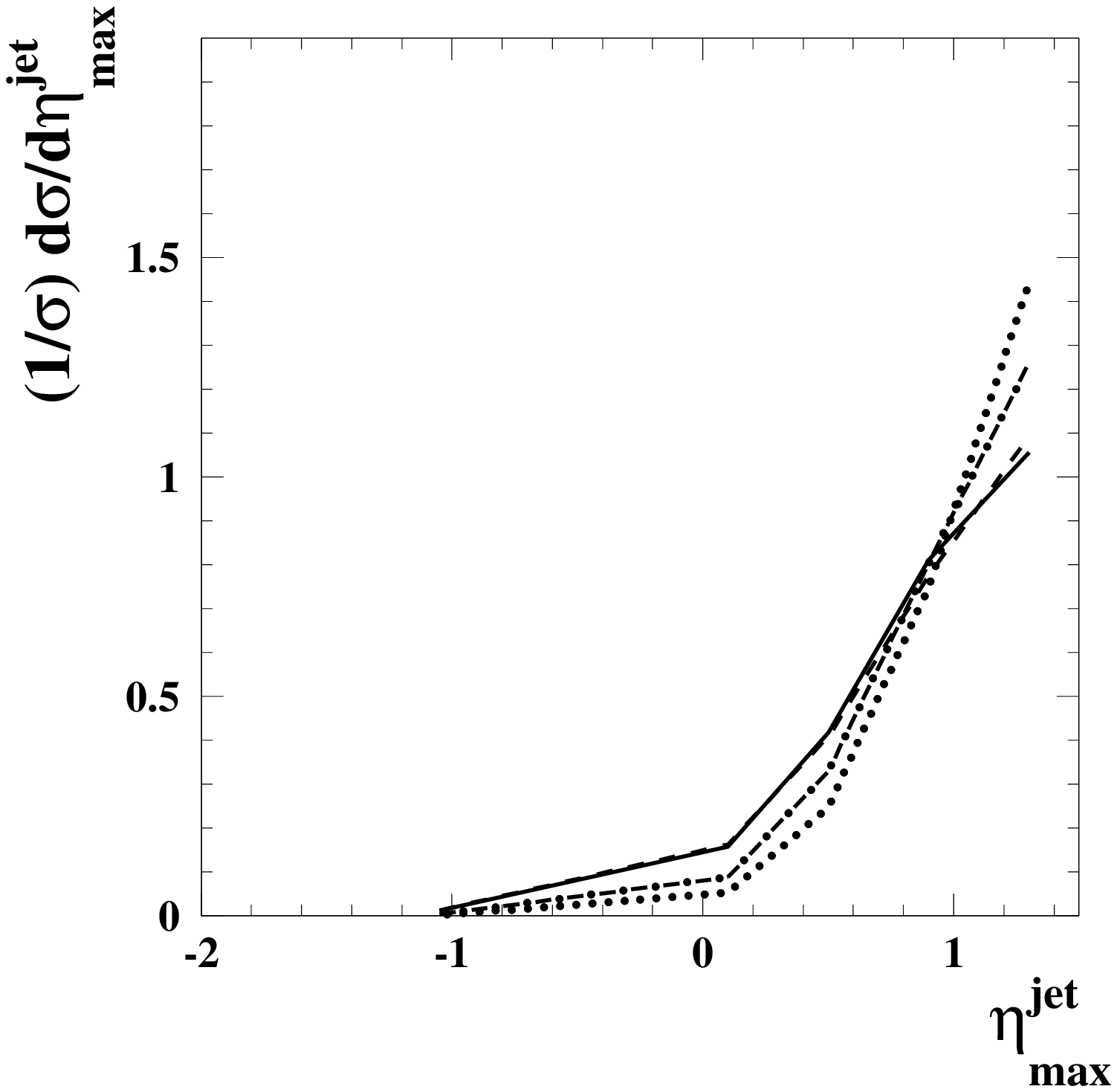,width=10cm}}
\put (6.3,15.0){\bf\small (a)}
\put (13.8,15.0){\bf\small (b)}
\put (6.3,7.0){\bf\small (c)}
\put (13.8,7.0){\bf\small (d)}
\end{picture}
\caption
{\it 
Predicted normalised differential $ep$ cross sections for three-jet
production in NC DIS at $\oass$ integrated over $\etjbj>8$~GeV,
$E^{\rm jet2,3}_{T,{\rm B}}>5$ GeV and $-2<\etajb<1.5$ in the
kinematic region given by $\q2>125$~\gf2\ and $|\cgh|<0.65$ as
functions of (a) $\th$, (b) $\cos(\a34)$, (c) $\cos(\pksw)$ and (d)
$\etajmax$. Other details as in the caption to Fig.~\ref{fig14}. These
calculations do not include corrections for hadronisation effects.
}
\label{fig15}
\vfill
\end{figure}

\begin{figure}[p]
\vfill
\setlength{\unitlength}{1.0cm}
\begin{picture} (18.0,17.0)
\put (-0.3,8.0){\centerline{\epsfig{figure=\figdir zeus.eps,width=10cm}}}
\put (-1.0,7.5){\epsfig{figure=\figdir 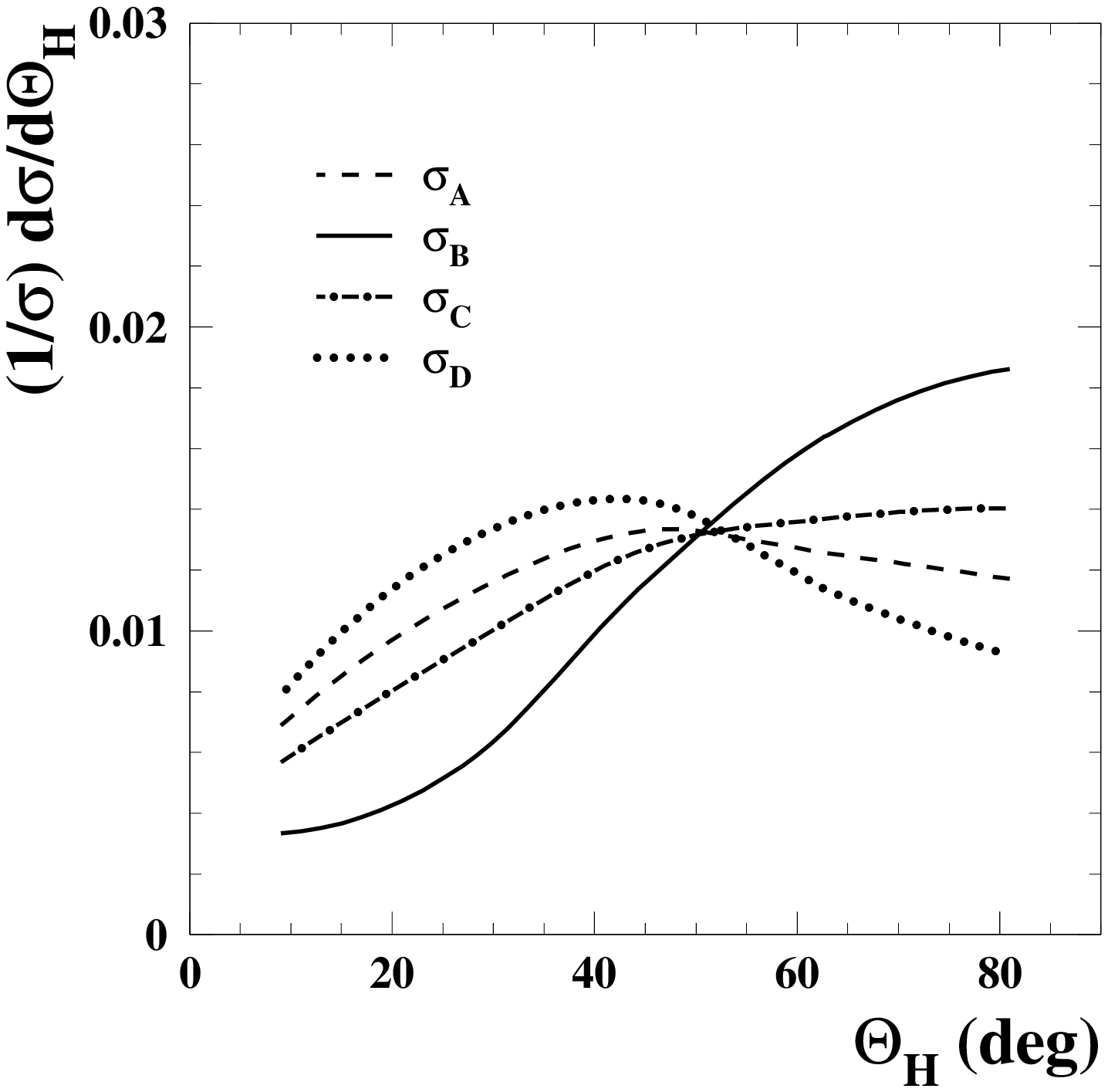,width=10cm}}
\put (6.5,7.5){\epsfig{figure=\figdir 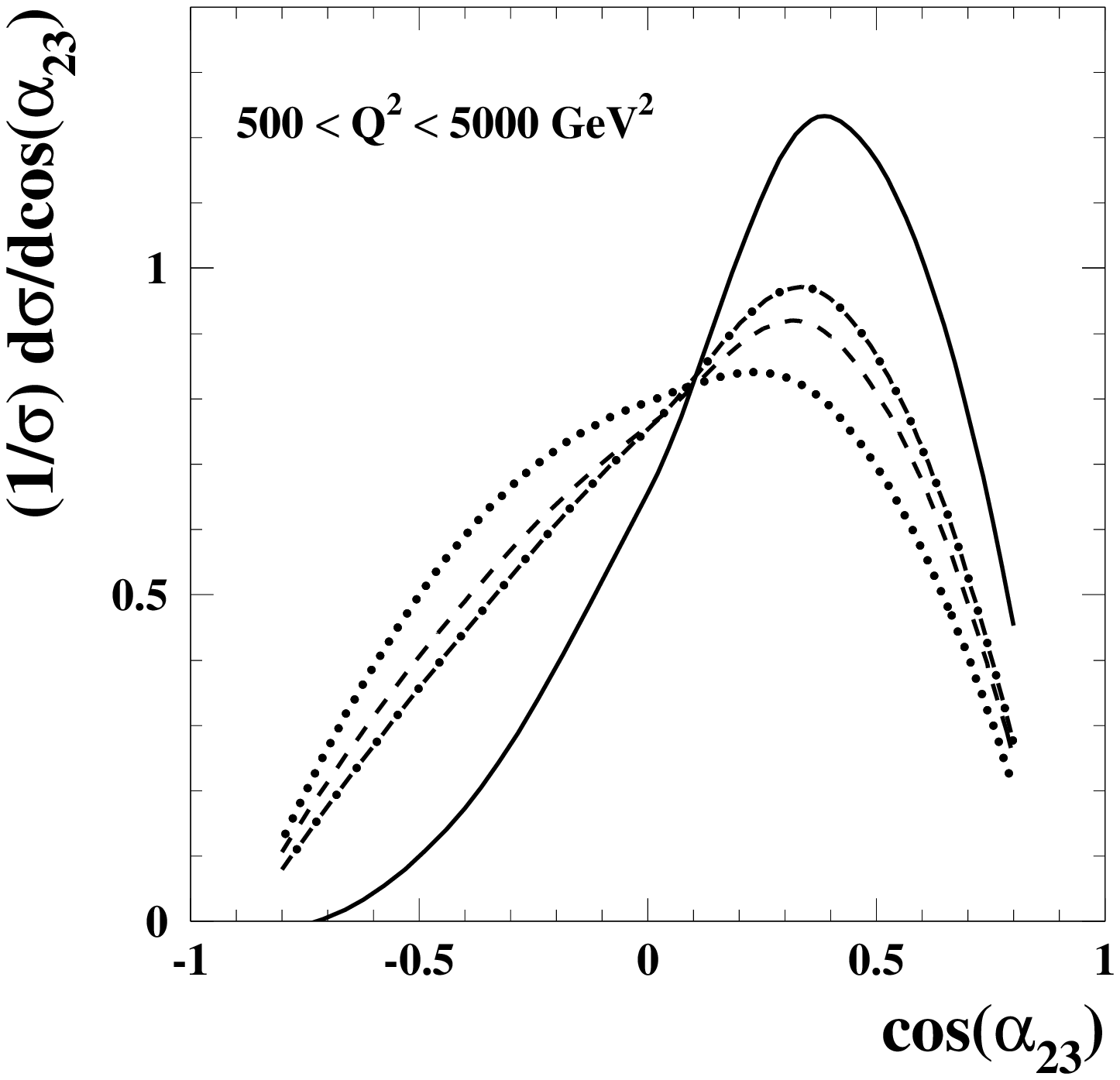,width=10cm}}
\put (-1.0,-0.5){\epsfig{figure=\figdir 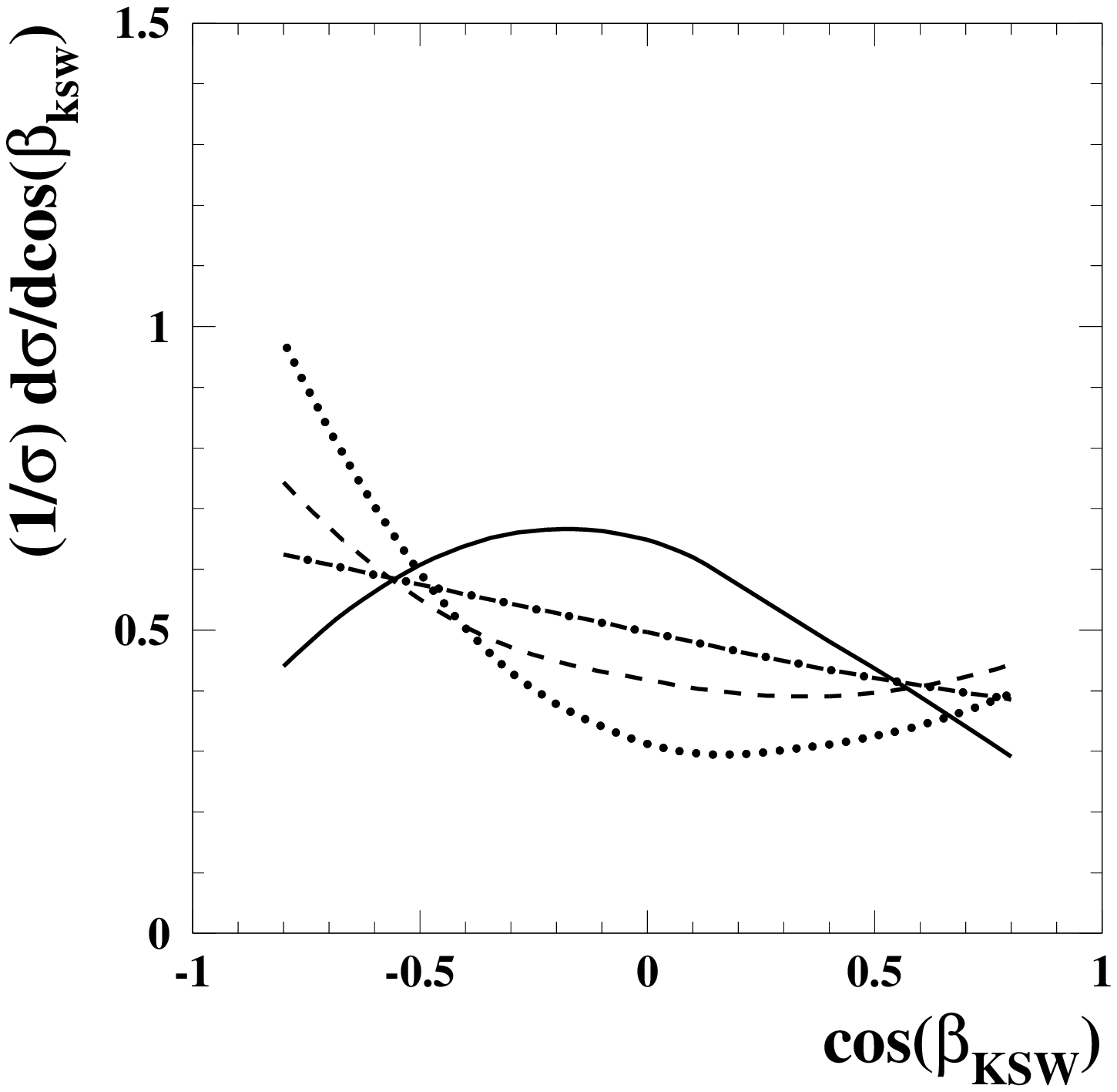,width=10cm}}
\put (6.5,-0.5){\epsfig{figure=\figdir 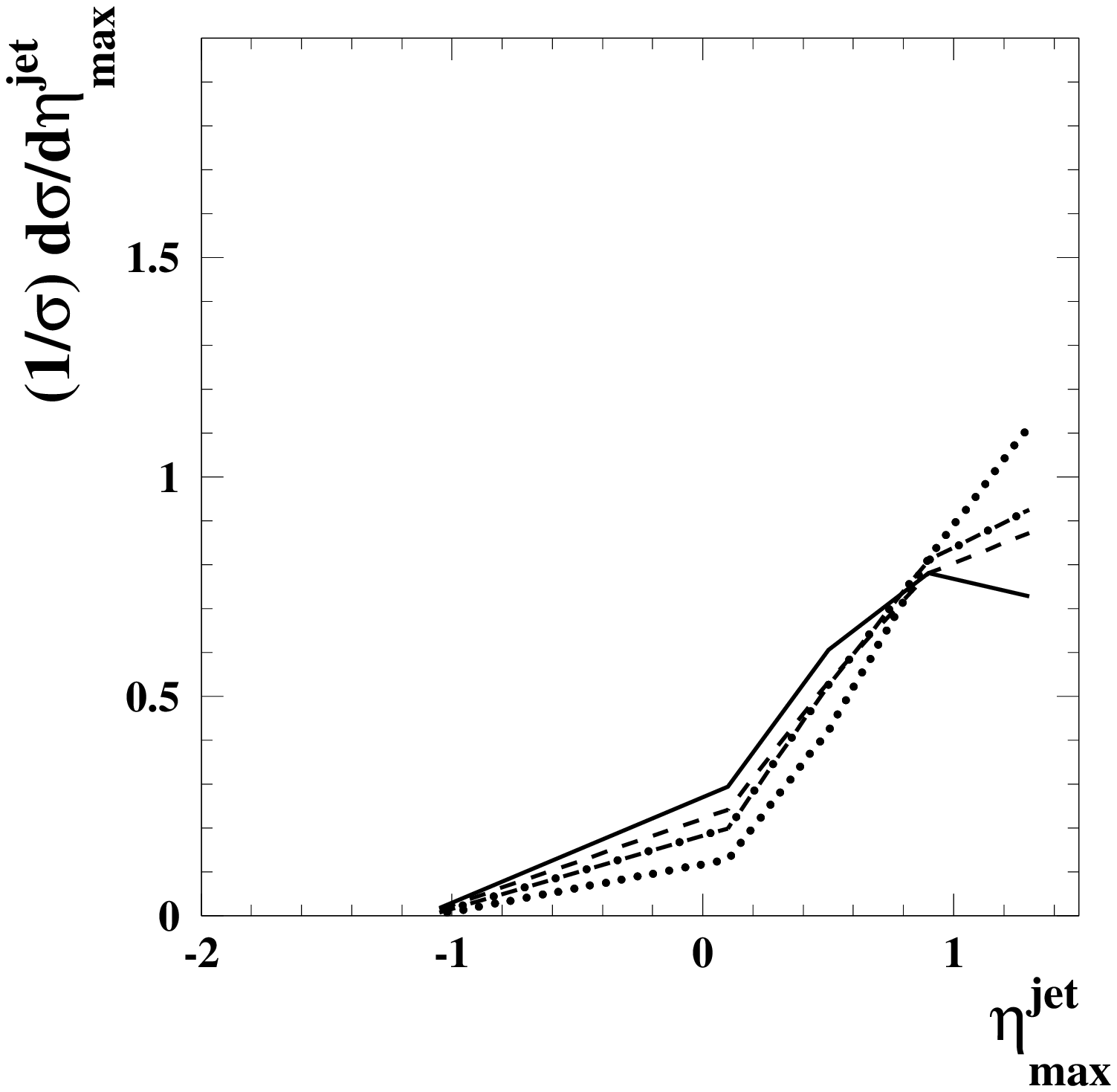,width=10cm}}
\put (6.3,15.0){\bf\small (a)}
\put (13.8,15.0){\bf\small (b)}
\put (6.3,7.0){\bf\small (c)}
\put (13.8,7.0){\bf\small (d)}
\end{picture}
\caption
{\it 
Predicted normalised differential $ep$ cross sections for three-jet
production in NC DIS at $\oass$ integrated over $\etjbj>8$~GeV,
$E^{\rm jet2,3}_{T,{\rm B}}>5$ GeV and $-2<\etajb<1.5$ in the
kinematic region given by $500<\q2<5000$~\gf2\ and $|\cgh|<0.65$ as
functions of (a) $\th$, (b) $\cos(\a34)$, (c) $\cos(\pksw)$ and (d)
$\etajmax$. Other details as in the caption to Fig.~\ref{fig14}. These
calculations do not include corrections for hadronisation effects.
}
\label{fig16}
\vfill
\end{figure}

Normalised differential three-jet cross sections at $\oass$ of the
individual colour components from Eq.~(\ref{one}), $\sigma_A$, ...,
$\sigma_D$, were calculated using the programs described in
Section~\ref{nlo} and are shown separately in Fig.~\ref{fig14} for
$\gp$ and in Fig.~\ref{fig15} (\ref{fig16}) for NC DIS with $\q2>125$
($500<\q2<5000$)~\g2\ as functions of the angular variables.
In these and subsequent figures, the predictions were obtained by
integrating over the same bins as for the data. The curves shown
join the points and are a result of a cubic spline interpolation,
except in the case of $\etajmax$, for which adjacent points are
connected by straight lines.

The component which contains the contribution from the TGV in
quark-induced processes, $\sigma_B$, has a very different shape than
the other components for all the angular variables considered. The
other components have distributions in $\pksw$ and $\th$ that are
similar and are best separated by the distribution of $\a34$ in
$\gp$. In NC DIS with $500<\q2<5000$~\g2, the different colour
components as functions of $\th$ and $\pksw$ also display different
shapes. In particular, the $\sigma_D$ component, which also contains a
TGV, shows a distinct shape for these distributions. This demonstrates
that the three-jet angular correlations studied show sensitivity to
the different colour components.

In $\gp$ (NC DIS: $\q2>125$~\g2, $500<\q2<5000$~\g2), the SU(3)-based
predictions for the relative contribution of each colour component are: 
(A): $0.13\ (0.23,\ 0.30)$, 
(B): $0.10\ (0.13,\ 0.14)$, 
(C): $0.45\ (0.39,\ 0.35)$ and 
(D): $0.32\ (0.25,\ 0.21)$.
Therefore, the overall contribution from the diagrams that involve a
TGV, B and D, amounts to $42\ (38,\ 35)\%$ in SU(3).

\subsection{Three-jet cross sections in {\boldmath $\gp$}}

The integrated three-jet cross section in $\gp$ in the
kinematic range considered was measured to be:

$$\sigma_{ep \rightarrow 3{\rm jets}}=14.59\pm 0.34\ ({\rm
  stat.})\ _{-1.31}^{+1.25}\ ({\rm syst.})\ {\rm pb}.$$
The predicted $\oass$ integrated cross section, which is the lowest
order for this process and contains only direct processes, is 
$8.90\ _{-2.92}^{+2.01}$~pb.

The measured normalised differential three-jet cross sections are
presented in Fig.~\ref{fig17} and Tables~\ref{tabone} to
\ref{tabthree} as functions of $\th$, $\cos(\a34)$ and
$\cos(\pksw)$. The measured cross section shows a peak at $\th\approx
60^{\circ}$, increases as $\cos(\a34)$ increases and shows a broad
peak in the range of $\cos(\pksw)$ between $-0.5$ to $0.1$.

\subsection{Three-jet cross sections in NC DIS}

The integrated three-jet cross sections in NC DIS for $\q2>125$~\g2\
and $500<\q2<5000$~\g2\ were measured to be:

$$\sigma_{ep \rightarrow 3{\rm jets}}=11.48\pm 0.35\ ({\rm
  stat.})\ \pm 1.98\ ({\rm syst.})\ {\rm pb}$$
and
$$\sigma_{ep \rightarrow 3{\rm jets}}=5.73\pm 0.26\ ({\rm
  stat.})\ \pm 0.60\ ({\rm syst.})\ {\rm pb}.$$
The predicted $\oasss$ integrated cross sections are
$14.14\pm 3.40$~pb and $6.86\pm 1.77$~pb for the
two kinematic regions, respectively.

The measured normalised differential three-jet cross sections in NC
DIS for $\q2>125$~\g2\ and $500<\q2<5000$~\g2\ are presented in
Figs.~\ref{fig18} and \ref{fig19}, respectively, as functions of
$\th$, $\cos(\a34)$, $\cos(\pksw)$ and $\etajmax$ (see
Tables~\ref{tabfour} to \ref{tabseven}). The measured cross sections
have similar shapes in the two kinematic regions considered, except
for the distribution as a function of $\cos(\pksw)$: the cross section
decreases as $\cos(\pksw)$ increases for $500<\q2<5000$~\g2\ whereas
for $\q2>125$~\g2\ it shows an approximately constant behaviour for
$-1<\cos(\pksw)<0.25$. The measured cross section as a function of
$\cos(\a34)$ peaks around $0.5$ and increases as $\th$ and $\etajmax$
increase.

\subsection{Comparison to fixed-order calculations}

\begin{figure}[p]
\vfill
\setlength{\unitlength}{1.0cm}
\begin{picture} (18.0,17.0)
\put (-0.3,11.0){\centerline{\epsfig{figure=\figdir zeus.eps,width=10cm}}}
\put (-1.0,9.5){\epsfig{figure=\figdir 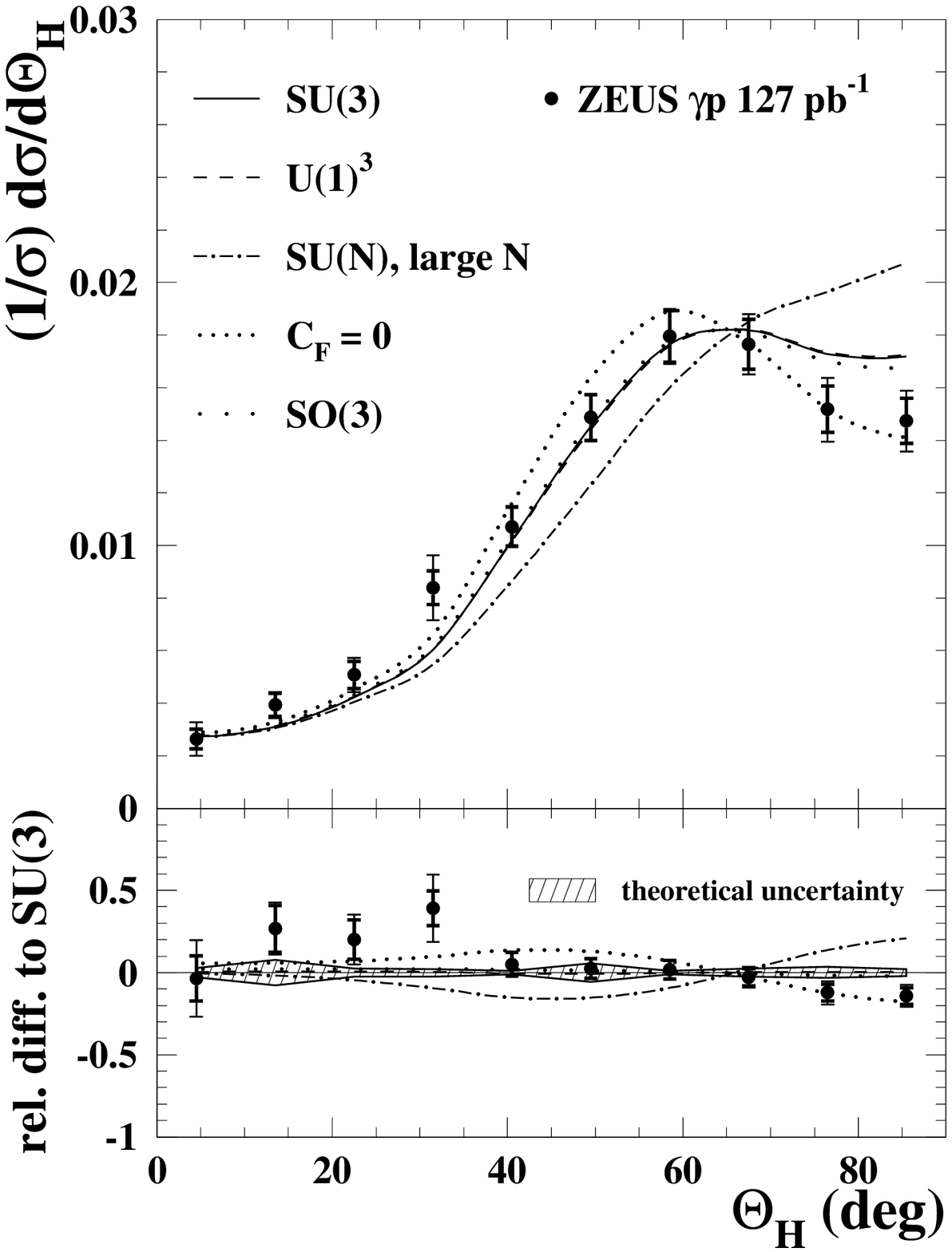,width=10cm}}
\put (6.5,9.5){\epsfig{figure=\figdir 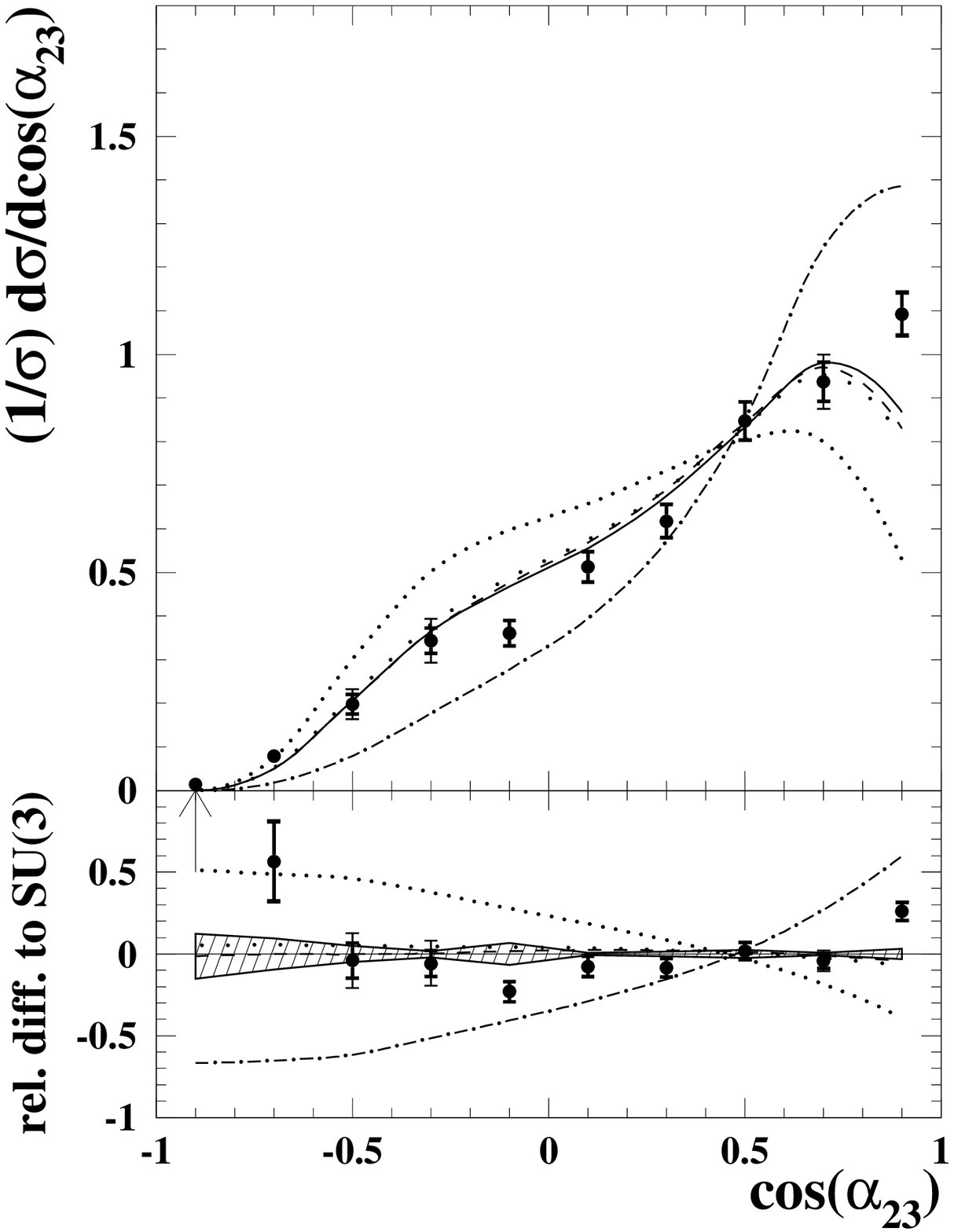,width=10cm}}
\put (-1.0,-0.5){\epsfig{figure=\figdir 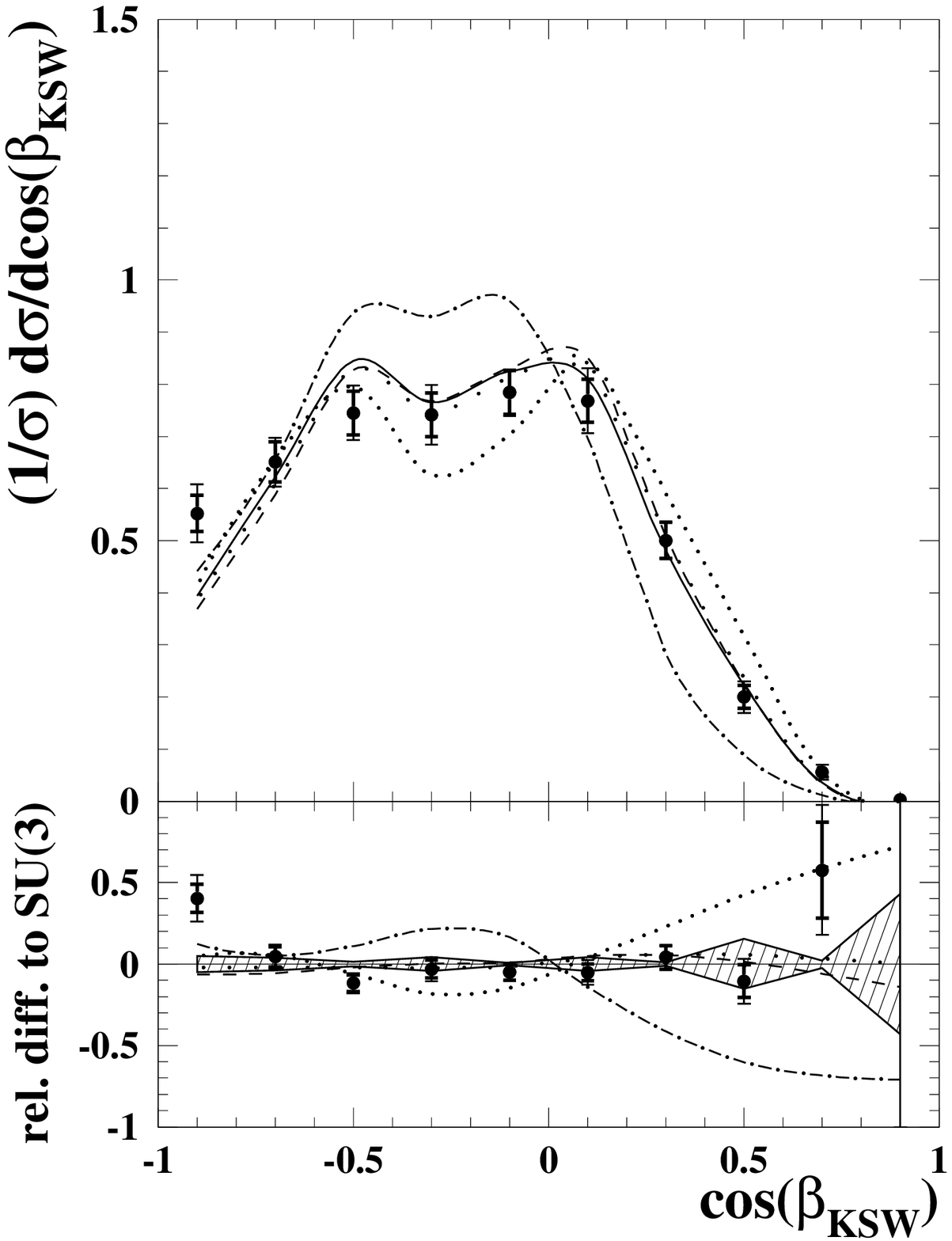,width=10cm}}
\put (6.3,17.5){\bf\small (a)}
\put (13.8,18.0){\bf\small (b)}
\put (6.3,8.0){\bf\small (c)}
\end{picture}
\caption
{\it 
Measured normalised differential $ep$ cross sections for three-jet
photoproduction (dots) integrated over $\etjet>14$ GeV and $\etar$ in
the kinematic region defined by $\q2<1$~\gf2, $0.2<y<0.85$ and
$\xo>0.8$ as functions of (a) $\th$, (b) $\cos(\a34)$ and (c)
$\cos(\pksw)$. The data points are plotted at the bin centres. The
inner error bars represent the statistical uncertainties of the data,
and the outer error bars show the statistical and systematic
uncertainties added in quadrature. For comparison, the $\oass$
calculations for direct-photon processes based on SU(3) (solid lines),
U(1)$^3$ (dashed lines), SU($N$) in the limit of large $N$ (dot-dashed
lines), $C_F=0$ (short-spaced dotted lines) and SO(3) (long-spaced
dotted lines) are included. The lower part of the figures displays the
relative difference to the calculations based on SU(3) and the hatched band
shows the relative uncertainty of this calculation.
}
\label{fig17}
\vfill
\end{figure}

\begin{figure}[p]
\vfill
\setlength{\unitlength}{1.0cm}
\begin{picture} (18.0,17.0)
\put (-0.3,11.0){\centerline{\epsfig{figure=\figdir zeus.eps,width=10cm}}}
\put (-1.0,9.5){\epsfig{figure=\figdir 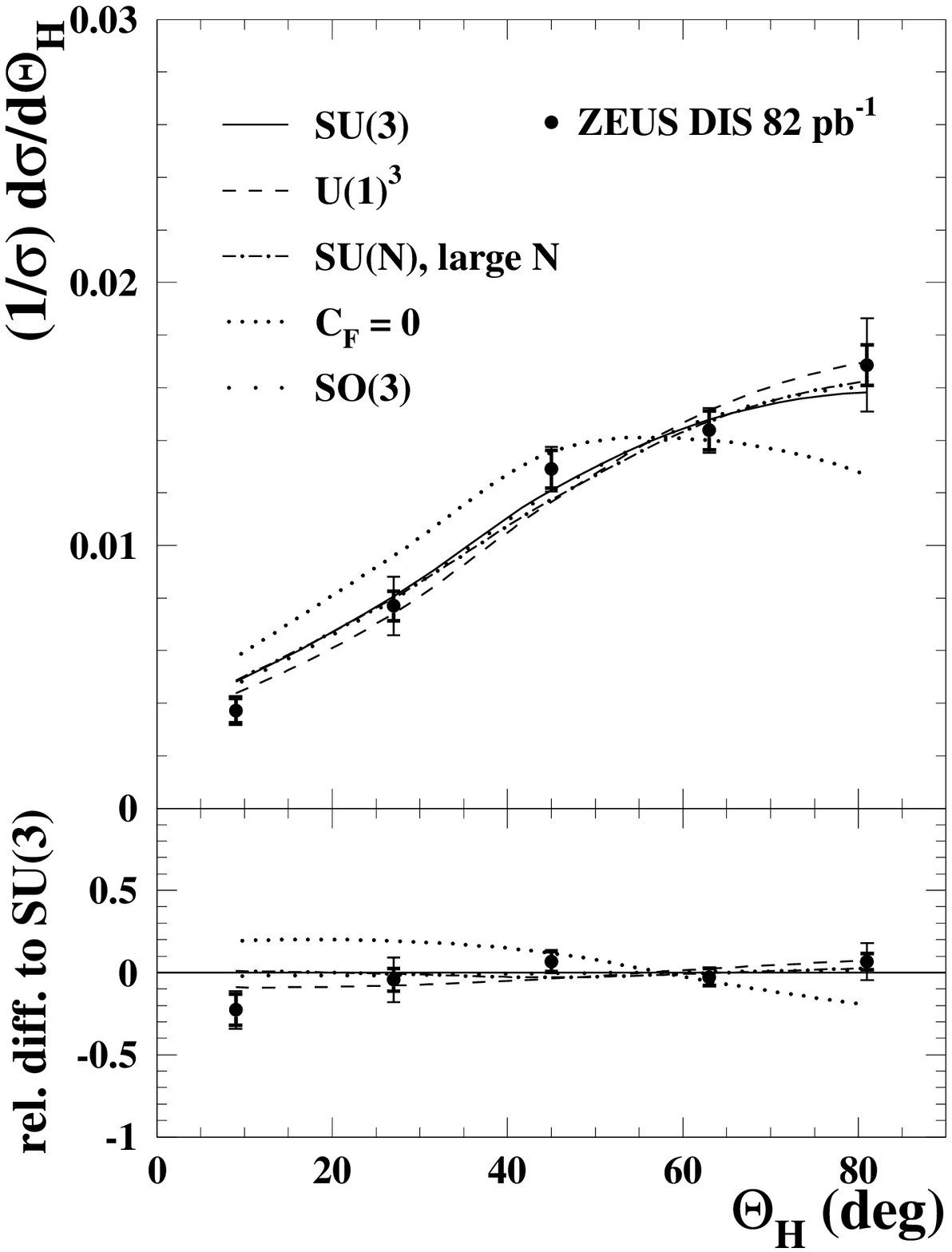,width=10cm}}
\put (6.5,9.5){\epsfig{figure=\figdir 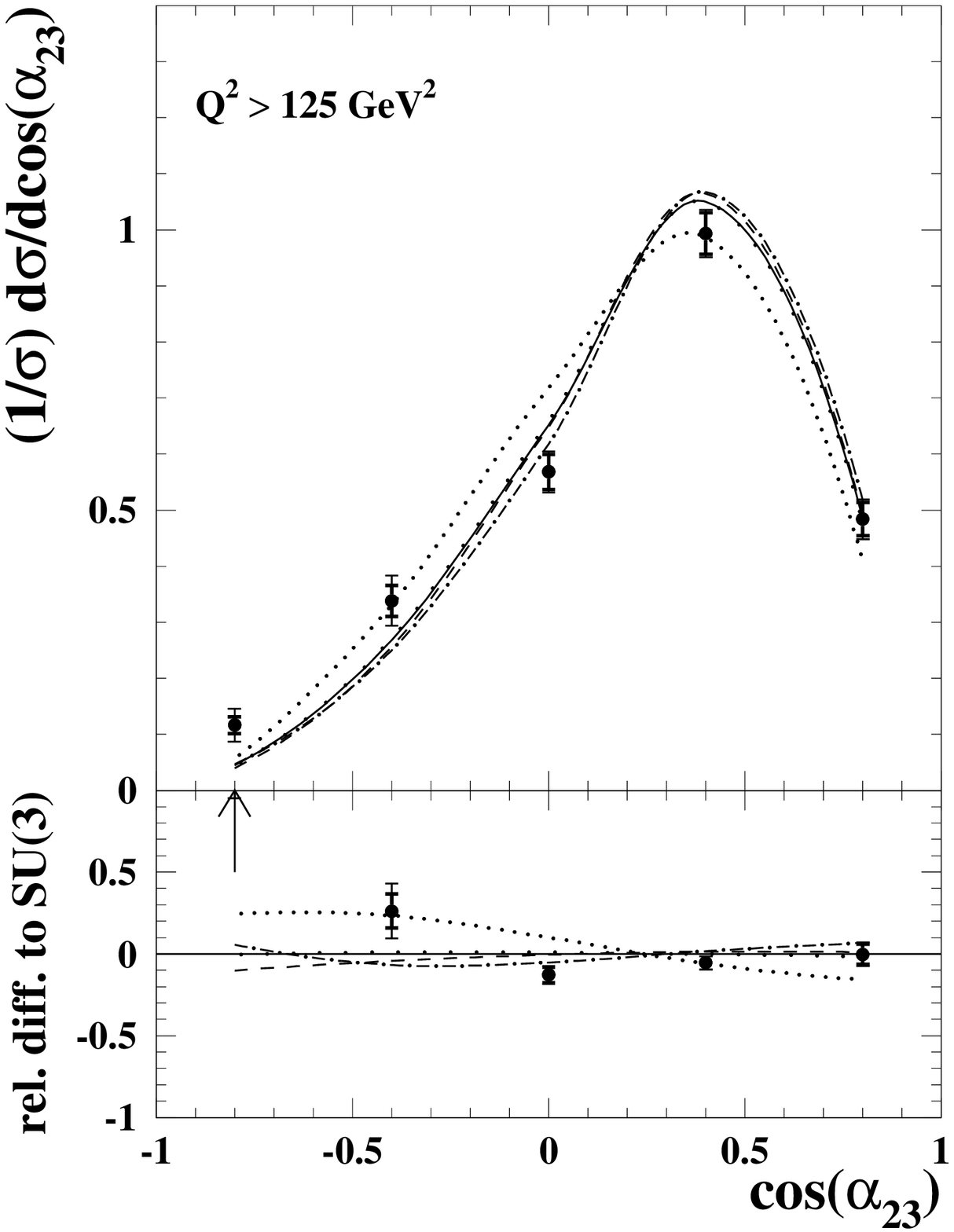,width=10cm}}
\put (-1.0,-0.5){\epsfig{figure=\figdir 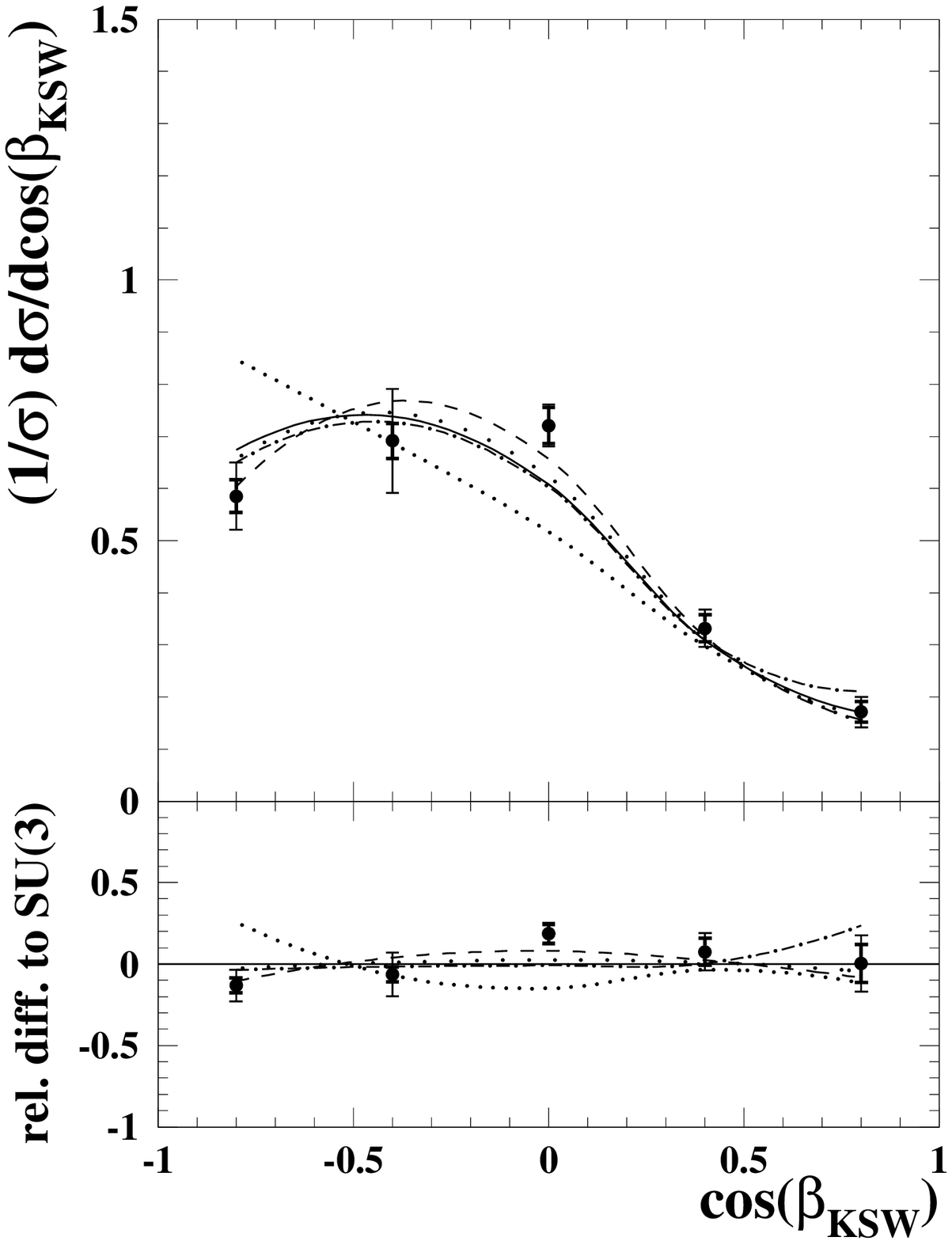,width=10cm}}
\put (6.5,-0.5){\epsfig{figure=\figdir 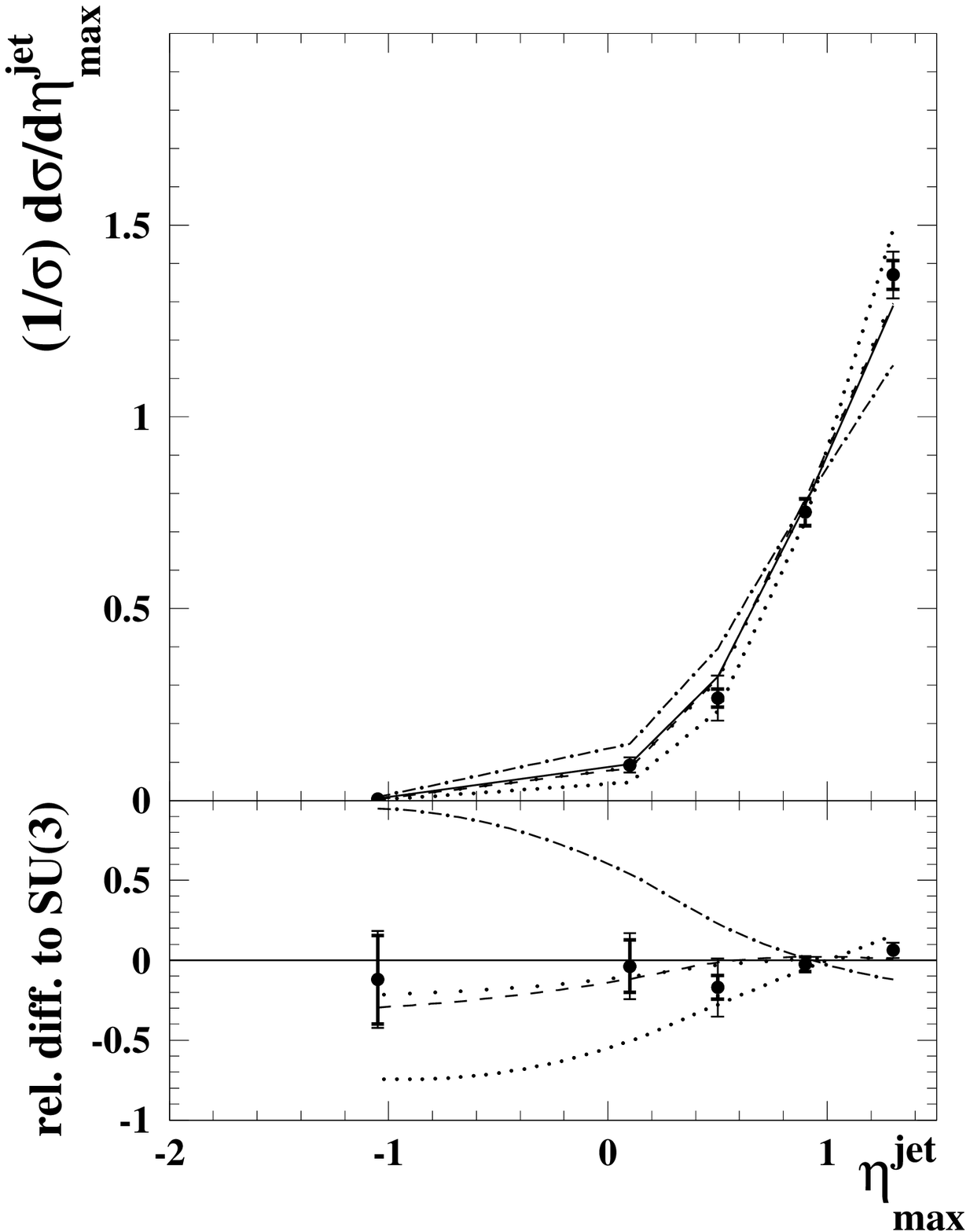,width=10cm}}
\put (6.3,17.5){\bf\small (a)}
\put (13.8,18.0){\bf\small (b)}
\put (6.3,8.0){\bf\small (c)}
\put (13.8,8.0){\bf\small (d)}
\end{picture}
\caption
{\it 
Measured normalised differential $ep$ cross sections for three-jet
production in NC DIS (dots) integrated over $\etjbj>8$~GeV,
$E^{\rm jet2,3}_{T,{\rm B}}>5$ GeV and $-2<\etajb<1.5$ in the
kinematic region given by $\q2>125$~\gf2\ and $|\cgh|<0.65$
as functions of (a) $\th$, (b) $\cos(\a34)$, (c) $\cos(\pksw)$ and (d)
$\etajmax$. Other details as in the caption to Fig.~\ref{fig17}.
}
\label{fig18}
\vfill
\end{figure}

\begin{figure}[p]
\vfill
\setlength{\unitlength}{1.0cm}
\begin{picture} (18.0,17.0)
\put (-0.3,11.0){\centerline{\epsfig{figure=\figdir zeus.eps,width=10cm}}}
\put (-1.0,9.5){\epsfig{figure=\figdir 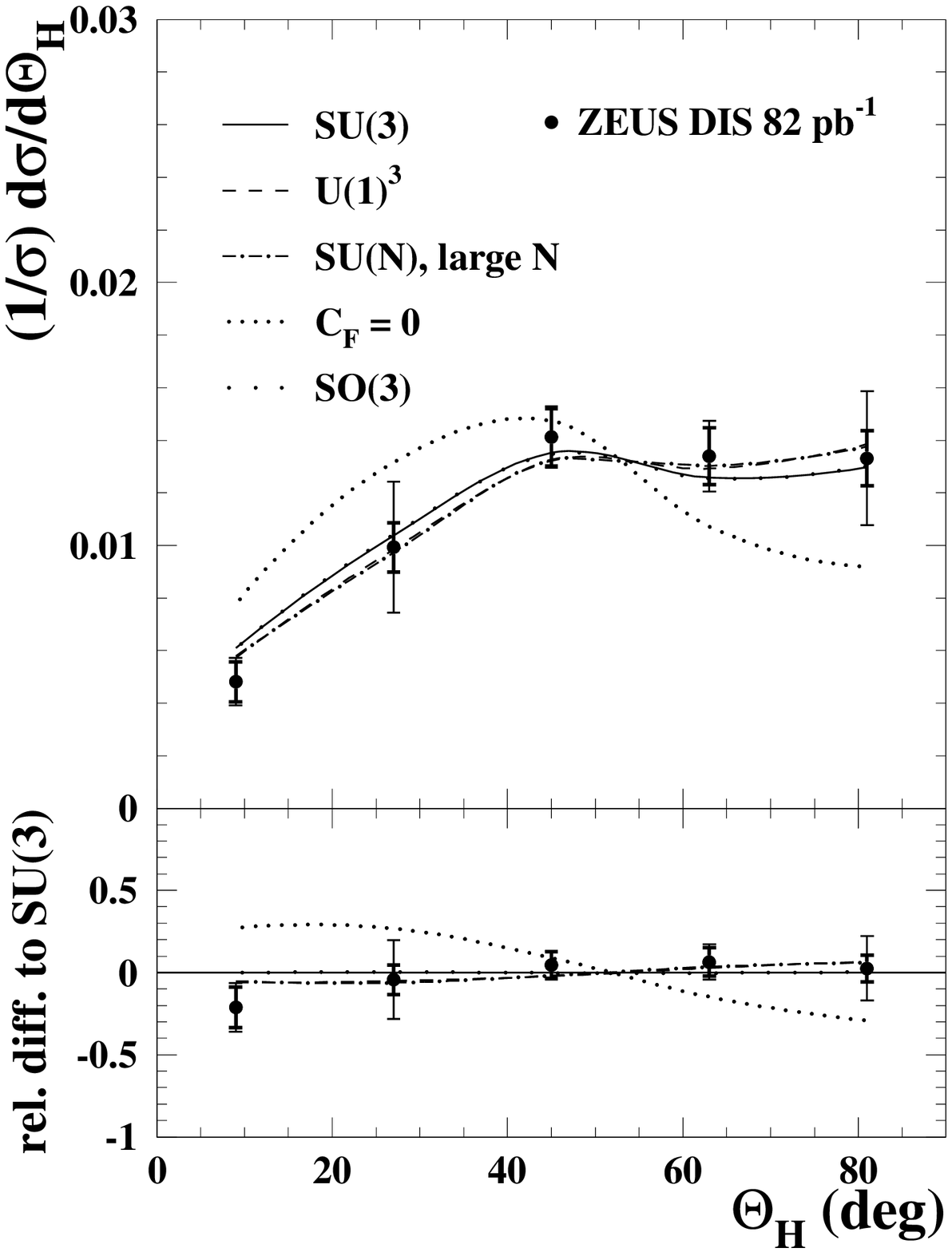,width=10cm}}
\put (6.5,9.5){\epsfig{figure=\figdir 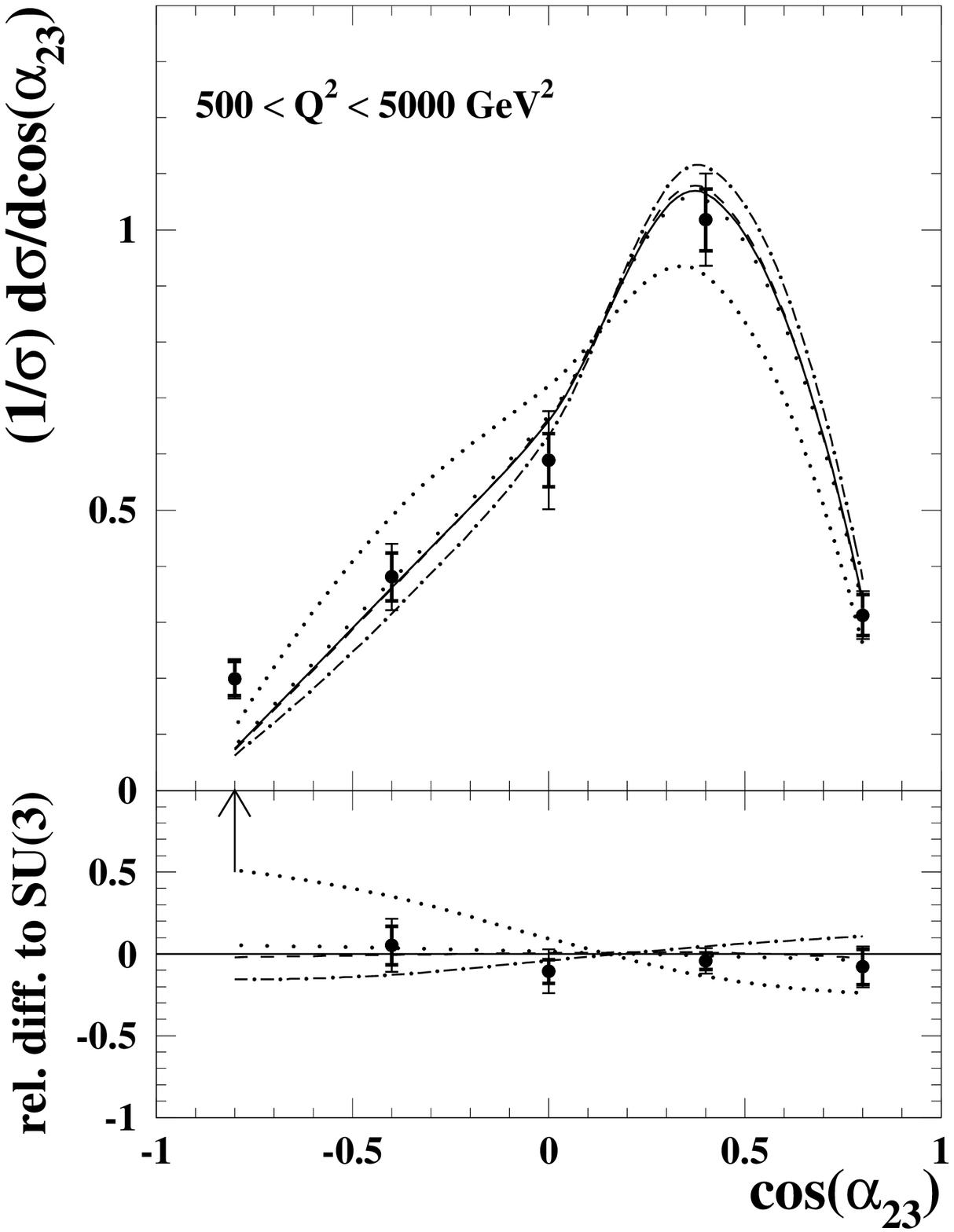,width=10cm}}
\put (-1.0,-0.5){\epsfig{figure=\figdir 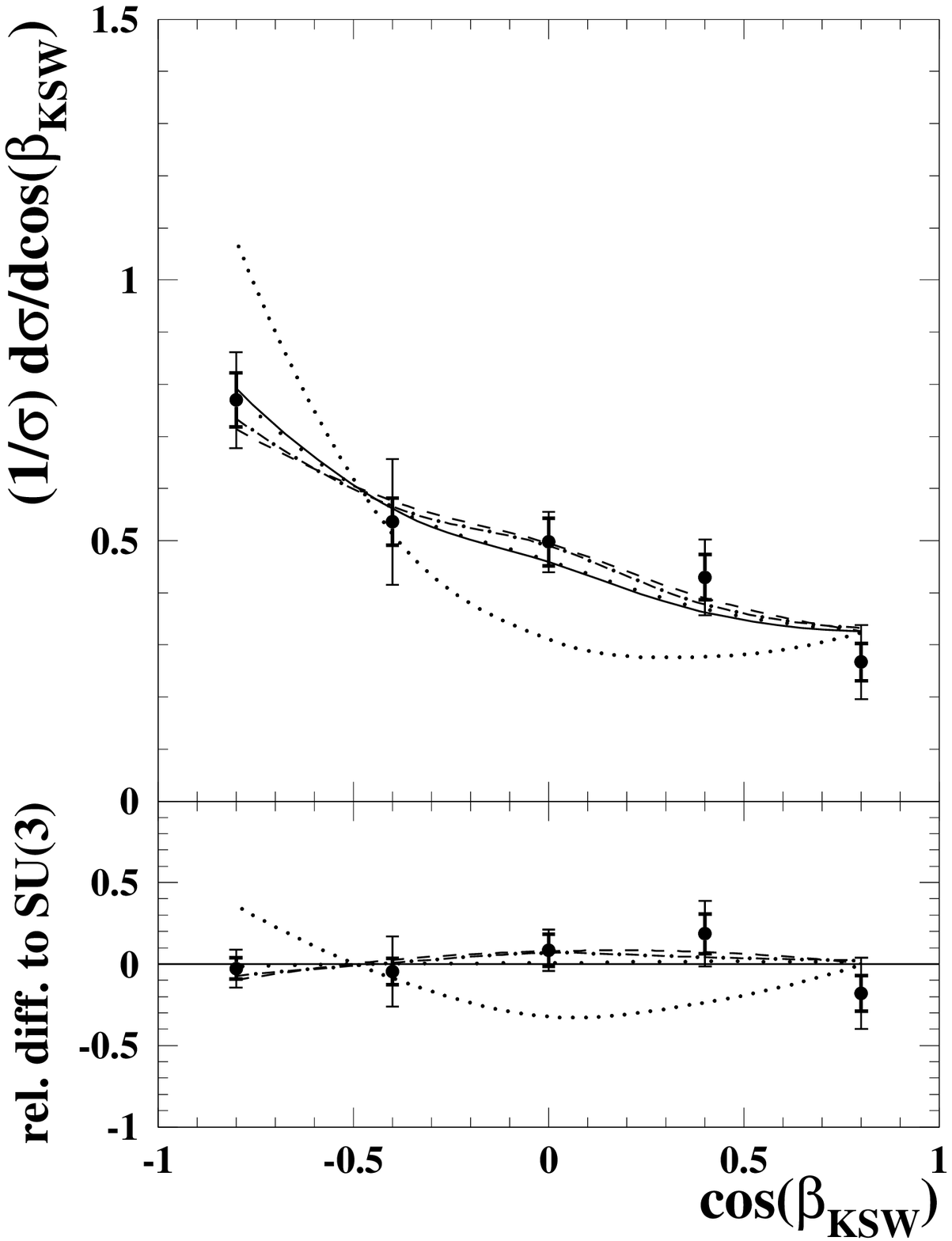,width=10cm}}
\put (6.5,-0.5){\epsfig{figure=\figdir 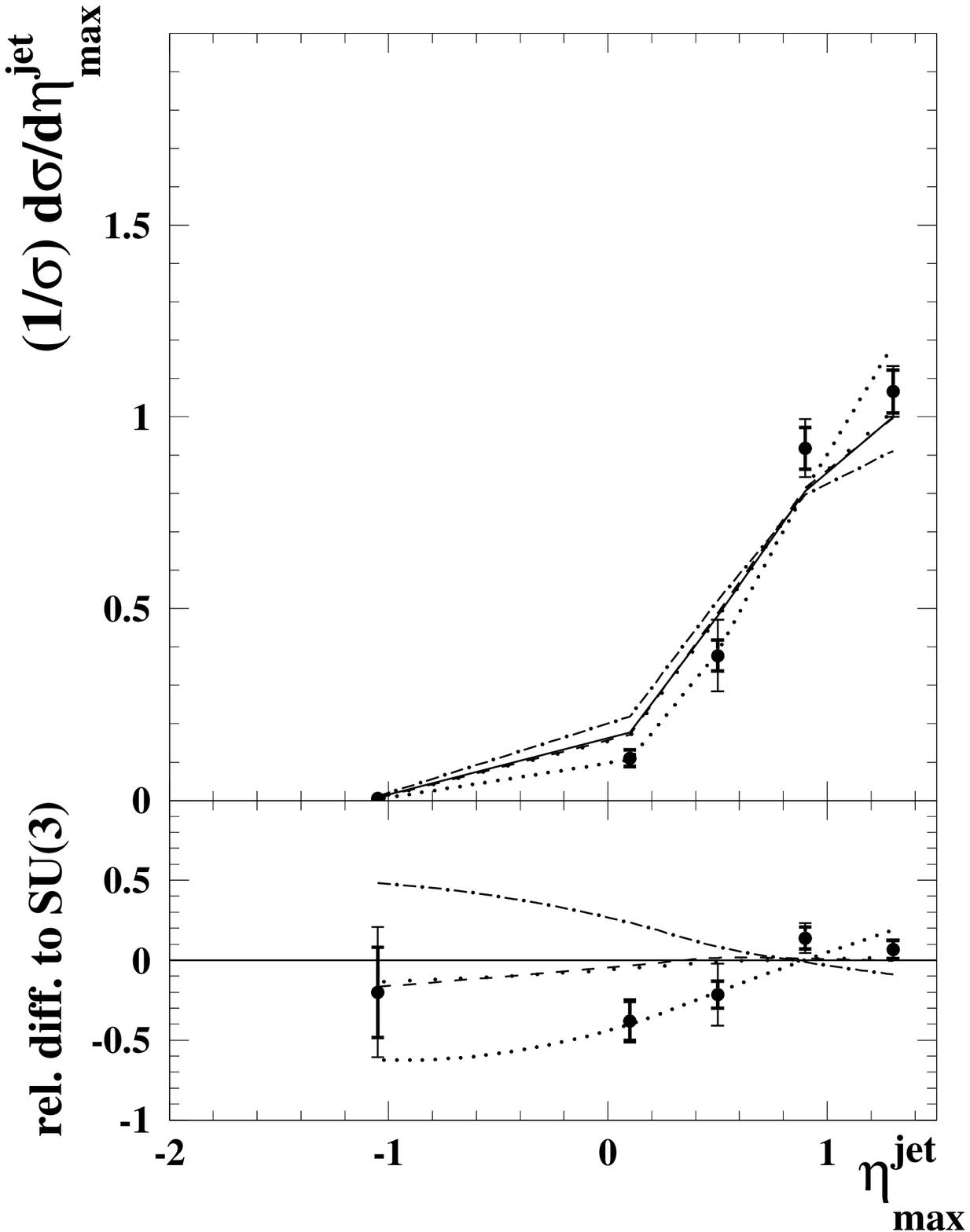,width=10cm}}
\put (6.3,17.5){\bf\small (a)}
\put (13.8,18.0){\bf\small (b)}
\put (6.3,8.0){\bf\small (c)}
\put (13.8,8.0){\bf\small (d)}
\end{picture}
\caption
{\it 
Measured normalised differential $ep$ cross sections for three-jet
production in NC DIS (dots) integrated over $\etjbj>8$~GeV,
$E^{\rm jet2,3}_{T,{\rm B}}>5$ GeV and $-2<\etajb<1.5$ in the
kinematic region given by $500<\q2<5000$~\gf2\ and $|\cgh|<0.65$
as functions of (a) $\th$, (b) $\cos(\a34)$, (c) $\cos(\pksw)$ and (d)
$\etajmax$. Other details as in the caption to Fig.~\ref{fig17}.
}
\label{fig19}
\vfill
\end{figure}

\begin{figure}[p]
\vfill
\setlength{\unitlength}{1.0cm}
\begin{picture} (18.0,17.0)
\put (-0.3,11.0){\centerline{\epsfig{figure=\figdir zeus.eps,width=10cm}}}
\put (-1.0,9.5){\epsfig{figure=\figdir 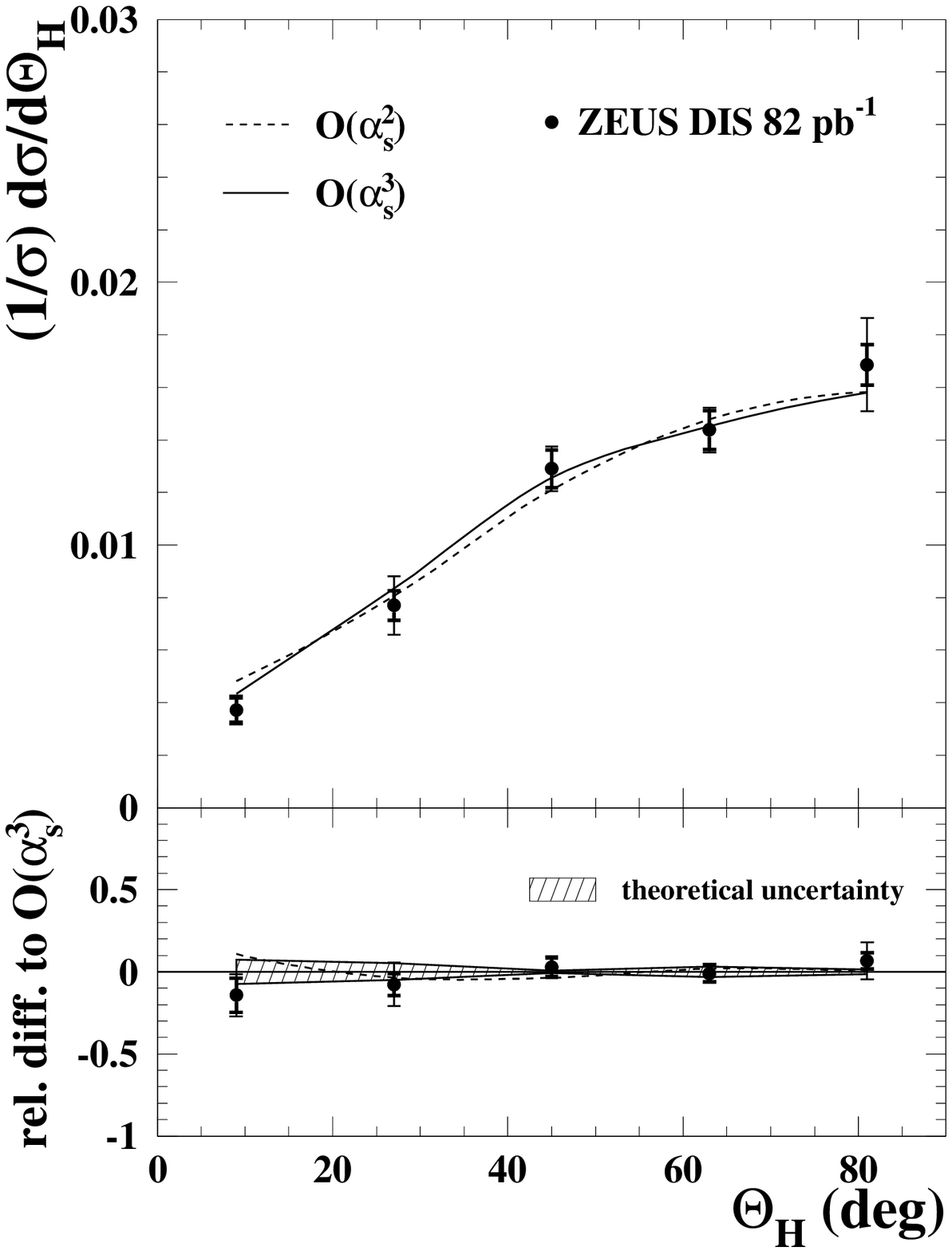,width=10cm}}
\put (6.5,9.5){\epsfig{figure=\figdir 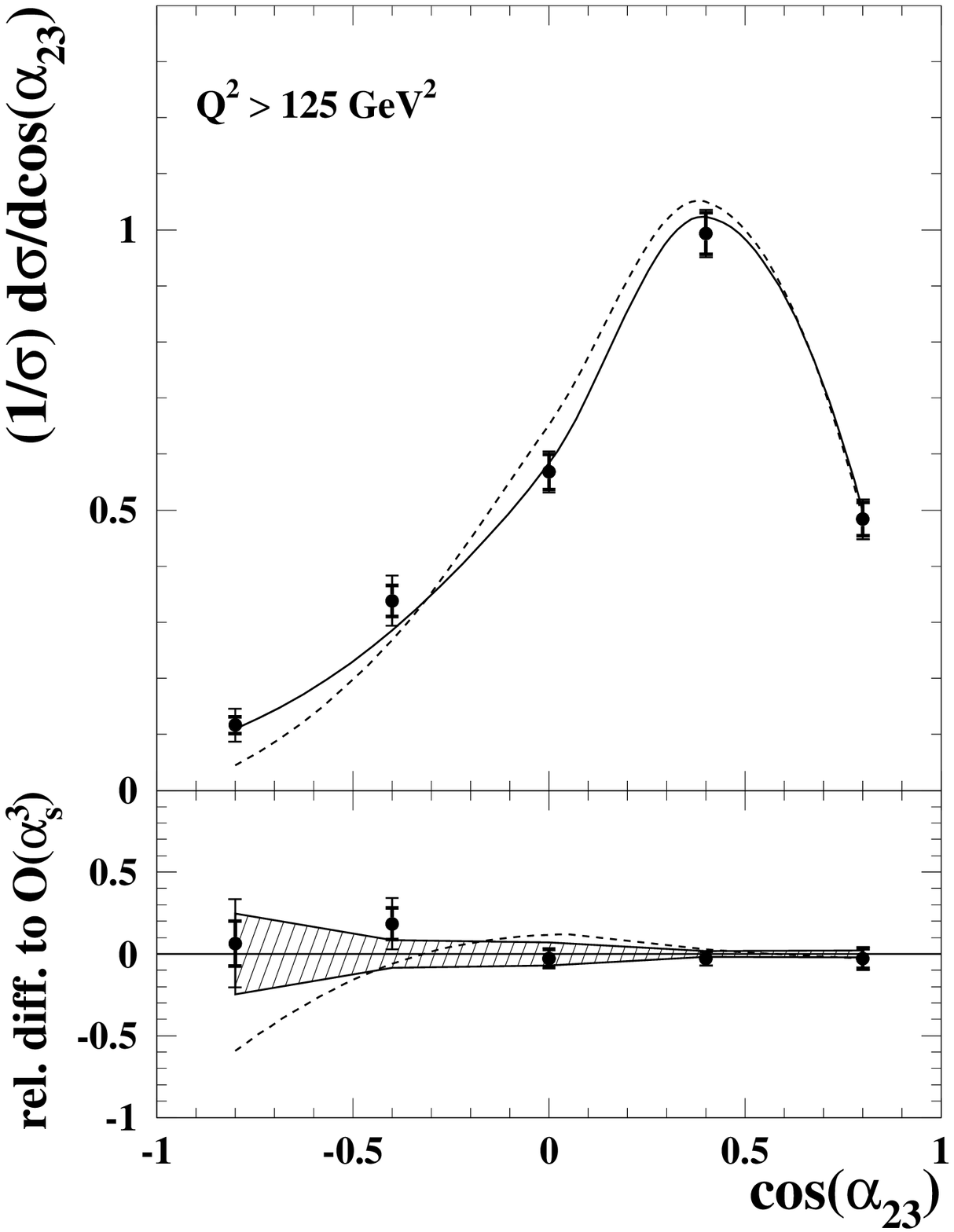,width=10cm}}
\put (-1.0,-0.5){\epsfig{figure=\figdir 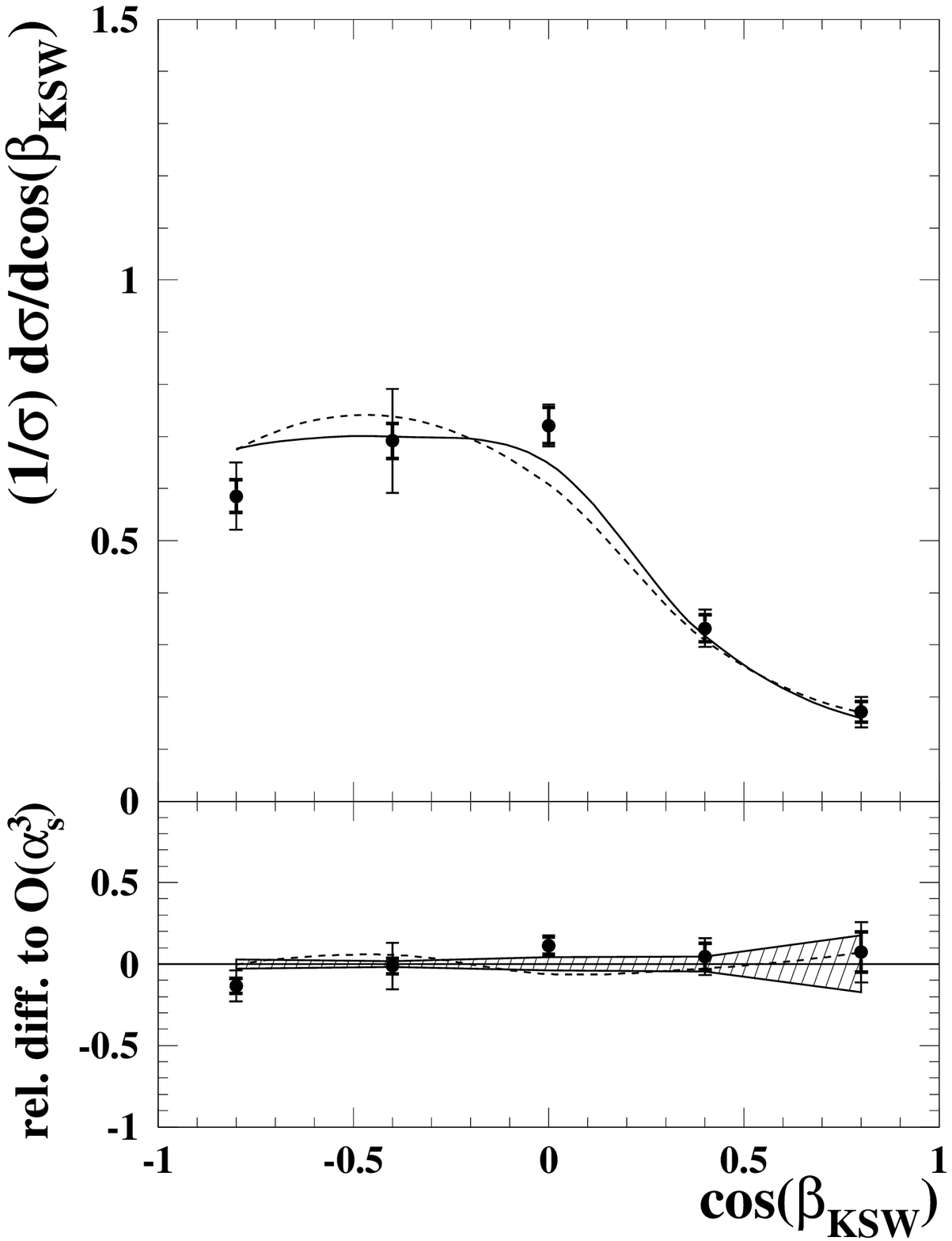,width=10cm}}
\put (6.5,-0.5){\epsfig{figure=\figdir 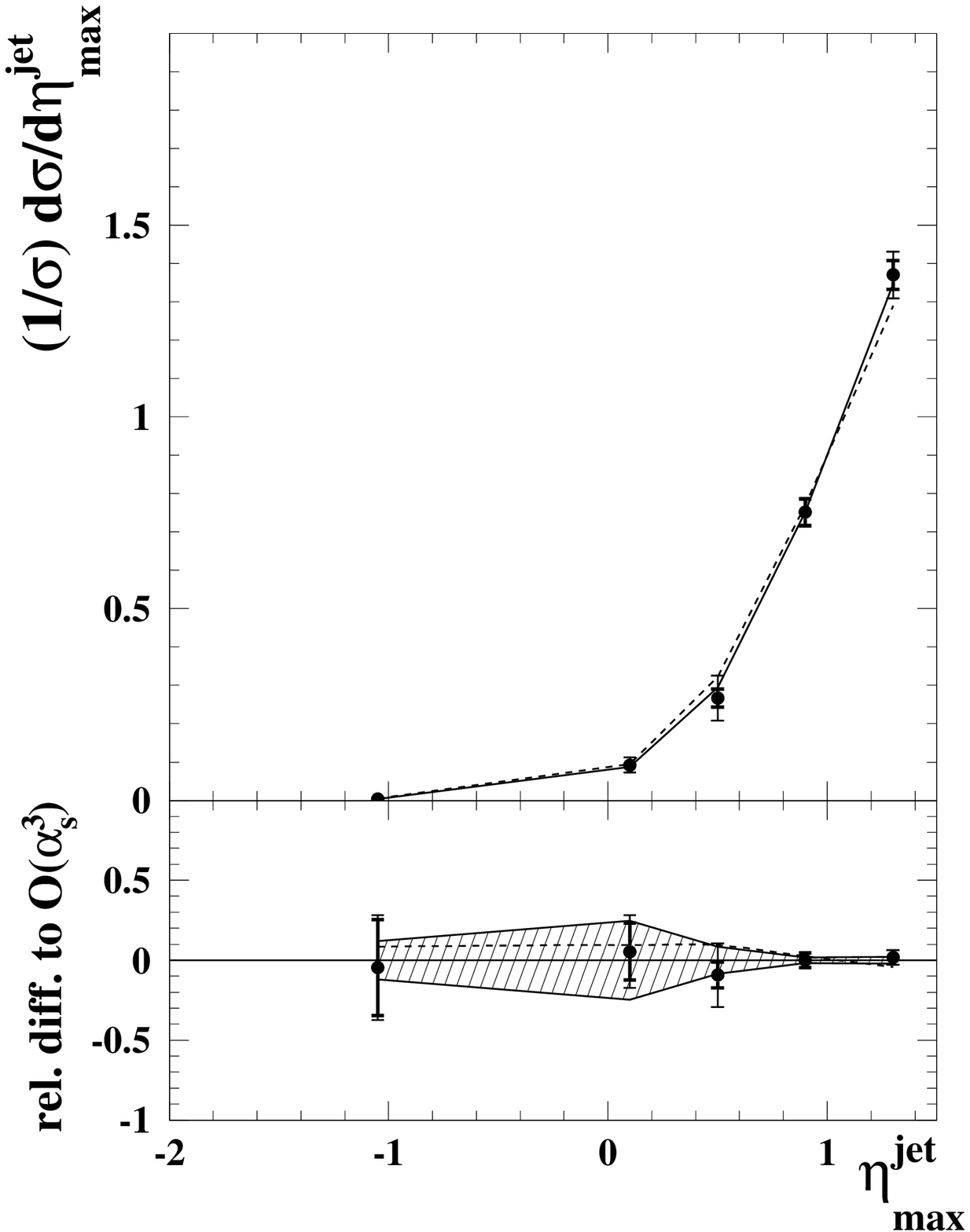,width=10cm}}
\put (6.3,17.5){\bf\small (a)}
\put (13.8,18.0){\bf\small (b)}
\put (6.3,8.0){\bf\small (c)}
\put (13.8,8.0){\bf\small (d)}
\end{picture}
\caption
{\it 
Measured normalised differential $ep$ cross sections for three-jet
production in NC DIS (dots) integrated over $\etjbj>8$~GeV,
$E^{\rm jet2,3}_{T,{\rm B}}>5$ GeV and $-2<\etajb<1.5$ in the
kinematic region given by $\q2>125$~\gf2\ and $|\cgh|<0.65$
as functions of (a) $\th$, (b) $\cos(\a34)$, (c) $\cos(\pksw)$ and (d)
$\etajmax$. For comparison, the $\oass$ (dashed lines) and $\oasss$
(solid lines) QCD calculations are also included. The hatched band
displays the relative theoretical uncertainty of the $\oasss$ 
calculation. Other details as in the caption to Fig.~\ref{fig17}.
}
\label{fig20}
\vfill
\end{figure}

\begin{figure}[p]
\vfill
\setlength{\unitlength}{1.0cm}
\begin{picture} (18.0,17.0)
\put (-0.3,11.0){\centerline{\epsfig{figure=\figdir zeus.eps,width=10cm}}}
\put (-1.0,9.5){\epsfig{figure=\figdir 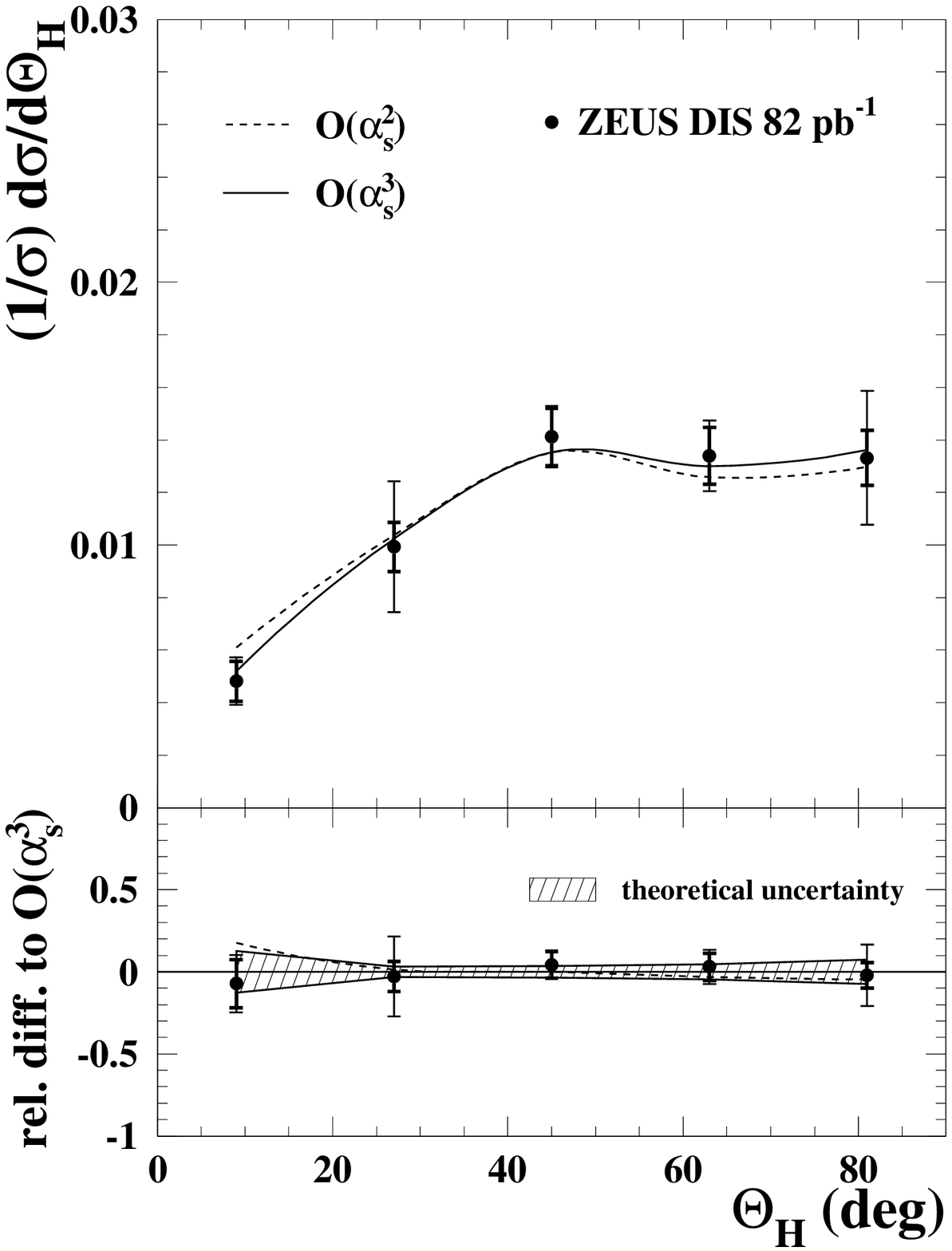,width=10cm}}
\put (6.5,9.5){\epsfig{figure=\figdir 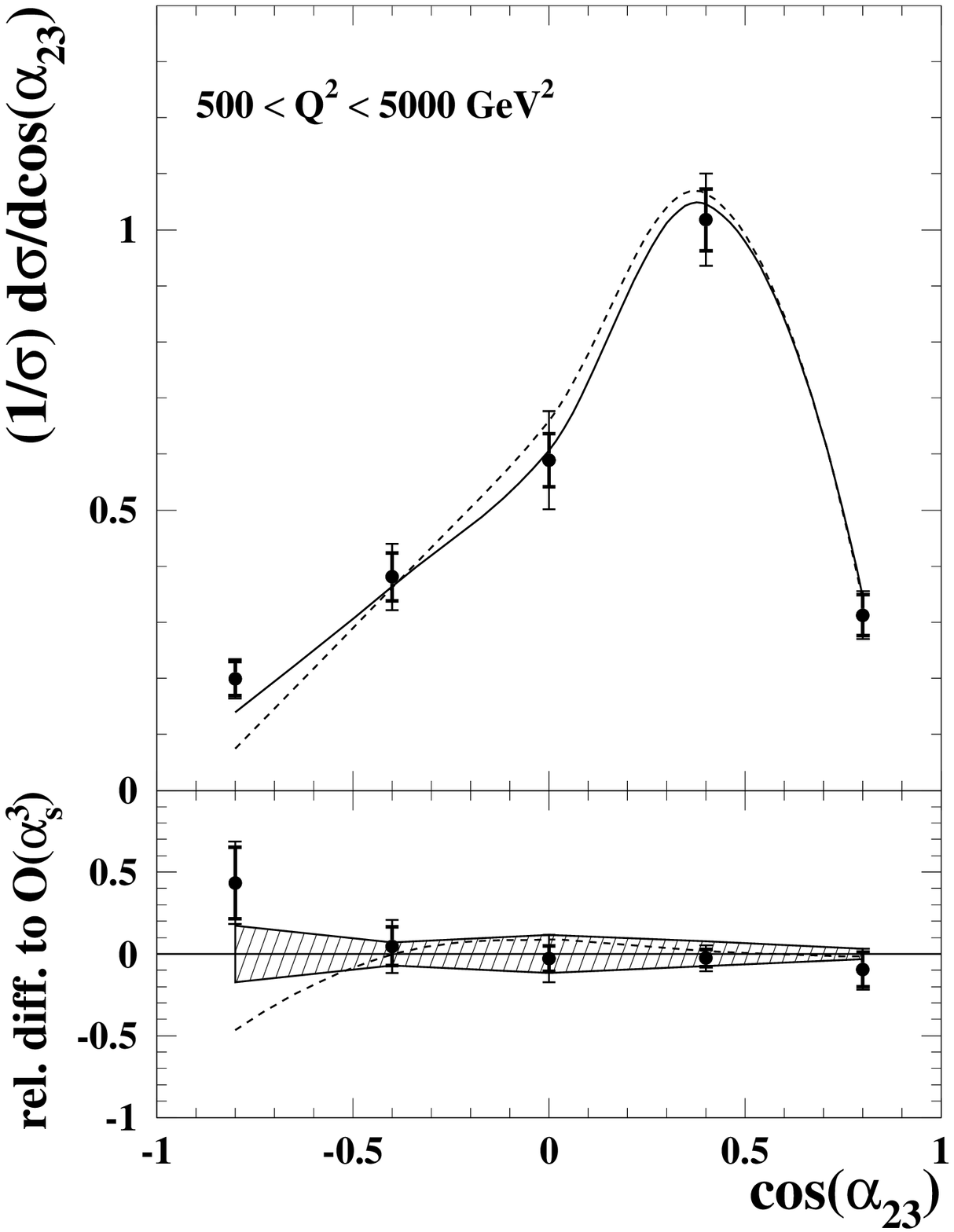,width=10cm}}
\put (-1.0,-0.5){\epsfig{figure=\figdir 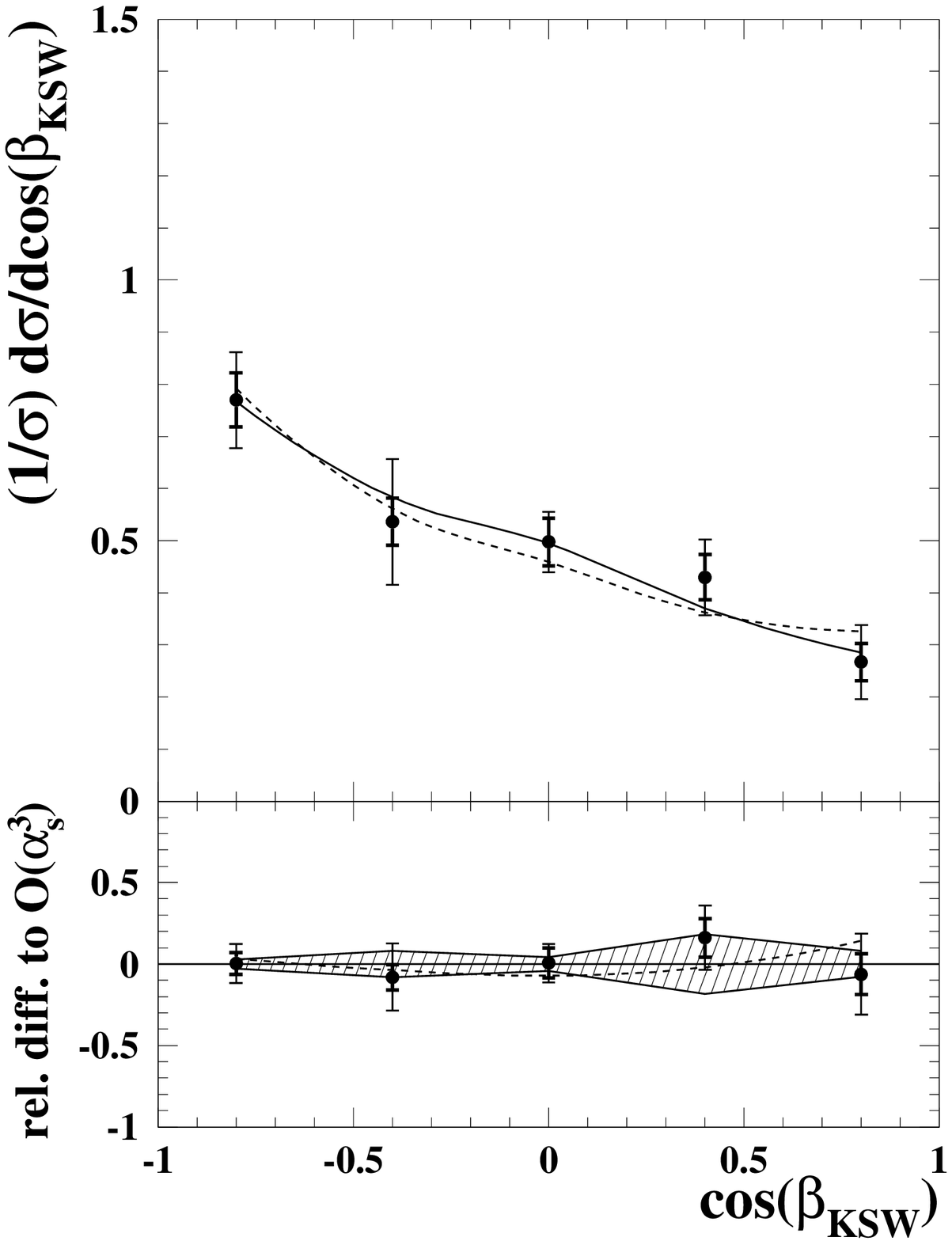,width=10cm}}
\put (6.5,-0.5){\epsfig{figure=\figdir 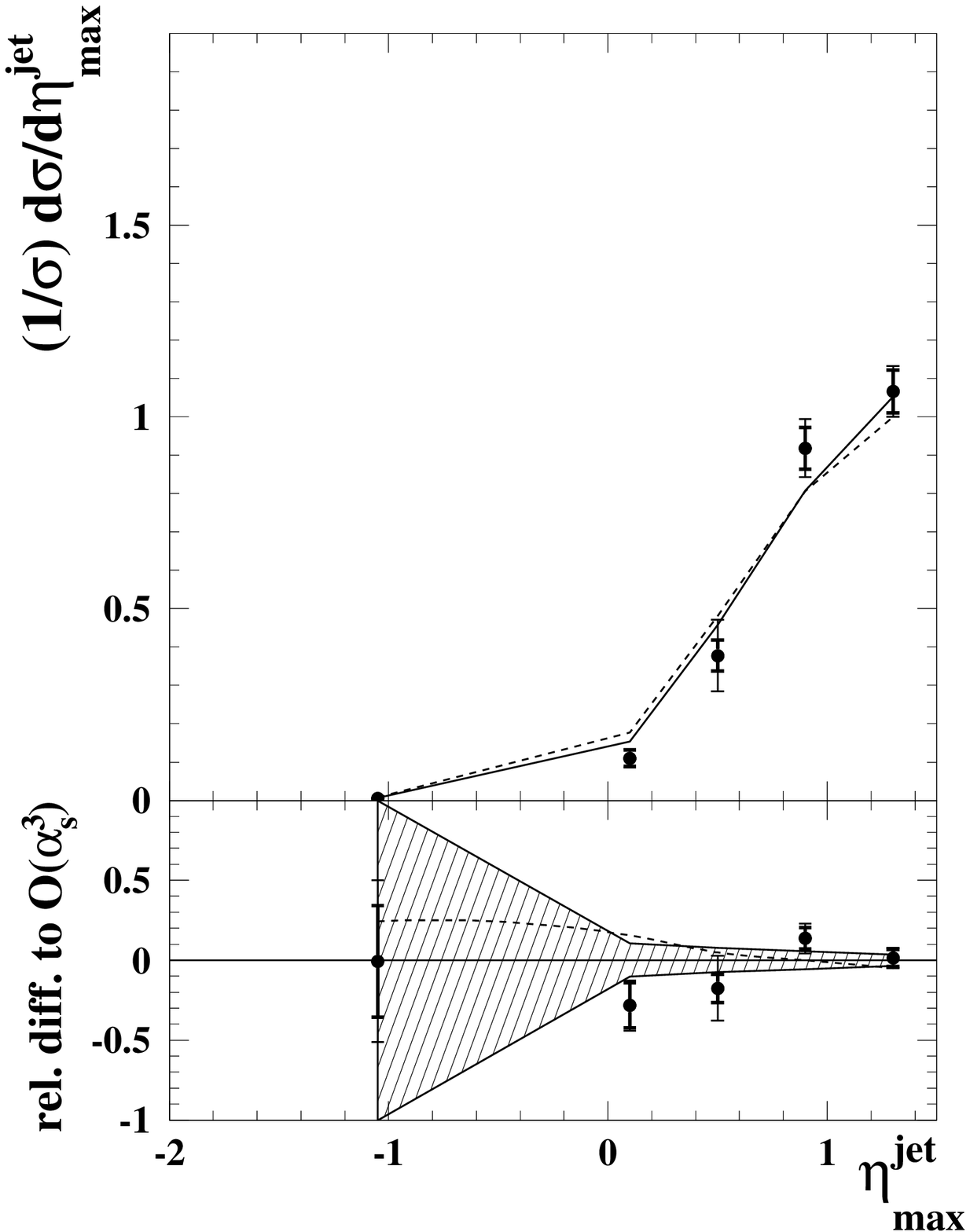,width=10cm}}
\put (6.3,17.5){\bf\small (a)}
\put (13.8,18.0){\bf\small (b)}
\put (6.3,8.0){\bf\small (c)}
\put (13.8,8.0){\bf\small (d)}
\end{picture}
\caption
{\it 
Measured normalised differential $ep$ cross sections for three-jet
production in NC DIS (dots) integrated over $\etjbj>8$~GeV,
$E^{\rm jet2,3}_{T,{\rm B}}>5$ GeV and $-2<\etajb<1.5$ in the
kinematic region given by $500<\q2<5000$~\gf2\ and $|\cgh|<0.65$
as functions of (a) $\th$, (b) $\cos(\a34)$, (c) $\cos(\pksw)$ and (d)
$\etajmax$. For comparison, the $\oass$ (dashed lines) and $\oasss$
(solid lines) QCD calculations are also included. The hatched band
displays the relative theoretical uncertainty of the $\oasss$ 
calculation. Other details as in the caption to Fig.~\ref{fig17}.
}
\label{fig21}
\vfill
\end{figure}

Calculations at $\oass$ in which each colour contribution in Eq. (1)
was weighted according to the colour factors predicted by SU(3)
($C_F=4/3$, $C_A=3$ and $T_F=1/2$) are compared to the measurements in
Figs.~\ref{fig17} to \ref{fig21}. The theoretical uncertainties are
shown in Figs.~\ref{fig17}, \ref{fig20} and \ref{fig21} as hatched
bands. Since the calculations are normalised to unity, the
uncertainties are correlated among the points; this correlation is
partially responsible for the pulsating pattern exhibited by the
theoretical uncertainties. The predictions based on SU(3) give a
reasonable description of the data for all angular correlations. For
$\gp$, the predictions do not include resolved processes (see
Section~\ref{nlo}), as calculations separated according to the
different colour factors are not available. Monte Carlo simulations of
such processes show that their contribution is most likely to be
different from that of direct processes in the fifth and last bin of
$(1/\sigma)(d\sigma/d\cos(\a34))$ (see Figs.~\ref{fig4}b and
\ref{fig17}b).

To illustrate the sensitivity of the measurements to the colour
factors, calculations based on different symmetry groups are also
compared to the data in Figs.~\ref{fig17} to \ref{fig19}. In these
calculations, the colour components were combined in such a way as
to reproduce the colour structure of a theory based on the non-Abelian
group SU($N$) in the limit of large $N$ ($C_F=1$, $C_A=2$ and 
$T_F=0$), the Abelian group U(1)$^3$ ($C_F=1$, $C_A=0$ and $T_F=3$),
the non-Abelian group SO(3) ($C_F=1/3$, $C_A=3$ and $T_F=1/3$) and, as
an extreme choice, a calculation with $C_F=0$. The shapes of the
distributions predicted by U(1)$^3$ in $\gp$ are very similar to those
by SU(3) due to the smallness of the component $\sigma_B$ and the
difficulty to distinguish the component $\sigma_D$. In NC DIS, the
predictions of U(1)$^3$ show differences of around $10\%$ with respect
to those of SU(3), which are of the same order as the statistical
uncertainties. In both regimes, the data clearly disfavour a theory
based on SU($N$) in the limit of large $N$ or on $C_F=0$.

Figures~\ref{fig20} and \ref{fig21} show the measurements in NC DIS
compared to the predictions of QCD at $\oass$ and $\oasss$. This
comparison provides a very stringent test of pQCD. The $\oasss$
calculations give a very good description of the data. In particular,
a significant improvement in the description of the data can be
observed for the first bin of the $\a34$ distribution
(Figs.~\ref{fig20}b and \ref{fig21}b).

\section{Summary and conclusions}

Measurements of angular correlations in three-jet $\gp$ and NC DIS
were performed in $ep$ collisions at HERA using $127$~\pb1\ of data
collected with the ZEUS detector. The cross sections refer to jets
identified with the $\kt$ cluster algorithm in the longitudinally
invariant inclusive mode and selected with $\etjet>14$~GeV and $\etar$
($\gp$) and $\etjbj>8$~GeV, $E^{\rm jet2,3}_{T,{\rm B}}>5$ GeV and
$-2<\etajb<1.5$ (NC DIS). The measurements were made in the kinematic
regions defined by $\q2<1$~\g2, $0.2<y<0.85$ and $\xo>0.8$ ($\gp$)
and $\q2>125$~\g2\ or $500<\q2<5000$~\g2\ and $|\cgh|<0.65$ (NC DIS).
Normalised differential three-jet cross sections were measured as
functions of $\th$, $\a34$, $\pksw$ and $\etajmax$. 

The colour configuration of the strong interaction was studied for the
first time in $ep$ collisions using the angular correlations in
three-jet events. While the extraction of the colour factors will
require the full analysis of all HERA data and complete $\oasss$
calculations, the studies presented in this paper demonstrate the
potential of the method.

Fixed-order calculations separated according to the colour
configurations were used to study the sensitivity of the angular
correlations to the underlying gauge structure. The predicted
distributions of $\th$, $\a34$ and $\pksw$ clearly isolate the
contribution from the triple-gluon coupling in quark-induced processes
while $\etajmax$ isolates the contribution from gluon-induced
processes. The variable $\a34$ provides additional separation for the
other contributions. Furthermore, the studies performed demonstrate
that normalised cross sections in three-jet $ep$ collisions have
reduced sensitivity to the assumed evolution of the PDFs and the
running of $\as$.

The data clearly disfavour theories based on SU($N$) in the limit of
large $N$ or $C_F=0$. Differences between SU(3) and U(1)$^3$ are
smaller than the current statistical uncertainties. The measurements
are found to be consistent with the admixture of colour configurations
as predicted by SU(3). The $\oasss$ calculations give a very good
description of the NC DIS data.

\vspace{0.5cm}
\noindent {\Large\bf Acknowledgements}
\vspace{0.3cm}

We thank the DESY Directorate for their strong support and encouragement.
We appreciate the contributions to the construction and maintenance of 
the ZEUS detector of many people who are not listed as authors. The HERA 
machine group and the DESY computing staff are especially acknowledged 
for their success in providing excellent operation of the collider and 
the data-analysis environment.
We would like to thank M. Fontannaz, M. Klasen and Z. Nagy for useful
discussions.

\vfill\eject

\providecommand{\etal}{et al.\xspace}
\providecommand{\coll}{Collab.\xspace}
\catcode`\@=11
\def\@bibitem#1{%
\ifmc@bstsupport
  \mc@iftail{#1}%
    {;\newline\ignorespaces}%
    {\ifmc@first\else.\fi\orig@bibitem{#1}}
  \mc@firstfalse
\else
  \mc@iftail{#1}%
    {\ignorespaces}%
    {\orig@bibitem{#1}}%
\fi}%
\catcode`\@=12

\clearpage
\newpage
 \begin{table}
 \begin{center}
     \begin{tabular}{||c|ccc||c||}
 \hline
   $\th$ bin (deg)
 & $(1/\sigma)\ d\sigma/d\th$ 
 & $\delta_{\rm stat}$
 & $\delta_{\rm syst}$
 & $C_{\rm had}$\\
 \hline
 \hline
  0, 9
 &  0.00264
 &  0.00038
 & $\pm 0.00052$
 &  0.93
 \\
  9, 18
 &  0.00393
 &  0.00044
 & $\pm 0.00021$
 &  0.94
 \\
  18, 27
 &  0.00507
 &  0.00051
 & ${}_{-  0.00039}^{+  0.00040}$
 &  1.00
 \\
  27, 36
 &  0.00838
 &  0.00064
 & ${}_{-  0.00104}^{+  0.00105}$
 &  0.93
 \\
  36, 45
 &  0.01071
 &  0.00075
 & $\pm 0.00023$
 &  0.96
 \\
  45, 54
 &  0.01486
 &  0.00087
 & ${}_{-  0.00016}^{+  0.00021}$
 &  0.94
 \\
  54, 63
 &  0.01795
 &  0.00098
 & ${}_{-  0.00035}^{+  0.00036}$
 &  0.95
 \\
  63, 72
 &  0.01765
 &  0.00095
 & $\pm 0.00062$
 &  0.94
 \\
  72, 81
 &  0.01517
 &  0.00088
 & ${}_{-  0.00084}^{+  0.00081}$
 &  0.94
 \\
  81, 90
 &  0.01473
 &  0.00086
 & ${}_{-  0.00077}^{+  0.00075}$
 &  0.96
 \\
 \hline

     \end{tabular}
 \caption{
   Normalised differential $ep$ cross section for three-jet photoproduction
   integrated over $\etjet>14$ GeV and $\etar$ in the kinematic region
   defined by $\q2<1$~\gf2, $0.2<y<0.85$ and $\xo>0.8$ as a function
   of $\th$. The statistical and systematic uncertainties are shown
   separately. The multiplicative corrections for hadronisation
   effects to be applied to the parton-level QCD differential cross
   section, $C_{\rm had}$, are shown in the last column.}
 \label{tabone}
\end{center}
\end{table}

 \begin{table}
 \begin{center}
     \begin{tabular}{||c|ccc||c||}
 \hline
   $\cos(\a34)$ bin
 & $(1/\sigma)\ d\sigma/d\cos(\a34)$
 & $\delta_{\rm stat}$
 & $\delta_{\rm syst}$
 & $C_{\rm had}$\\
 \hline
 \hline
  -1, -0.8
 &  0.0138
 &  0.0046
 & $\pm 0.00042$
 &  1.04
 \\
  -0.8, -0.6
 &  0.078
 &  0.012
 & ${}_{-  0.003}^{+  0.004}$
 &  0.96
 \\
  -0.6, -0.4
 &  0.198
 &  0.022
 & ${}_{-  0.027}^{+  0.026}$
 &  0.95
 \\
  -0.4, -0.2
 &  0.343
 &  0.029
 & ${}_{-  0.040}^{+  0.041}$
 &  0.93
 \\
  -0.2, 0
 &  0.360
 &  0.029
 & $\pm 0.010$
 &  0.97
 \\
  0, 0.2
 &  0.512
 &  0.034
 & ${}_{-  0.013}^{+  0.014}$
 &  0.98
 \\
  0.2, 0.4
 &  0.618
 &  0.037
 & ${}_{-  0.016}^{+  0.015}$
 &  1.00
 \\
  0.4, 0.6
 &  0.847
 &  0.044
 & $\pm 0.013$
 &  0.99
 \\
  0.6, 0.8
 &  0.937
 &  0.045
 & ${}_{-  0.042}^{+  0.043}$
 &  0.99
 \\
  0.8, 1
 &  1.092
 &  0.049
 & ${}_{-  0.018}^{+  0.019}$
 &  1.02
 \\
 \hline

     \end{tabular}
 \caption{
   Normalised differential $ep$ cross section for three-jet photoproduction
   integrated over $\etjet>14$ GeV and $\etar$ in the kinematic region
   defined by $\q2<1$~\gf2, $0.2<y<0.85$ and $\xo>0.8$ as a function
   of $\cos(\a34)$. Other details as in the caption to
   Table~\ref{tabone}.}
 \label{tabtwo}
\end{center}
\end{table}

 \begin{table}
 \begin{center}
     \begin{tabular}{||c|ccc||c||}
 \hline
   $\cos(\pksw)$ bin
 & $(1/\sigma)\ d\sigma/d\cos(\pksw)$ 
 & $\delta_{\rm stat}$
 & $\delta_{\rm syst}$
 & $C_{\rm had}$\\
 \hline
 \hline
  -1, -0.8
 &  0.552
 &  0.035
 & $\pm 0.044$
 &  0.97
 \\
  -0.8, -0.6
 &  0.651
 &  0.039
 & $\pm 0.026$
 &  0.99
 \\
  -0.6, -0.4
 &  0.745
 &  0.042
 & ${}_{-  0.031}^{+  0.032}$
 &  0.97
 \\
  -0.4, -0.2
 &  0.741
 &  0.042
 & $\pm  0.039$
 &  0.93
 \\
  -0.2, 0
 &  0.784
 &  0.042
 & ${}_{-  0.016}^{+  0.014}$
 &  0.96
 \\
  0, 0.2
 &  0.768
 &  0.042
 & $\pm 0.046$
 &  0.95
 \\
  0.2, 0.4
 &  0.500
 &  0.034
 & $\pm 0.005$
 &  0.94
 \\
  0.4, 0.6
 &  0.200
 &  0.022
 & $\pm 0.021$
 &  0.95
 \\
  0.6, 0.8
 &  0.056
 &  0.010
 & ${}_{-  0.009}^{+  0.010}$
 &  0.85
 \\
  0.8, 1
 &  0.0029
 &  0.0015
 & $\pm  0.0037$
 &  0.74
 \\
 \hline

     \end{tabular}
 \caption{
   Normalised differential $ep$ cross section for three-jet photoproduction
   integrated over $\etjet>14$ GeV and $\etar$ in the kinematic region
   defined by $\q2<1$~\gf2, $0.2<y<0.85$ and $\xo>0.8$ as a function
   of $\cos(\pksw)$. Other details as in the caption to
   Table~\ref{tabone}.}
 \label{tabthree}
\end{center}
\end{table}

 \begin{table}
 \begin{center}
     \begin{tabular}{||c|ccc||c||c||}
 \hline
   $\th$ bin (deg)
 & $(1/\sigma)\ d\sigma/d\th$ 
 & $\delta_{\rm stat}$
 & $\delta_{\rm syst}$
 & $C_{\rm QED}$
 & $C_{\rm had}$\\
 \hline
 \multicolumn{6}{||c||}{$Q^2>125$ GeV$^2$} \\
 \hline
  0, 18
 &  0.00372
 &  0.00046
 & $\pm 0.00031$
 &  0.92
 &  0.89
 \\
  18, 36
 &  0.00770
 &  0.00056
 & $\pm 0.00095$
 &  0.88
 &  0.90
 \\
  36, 54
 &  0.01291
 &  0.00072
 & $\pm 0.00045$
 &  0.96
 &  0.84
 \\
  54, 72
 &  0.01438
 &  0.00074
 & $\pm 0.00042$
 &  1.00
 &  0.84
 \\
  72, 90
 &  0.01686
 &  0.00077
 & $\pm 0.00160$
 &  0.99
 &  0.84
 \\
 \hline
 \multicolumn{6}{||c||}{$500<Q^2<5000$ GeV$^2$} \\
 \hline

  0, 18
 &  0.00481
 &  0.00076
 & $\pm 0.00048$
 &  0.88
 &  0.92
 \\
  18, 36
 &  0.00993
 &  0.00094
 & $\pm 0.00231$
 &  0.95
 &  0.96
 \\
  36, 54
 &  0.0141
 &  0.0011
 & $\pm 0.0004$
 &  0.92
 &  0.97
 \\
  54, 72
 &  0.0134
 &  0.0011
 & $\pm 0.0008$
 &  1.03
 &  0.89
 \\
  72, 90
 &  0.0133
 &  0.0011
 & $\pm 0.0023$
 &  0.96
 &  0.94
 \\
 \hline

     \end{tabular}
 \caption{
   Normalised differential $ep$ cross section for three-jet 
   production in NC DIS integrated over $\etjbj>8$~GeV,
   $E^{\rm jet2,3}_{T,{\rm B}}>5$ GeV and $-2<\etajb<1.5$ in the
   kinematic region given by $|\cgh|<0.65$ and $\q2>125$~\gf2\ or
   $500<\q2<5000$~\gf2\ as a function of $\th$. The multiplicative
   corrections applied to the differential measured cross section to
   correct for QED radiative effects, $C_{\rm QED}$, is also
   shown.  The multiplicative corrections for hadronisation effects
   and the $\z0$-exchange contribution to be applied to the
   parton-level QCD differential cross section, $C_{\rm had}$, are
   shown in the last column. Other details as in the caption to
   Table~\ref{tabone}.}
 \label{tabfour}
\end{center}
\end{table}

 \begin{table}
 \begin{center}
     \begin{tabular}{||c|ccc||c||c||}
 \hline
   $\cos(\a34)$ bin
 & $(1/\sigma)\ d\sigma/d\cos(\a34)$ 
 & $\delta_{\rm stat}$
 & $\delta_{\rm syst}$
 & $C_{\rm QED}$
 & $C_{\rm had}$\\
 \hline
 \multicolumn{6}{||c||}{$Q^2>125$ GeV$^2$} \\
 \hline
  -1, -0.6
 &  0.117
 &  0.015
 & $\pm 0.025$
 &  0.96
 &  0.90
 \\
  -0.6, -0.2
 &  0.338
 &  0.028
 & $\pm 0.035$
 &  1.01
 &  0.70
 \\
  -0.2, 0.2
 &  0.568
 &  0.032
 & $\pm 0.018$
 &  0.90
 &  0.78
 \\
  0.2, 0.6
 &  0.993
 &  0.037
 & $\pm 0.021$
 &  0.95
 &  0.88
 \\
  0.6, 1
 &  0.484
 &  0.030
 & $\pm 0.020$
 &  1.02
 &  1.01
 \\
 \hline
 \multicolumn{6}{||c||}{$500<Q^2<5000$ GeV$^2$} \\
 \hline
 -1, -0.6
 &  0.199
 &  0.030
 & $\pm 0.018$
 &  1.04
 &  0.83
 \\
 -0.6, -0.2
 &  0.381
 &  0.043
 & $\pm 0.041$
 &  0.97
 &  0.75
 \\
 -0.2,  0.2
 &  0.589
 &  0.047
 & $\pm 0.074$
 &  0.92
 &  0.83
 \\
  0.2,  0.6
 &  1.018
 &  0.055
 & $\pm 0.061$
 &  0.95
 &  1.07
 \\
  0.6,  1
 &  0.313
 &  0.036
 & $\pm 0.022$
 &  0.97
 &  1.16
 \\
 \hline
     \end{tabular}
 \caption{
   Normalised differential $ep$ cross section for three-jet 
   production in NC DIS integrated over $\etjbj>8$~GeV,
   $E^{\rm jet2,3}_{T,{\rm B}}>5$ GeV and $-2<\etajb<1.5$ in the
   kinematic region given by $|\cgh|<0.65$ and $\q2>125$~\gf2\ or
   $500<\q2<5000$~\gf2\ as a function of $\cos(\a34)$. Other details
   as in the caption to Table~\ref{tabfour}.}
 \label{tabfive}
\end{center}
\end{table}

 \begin{table}
 \begin{center}
     \begin{tabular}{||c|ccc||c||c||}
 \hline
   $\cos(\pksw)$ bin
 & $(1/\sigma)\ d\sigma/d\cos(\pksw)$ 
 & $\delta_{\rm stat}$
 & $\delta_{\rm syst}$
 & $C_{\rm QED}$
 & $C_{\rm had}$\\
 \hline
 \multicolumn{6}{||c||}{$Q^2>125$ GeV$^2$} \\
 \hline
 -1, -0.6
 &  0.585
 &  0.031
 & $\pm 0.057$
 &  0.92
 &  0.95
 \\
 -0.6, -0.2
 &  0.691
 &  0.034
 & $\pm 0.094$
 &  0.99
 &  0.88
 \\
 -0.2, 0.2
 &  0.721
 &  0.035
 & $\pm 0.020$
 &  1.01
 &  0.85
 \\
 0.2, 0.6
 &  0.332
 &  0.026
 & $\pm 0.025$
 &  0.92
 &  0.74
 \\
 0.6, 1
 &  0.171
 &  0.020
 & $\pm 0.022$
 &  0.93
 &  0.71
 \\
 \hline
 \multicolumn{6}{||c||}{$500<Q^2<5000$ GeV$^2$} \\
 \hline
 -1, -0.6
 &  0.770
 &  0.052
 & $\pm 0.076$
 &  0.94
 &  1.04
 \\
 -0.6, -0.2
 &  0.536
 &  0.045
 & $\pm 0.112$
 &  0.93
 &  0.97
 \\
 -0.2,  0.2
 &  0.497
 &  0.045
 & $\pm 0.037$
 &  1.01
 &  0.94
 \\
  0.2,  0.6
 &  0.430
 &  0.044
 & $\pm 0.058$
 &  1.01
 &  0.84
 \\
  0.6,  1
 &  0.267
 &  0.036
 & $\pm 0.061$
 &  0.89
 &  0.78
 \\
 \hline
     \end{tabular}
 \caption{
   Normalised differential $ep$ cross section for three-jet 
   production in NC DIS integrated over $\etjbj>8$~GeV,
   $E^{\rm jet2,3}_{T,{\rm B}}>5$ GeV and $-2<\etajb<1.5$ in the
   kinematic region given by $|\cgh|<0.65$ and $\q2>125$~\gf2\ or
   $500<\q2<5000$~\gf2\ as a function of $\cos(\pksw)$. Other details
   as in the caption to Table~\ref{tabfour}.}
 \label{tabsix}
\end{center}
\end{table}

 \begin{table}
 \begin{center}
     \begin{tabular}{||c|ccc||c||c||}
 \hline
   $\etajmax$ bin
 & $(1/\sigma)\ d\sigma/d\etajmax$ 
 & $\delta_{\rm stat}$
 & $\delta_{\rm syst}$
 & $C_{\rm QED}$
 & $C_{\rm had}$\\
 \hline
 \multicolumn{6}{||c||}{$Q^2>125$ GeV$^2$} \\
 \hline
 -2, -0.1
 &  0.0042
 &  0.0013
 & $\pm 0.0006$
 &  1.07
 &  0.61
 \\
 -0.1, 0.3
 &  0.092
 &  0.016
 & $\pm 0.012$
 &  1.17
 &  0.77
 \\
 0.3, 0.7
 &  0.267
 &  0.024
 & $\pm 0.054$
 &  0.96
 &  0.81
 \\
 0.7, 1.1
 &  0.751
 &  0.034
 & $\pm 0.016$
 &  0.93
 &  0.83
 \\
 1.1, 1.5
 &  1.370
 &  0.038
 & $\pm 0.048$
 &  0.96
 &  0.88
 \\
 \hline
 \multicolumn{6}{||c||}{$500<Q^2<5000$ GeV$^2$} \\
 \hline
 -2, -0.1
 &  0.0059
 &  0.0021
 & $\pm 0.0022$
 &  1.14
 &  0.62
 \\
 -0.1,  0.3
 &  0.110
 &  0.022
 & $\pm 0.011$
 &  0.96
 &  0.77
 \\
  0.3,  0.7
 &  0.378
 &  0.040
 & $\pm 0.084$
 &  0.96
 &  0.86
 \\
  0.7,  1.1
 &  0.918
 &  0.054
 & $\pm 0.052$
 &  0.93
 &  0.93
 \\
  1.1,  1.5
 &  1.066
 &  0.056
 & $\pm 0.035$
 &  0.98
 &  1.00
 \\
 \hline
     \end{tabular}
 \caption{
   Normalised differential $ep$ cross section for three-jet 
   production in NC DIS integrated over $\etjbj>8$~GeV,
   $E^{\rm jet2,3}_{T,{\rm B}}>5$ GeV and $-2<\etajb<1.5$ in the
   kinematic region given by $|\cgh|<0.65$ and $\q2>125$~\gf2\ or
   $500<\q2<5000$~\gf2\ as a function of $\etajmax$. Other details
   as in the caption to Table~\ref{tabfour}.}
 \label{tabseven}
\end{center}
\end{table}

\newpage
\clearpage

\newpage
\clearpage

\newpage
\clearpage

\newpage
\clearpage

\newpage
\clearpage

\newpage
\clearpage

\newpage
\clearpage

\newpage
\clearpage

\newpage
\clearpage

\newpage
\clearpage

\newpage
\clearpage

\newpage
\clearpage

\newpage
\clearpage

\newpage
\clearpage

\newpage
\clearpage

\newpage
\clearpage

\newpage
\clearpage

\newpage
\clearpage

\newpage
\clearpage

\end{document}